\documentclass[a4paper,11pt]{article}
\pdfoutput=1
\usepackage{jheppub}
\usepackage[T1]{fontenc}
\usepackage{longfigure}
\usepackage{graphicx}
\usepackage{adjustbox}
\usepackage{amssymb}
\usepackage{extarrows}
\usepackage{amsmath}
\usepackage{lineno}
\usepackage{subfigure}
\usepackage{booktabs}
\usepackage{pdflscape}
\usepackage{multirow}
\usepackage{bbm}
\usepackage{CJKutf8}
\usepackage{listings}
\usepackage[dvipsnames]{xcolor}
\lstdefinestyle{mystyle}{
  language=Mathematica,
  backgroundcolor=\color{white},
  basicstyle=\ttfamily\small,
  keywordstyle=\color{blue}\bfseries,
  commentstyle=\color{TealBlue},
  stringstyle=\color{gray},
  frame=tb, 
  numbers=left,
  numberstyle=\tiny\color{gray},
  stepnumber=1,
  numbersep=5pt,
  showstringspaces=false,
  tabsize=2,
  breaklines=true,
  breakatwhitespace=false,
  captionpos=b
}
\allowdisplaybreaks

\title{\boldmath Study of the $D_s \to \phi \ell \nu_\ell$ semileptonic decay with (2+1)-flavor lattice QCD 
}

\author{
\begin{center} \includegraphics[scale=0.25]{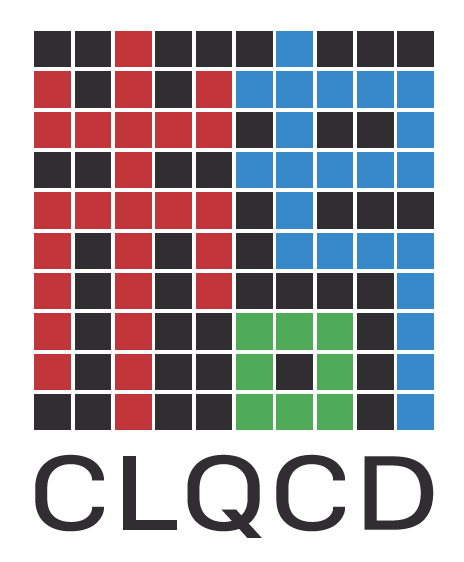} \end{center}}
\author[a,b]{\\Gaofeng Fan,}
\author[c,1]{Yu Meng,\note{Corresponding author.}}
\author[d,e,f]{Chuan Liu,}
\author[a,1]{Zhaofeng Liu,}
\author[a,g]{Tinghong Shen,}
\author[a,h]{Ting-Xiao Wang,}
\author[i]{Ke-Long Zhang,}
\author[b]{Lei Zhang}

\affiliation[a]{Institute of High Energy Physics, Chinese Academy of Sciences,Beijing 100049, China}
\affiliation[b]{School of Physics, Nanjing University, Nanjing 210093, China}
\affiliation[c]{School of Physics, Zhengzhou University, Zhengzhou 450001, China}
\affiliation[d]{School of Physics, Peking University, Beijing 100871, China}
\affiliation[e]{Center for High Energy Physics, Peking University, Beijing 100871, China}
\affiliation[f]{Collaborative Innovation Center of Quantum Matter, Beijing 100871, China}
\affiliation[g]{Hubei Nuclear Solid Physics Key Laboratory, School of Physics and Technology,\\ Wuhan University, Wuhan 430072, China}
\affiliation[h]{School of Physical Sciences, University of Chinese Academy of Sciences, Beijing 101408, China}
\affiliation[i]{Key Laboratory of Quark \& Lepton Physics (MOE) and Institute of Particle Physics,\\ Central China Normal University, Wuhan 430079, China}

\emailAdd{yu\_meng@zzu.edu.cn}
\emailAdd{liuzf@ihep.ac.cn}

\abstract{We present a systematic lattice calculation of the $D_s \to \phi \ell \nu_\ell$ semileptonic decay using (2+1)-flavor Wilson-clover fermion configurations generated by the CLQCD collaboration. Seven gauge ensembles with different lattice spacings, from $0.052~\text{fm}$ to $0.105~\text{fm}$, and different pion masses, from about $210~\text{MeV}$ to $320~\text{MeV}$ are utilized, enabling us to take both the continuum limit and physical pion mass extrapolation. The ratios of form factors are obtained to be $r_V=1.614(19)$ and $r_2=0.741(31)$, with the precision improved by up to
an order of magnitude compared to previous lattice studies. The branching fractions are given as $\mathcal{B}(D_s\to\phi e\nu_e)=2.493(66)_{\text{stat}}(31)_{|V_{cs}|}\times 10^{-2}$ and $\mathcal{B}(D_s\to\phi\mu\nu_\mu)=2.351(60)_{\text{stat}}(29)_{|V_{cs}|}\times 10^{-2}$. The corresponding ratio of the branching fractions between the lepton $\mu$ and $e$ is given by $\mathcal{R}_{\mu/e}=0.9432(13)$, which provides essential theoretical support for future high-precision experimental tests of the lepton flavor universality. The CKM matrix element $|V_{cs}|$ is also extracted to be $0.952(12)_{\text{stat}}(23)_{\text{PDG}}$ and $0.945(12)_{\text{stat}}(24)_{\text{PDG}}$ for the $\mu$ and $e$ channels, respectively.}

\begin{document} 
\maketitle
\flushbottom

\section{Introduction}
\label{sec:intro}

The semileptonic decays of charmed mesons, where one
meson decays into another and emits a $W~(\to \ell\nu_\ell)$ boson, provide an excellent opportunity to determine the Cabibbo-Kobayashi-Maskawa (CKM) matrix elements and can be a strong test of QCD~\cite{FlavourLatticeAveragingGroupFLAG:2024oxs}. The semileptonic decay rates involve the elements of the CKM matrix, especially $|V_{cs}|$ or $|V_{cd}|$, as well as the hadronic form factors, which describe the non-perturbative strong interactions. Precise calculations of these form factors can not only rigorously test the Standard Model, but also point to potential new physics. The semileptonic decays also provide support for testing the lepton flavor universality by calculating the ratios of the branching fractions between the $\mu$ and $e$ lepton final state~\cite{Ke:2023qzc}. 

Charmed meson semileptonic decay is the golden channel for extracting the CKM matrix elements $|V_{cs}|$ or $|V_{cd}|$. The current most stringent constraints of the CKM unitarity come from decays of charmed mesons into pseudoscalar final states, for which both experimental measurements and lattice QCD calculations have already reached unprecedented precision~\cite{BESIII:2018ccy,BESIII:2018nzb,Chakraborty:2021qav,FermilabLattice:2022gku}. In contrast, the process with a vector final particle is much less studied. Experimentally, the vector meson could subsequently decay into two light pseudoscalar particles. It therefore leads to a four-body final state, which introduces more degrees of freedom and complicates a precise measurement. Theoretically, the decay of the unstable vector particle also poses further non-perturbative challenges that must be controlled in the lattice calculation. The motivation of this subject is two-fold. (i) Although the experimental precision for vector-meson semileptonic decays is still inferior to that of pseudoscalar ones, these decays provide an independent and complementary determination of the CKM matrix elements. Future measurements with higher statistics are able to significantly improve the precision. (ii) The four-body decay carries richer angular information. Compared with semileptonic decays with a pseudoscalar final-state meson, those with a vector meson have more degrees of freedom, thereby leading to much richer differential distributions that can eventually be utilized and compared with the experiments.
For these reasons, a systematic lattice QCD study of charmed meson decay into a vector final state is essential.

In recent years, experiments have made significant progress in studying the semileptonic decay of the $D_s\to\phi\ell\nu_\ell$ process, including different lepton channels and decay dynamics analysis in BaBar~\cite{BaBar:2008gpr}, CLEO~\cite{Hietala:2015jqa}, and BESIII~\cite{BESIII:2017ikf,BESIII:2023opt} experiments. From the theoretical point of view, different methods are implemented to investigate this process, including the constituent quark model (CQM)~\cite{Melikhov:2000yu}, the heavy meson chiral theory (HM$\chi$T)~\cite{Fajfer:2005ug}, the heavy quark effective field theory (HQEFT)~\cite{Wu:2006rd}, the covariant light-front quark model (CLFQM)~\cite{Verma:2011yw,Cheng:2017pcq}, the lattice QCD (LQCD) method~\cite{Donald:2013pea}, the covariant confining quark model (CCQM)~\cite{Soni:2018adu}, the light-front quark model (LFQM)~\cite{Chang:2019mmh}, the relativistic quark model (RQM)~\cite{Faustov:2019mqr}, the symmetry-preserving contact interaction (SCI)~\cite{Xing:2022sor}, and the light-cone sum rule (LSCR)~\cite{Wang:2025hiv}. These experimental and theoretical results demonstrate the great attention and importance of the $D_s\to\phi\ell\nu_\ell$ decay. Thus, a more accurate and reliable non-perturbative lattice calculation is needed to improve previous studies and assist in future experiments.

In this work, we develop the scalar function method to extract the form factors at different transfer momenta. The core idea of the method is to extract the form factors by constructing a suitable scalar function. The scalar function is usually built by properly contracting the Lorentzian tensor structure that appears in the parameterization of the hadronic matrix element. The approach inherently has two advantages. First, the scalar function possesses exact rotational invariance, so it drastically reduces the systematic error caused by the broken rotational symmetry on the lattice. Traditional extractions of these matrix elements were based on a specific component. These components usually do not belong to the $A_1$ representation, they would suffer from larger contamination due to higher angular momenta. A scalar function projected onto the $A_1$ sector only receives rotational symmetry breaking effects from the angular momentum $I=4$, whereas for a specific component, the contamination arises from momenta separated by $\Delta I = 2$. For example, tensor components ($T_2$ and $E$ representations) both appear in $I = 2,4,\cdots$ and therefore start to mix in a lower order. Second, the scalar function is a coherent combination of all components of the hadronic matrix element rather than a single fixed component. Consequently, the statistical precision can be improved as the total statistics are effectively increased.

In this lattice calculation, seven gauge ensembles are used to investigate lattice spacing and pion mass dependence. Then the extrapolation to the continuum limit and physical pion mass is performed, where the accuracy at zero transfer momentum is well controlled. The finite-volume effect is also examined using two gauge ensembles with the same lattice spacing and pion mass. In addition, we take a series of large time separations between the initial and final particles, which leads to the ground-state dominance and clear plateaus. Thus, excited-state contamination can be effectively avoided. Finally, the differential decay width distributions and branching fractions of the $\mu$ and $e$ final states are calculated and the CKM matrix element $|V_{cs}|$ is extracted.

This paper is organized as follows. Sec.~\ref{sec:method} describes the methodology to calculate the form factors and differential decay width on the lattice. This section is further divided into three parts: Sec.~\ref{method:C} gives the differential width formula of the $D_s\to\phi\ell\nu_\ell$ decay; Sec.~\ref{method:A} introduces the scalar function method to obtain the form factors; 
Sec.~\ref{method:B} discusses the relation between the hadronic function and the correlation function. Sec.~\ref{sec:res} gives the simulation details and the main results. This section is further divided into four parts: Sec.~\ref{res:A} presents the lattice sets utilized in this work; Sec.~\ref{res:B} gives the numerical values of the mass spectra, dispersion relations, and decay constants of the $D_s$ and $\phi$ mesons; Sec.~\ref{res:C} gives the numerical values of the form factors on each ensemble and a continuum limit and physical pion mass extrapolation; Sec.~\ref{res:D} shows the differential decay width distributions of the $D_s\to\phi\ell\nu_\ell$ process and a comparison with the BESIII experimental data. Sec.~\ref{sec:dis} presents a detailed discussion on the results. This section is further divided into four parts: Sec.~\ref{dis:D} and Sec.~\ref{dis:A} discuss the systematic effects of different parameterization schemes and finite-volume effects; Sec.~\ref{dis:B} gives a comparison of our results and previous theory/experiment results; 
Sec.~\ref{dis:C} determines the CKM matrix element $|V_{cs}|$ by combining the experiment data. Finally, the conclusions are made in Sec.~\ref{sec:conclu}.

\section{Methodology}
\label{sec:method}

\subsection{Differential decay width}
\label{method:C}
\begin{figure}[!htb]
\renewcommand{\figurename}{Figure}
      \centering
      \includegraphics[width=9cm]
      {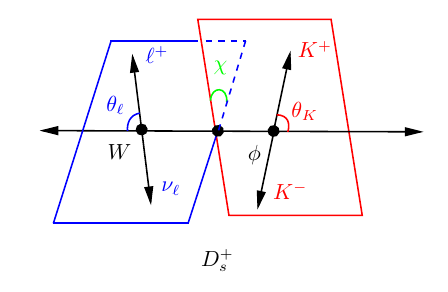}
      \caption{Definition of the angular variables in the $D_s^+\to K^+K^-\ell^+\nu_\ell$ decay.}
      \label{4body}
    \end{figure}

The semileptonic decay amplitude of the $D_s$ to a vector $\phi$ meson is
\begin{flalign}
    \mathcal{M}=\frac{G_F}{\sqrt{2}}|V_{cs}| L^\mu H_\mu ,
\end{flalign}
where $G_F$ is the Fermi constant. The leptonic and hadronic currents are 
\begin{flalign}
    L^\mu&=\bar{\nu}_\ell \gamma^\mu\left(1-\gamma_5\right)\ell,\nonumber\\
    H_\mu&=\langle \phi\left(\vec{p}\right)|J_\mu^W|D_s\left(p^\prime\right)\rangle,
\end{flalign}
where $J^W_\mu=\bar{s}\gamma_\mu\left(1-\gamma_5\right)c$. The hadronic matrix element $H_{\mu}$ is traditionally parameterized by four form factors $V,A_0,A_1,A_2$~\cite{Richman:1995wm}. Since the $\phi$ meson is unstable, it can further decay into a $K^+K^-$ pair. Hence, this is a four-body final-state process. The differential decay width that includes the leptonic mass $m_\ell$ is known as follows~\cite{Ivanov:2019nqd}
\begin{flalign}\label{eq:four_body_br}
    \frac{\mathrm{d}\Gamma(D_s^+\to \phi\ell^+\nu_\ell)}{\mathrm{d}q^2 \mathrm{d}\chi\mathrm{d}\cos\theta_\ell\mathrm{d}\cos\theta_K}=\frac{G_F^2|V_{cs}|^2|\vec{p}|q^2}{12(2\pi)^4M^2}\left(1-\frac{m_\ell^2}{q^2}\right)^2W\left(\theta_K,\theta_\ell,\chi\right),
\end{flalign}
where
\begin{flalign}\label{eq:W_function}
    \begin{aligned}
W\left(\theta_K,\theta_\ell,\chi\right) & =\frac{9}{32}\left(1+\cos^{2}\theta_\ell\right)\sin^{2}\theta_K\left(|H_+|^2+|H_-|^2\right)+\frac{9}{8}\sin^{2}\theta_\ell \cos^{2}\theta_K|H_0|^2\\
&+\frac{9}{16}\cos\theta_\ell \sin^{2}\theta_K\left(|H_+|^2-|H_-|^2\right) -\frac{9}{16}\sin^{2}\theta_\ell\sin^{2}\theta_K\cos2\chi\left(H_+H_- \right)\\
&+\frac{9}{16}\sin\theta_\ell\sin2\theta_K\cos\chi\left(H_+H_0 -H_-H_0 \right) \\
&+\frac{9}{32}\sin2\theta_\ell\sin2\theta_K\cos\chi\left(H_+H_0 +H_-H_0 \right) \\
 & +\frac{m^2_\ell}{2q^2}\left[\frac{9}{4}\cos^{2}\theta_K|H_t|^2-\frac{9}{2}\cos\theta_\ell\cos^{2}\theta_K\left(H_0H_t \right)+\frac{9}{4}\cos^{2}\theta_\ell\cos^{2}\theta_K|H_0|^2\right.\\
 & +\frac{9}{16}\sin^{2}\theta_\ell\sin^{2}\theta_K\left(|H_+|^2+|H_-|^2\right)+\frac{9}{8}\sin^{2}\theta_\ell\sin^{2}\theta_K\cos2\chi\left(H_+H_- \right) \\
 & +\frac{9}{8}\sin\theta_\ell\sin2\theta_K\cos\chi\left(H_+H_t +H_-H_t \right)\\
 &\left.-\frac{9}{16}\sin2\theta_\ell\sin2\theta_K\cos\chi\left(H_+H_0 +H_-H_0 \right) \right],
\end{aligned}
\end{flalign}
and the helicity amplitudes are
\begin{flalign}\label{eq:V_A_FFS}
&H_\pm\left(q^2\right)=\left(M+m\right)A_1\left(q^2\right)\mp \frac{2M|\vec{p}|}{M+m}V\left(q^2\right),\nonumber\\
   & H_0\left(q^2\right)=\frac{1}{2M\sqrt{q^2}}\times\left[\left(M^2-m^2-q^2\right)\left(M+m\right)A_1\left(q^2\right)-4\frac{M^2|\vec{p}|^2}{M+m}A_2\left(q^2\right)\right],\nonumber\\
   & 
H_t\left(q^2\right)=\frac{2M|\vec{p}|}{\sqrt{q^2}}A_0\left(q^2\right),
\end{flalign}
where $M$, $m$ are masses of the $D_s$ and $\phi$ meson, and $q^2=\left(M-E\right)^2-|\vec{p}|^2$ is the transfer momentum square as $D_s$ is at rest. Here, $E$ and $\vec{p}$ are the energy and momentum of the $\phi$ meson. For the $\phi$ ground state at rest, we have $E=m$. The decay angle $\theta_\ell$ ($\theta_K$) is the angle between the momentum of the charged lepton (kaon) in the rest frame of the $\ell\nu_\ell$ ($K^+K^-$) system with respect to the $\ell\nu_\ell$ ($K^+K^-$) flight direction as seen from the rest frame of the $D_s$ particle. The angle $\chi$ is the angle between the two decay planes of the $\ell\nu_\ell$ and $K^+K^-$ systems. The angles are defined in $-1\leq \cos\theta_\ell\leq 1$, $-1\leq \cos\theta_K\leq 1$ and $-\pi\leq \chi\leq\pi$. The decay planes are shown in Fig.~\ref{4body}.

After integrating the distributions over the angles $\chi$, $\cos\theta_\ell$, and $\cos\theta_K$, the differential decay width over $q^2$ is
\begin{flalign}
\frac{\mathrm{d}\Gamma(D_s\to \phi\ell\nu) }{\mathrm{d}q^2} =&\frac{G_F^{2}|V_{cs}|^{2}|\vec{p}|q^{2}v^2}{96\pi^{3}M^{2}}\left[\left(1+\delta\right)\left(|H_{+}|^{2}+|H_{-}|^{2}+|H_{0}|^{2}\right)+3\delta|H_{t}|^{2}\right],
\end{flalign}
where $v=1-m_\ell^2/q^2$ and $\delta=m_\ell^2/\left(2q^2\right)$.

\subsection{Scalar function method}
\label{method:A}

In this section, we briefly introduce the scalar function method for extracting the form factors, which has been applied to various physical processes~\cite{Tuo:2021ewr,Meng:2021ecs,Meng:2024gpd,Meng:2024nyo,Lin:2024khg,Meng:2024axn}, and has achieved great success.

We start the discussion with a Euclidean hadronic function in the finite volume 
   \begin{flalign}
  H_{\mu\nu}\left(x\right)=\langle \phi_{\nu}\left(\vec{x},t\right)J_{\mu}^{W}\left(0\right)|D_s\left(p^\prime\right)\rangle,~t>0,
    \end{flalign}
where $\phi_\nu\left(\vec{x},t\right)$ is the interpolating operator of the $\phi$ meson and $|D_s\left(p^\prime\right)\rangle$ is the $D_s$ state with momentum $p^\prime=(iM,\vec{0})$. At large time $t$, the hadronic function is saturated by the single $\phi$ state
\begin{flalign}
H_{\mu\nu}\left(x\right)&=\sum_{p,\varepsilon}\frac{1}{2E\mathcal{V}}e^{-Et+i\vec{p}\cdot\vec{x}}\langle0|\phi_\nu\left(0\right)|\phi\left(\vec{p},\varepsilon\right)\rangle\langle \phi\left(\vec{p},\varepsilon\right)|J_\mu^W\left(0\right)|D_s\left(p^\prime\right)\rangle \nonumber\\
&=\sum_{p,\varepsilon}\frac{1}{2E\mathcal{V}}e^{-Et+i\vec{p}\cdot\vec{x}}\langle0|\phi_\nu\left(0\right)|\phi_\rho\left(\vec{p}\right)\rangle \varepsilon_{\rho} \varepsilon^{*}_\sigma\langle  \phi_\sigma\left(\vec{p}\right)|J_\mu^W\left(0\right)|D_s\left(p^\prime\right)\rangle \nonumber\\
&=\sum_{p,\varepsilon}\frac{1}{2E\mathcal{V}}e^{-Et+i\vec{p}\cdot\vec{x}}\langle0|\phi\left(0\right)|\phi\left(\vec{p}\right)\rangle\delta_{\nu\rho}\varepsilon_{\rho} \varepsilon_\sigma^{*}\langle  \phi_\nu\left(\vec{p}\right)|J_\mu^W\left(0\right)|D_s\left(p^\prime\right)\rangle \nonumber\\
&=\sum_{p}\frac{1}{2E\mathcal{V}}e^{-Et+i\vec{p}\cdot\vec{x}}\langle0|\phi\left(0\right)|\phi\left(\vec{p}\right)\rangle\left(-\delta_{\nu\sigma}-\frac{p_\nu p_\sigma}{m^2}\right)\langle \phi_\sigma\left(\vec{p}\right)|J_\mu^W\left(0\right)|D_s\left(p^\prime\right)\rangle,
\end{flalign}
where $p=(iE,\vec{p})$ and $\varepsilon_\mu$ are the momentum and polarization vector of the $\phi$ meson. $\delta_{\mu\nu}=\mathrm{diag}\left(1,1,1,1\right)$ is the Euclidean metric and $\mathcal{V}$ is the spatial volume. Considering the following parametrizations~\cite{Korner:1989qb}
\begin{flalign}\label{eq:F_FF}
\langle \phi_\sigma\left(\vec{p}\right)|J_\mu^W\left(0\right)|D_s\left(p^\prime\right)\rangle
&=\frac{{F_{0}}\left(q^{2}\right)}{Mm}\epsilon_{\mu\sigma\alpha\beta}p^{\prime}_{\alpha}p_{\beta}+{F_{1}}\left(q^{2}\right)\delta_{\mu\sigma}+\frac{{F_{2}}\left(q^{2}\right)}{Mm}p_{\mu}p_{\sigma}^{\prime}+\frac{{F_{3}}\left(q^{2}\right)}{M^{2}}p_{\mu}^{\prime}p_{\sigma}^{\prime},
\end{flalign}
then, the spatial Fourier transform of $H_{\mu\nu}\left(\vec{x},t\right)\equiv V_{\mu\nu}\left(\vec{x},t\right)-A_{\mu\nu}\left(\vec{x},t\right)$ yields
\begin{flalign}
\tilde{V}_{\mu\nu}&=-\frac{F_{0}\left(q^{2}\right)}{Mm}\epsilon_{\mu\nu\alpha\beta}p^{\prime}_{\alpha}p_{\beta}\times \frac{Z_{\phi}e^{-Et}}{2E},\nonumber\\
\tilde{A}_{\mu\nu}&=\left[-F_{1}\left(q^{2}\right)\delta_{\mu\sigma}-\frac{F_{2}\left(q^{2}\right)}{Mm}p_{\mu}p_{\sigma}^{\prime}-\frac{F_{3}\left(q^{2}\right)}{M^{2}}p_{\mu}^{\prime}p_{\sigma}^{\prime}\right]\left(-\delta_{\nu\sigma}-\frac{p_\nu p_\sigma}{m^2}\right)\times\frac{Z_{\phi}e^{-Et}}{2E},
\end{flalign}
where $Z_{H}=\langle 0|H\left(0\right)|H\left(\vec{p}\right)\rangle$ with $H= D_s,\phi$ particles in this work.
To extract $F_i~(i=0,1,2,3)$, we construct the following scalar functions
\begin{flalign}
\mathcal{I}_0&=\frac{1}{M|\vec{p}|^2}\epsilon_{{\mu\nu\alpha^{\prime}\beta^{\prime}}}p^{\prime}_{\alpha^{\prime}}p_{{\beta^{\prime}}}\tilde{V}_{\mu\nu}=\frac{1}{|\vec{p}|^{2}}\epsilon_{\mu\nu0\beta}p_{\beta}\int \mathrm{d}^{3}\vec{x}\sin\left(\vec{p}\cdot\vec{x}\right)V_{\mu\nu}\left(\vec{x},t\right),\nonumber\\
\mathcal{I}_1&=\delta_{\mu\nu}\tilde{A}_{\mu\nu}=\delta_{\mu\nu}\int\mathrm{d}^3\vec{x}e^{-i\vec{p}\cdot\vec{x}}A_{\mu\nu}\left(\vec{x},t\right),\nonumber\\
\mathcal{I}_2&=\frac{E}{M}\frac{p_{\mu}p^{\prime}_{\nu}}{|\vec{p}|^2}\tilde{A}_{\mu\nu}=-\frac{E^2}{|\vec{p}|^2}\int \mathrm{d}^3\vec{x}\cos\left(\vec{p}\cdot\vec{x}\right)A_{00}\left(\vec{x},t\right)+\frac{E}{|\vec{p}|^2}\int \mathrm{d}^3\vec{x}\sin\left(\vec{p}\cdot\vec{x}\right)p_iA_{i0}\left(\vec{x},t\right),\nonumber\\
\mathcal{I}_3&=\frac{p^{\prime}_{\mu} p^{\prime}_{\nu}}{|\vec{p}|^2}\tilde{A}_{\mu\nu}=-\frac{M^2}{|\vec{p}|^2}\int \mathrm{d}^3\vec{x}\cos\left(\vec{p}\cdot\vec{x}\right)A_{00}\left(\vec{x},t\right).
\end{flalign}
The form factors $F_i~(i=0,1,2,3)$ then can be accessed via
\begin{flalign}\label{eq:F_scalar}
F_0&=\frac{m}{2}\tilde{\mathcal{I}}_0,
\nonumber\\
F_1&=\frac{1}{2}\tilde{\mathcal{I}}_1+\frac{1}{2}\tilde{\mathcal{I}}_2-\frac{m^2}{2M^2}\tilde{\mathcal{I}}_3,\nonumber\\
F_2&=\frac{mE}{2\left(E^2-m^2\right)}\tilde{\mathcal{I}}_1+\frac{mE^2+2m^3}{2\left(E^3-Em^2\right)}\tilde{\mathcal{I}}_2-\frac{3Em^3}{2M^2\left(E^2-m^2\right)}\tilde{\mathcal{I}}_3,\nonumber\\
F_3&=-\frac{m^2}{2\left(E^2-m^2\right)}\tilde{\mathcal{I}}_1-\frac{3m^2}{2\left(E^2-m^2\right)}\tilde{\mathcal{I}}_2+\frac{3m^4}{2M^2\left(E^2-m^2\right)}\tilde{\mathcal{I}}_3,
\end{flalign}
where 
\begin{flalign}
\tilde{\mathcal{I}}_j=\mathcal{I}_j\times \frac{2Ee^{Et}}{Z_{\phi}}~(j=0,1,2,3).
\end{flalign}
The details of the scalar functions can be found in Appendix~\ref{sec:ap1}. 

The helicity amplitudes are traditionally related to form factors $V,A_0,A_1,A_2$ directly, as given in Eq.~(\ref{eq:V_A_FFS}). We would extract these form factors using the above form factors $F_i~(i=0,1,2,3)$. Such matching is straightforward, as both are derived from the same hadronic matrix element. For the traditional parametrization scheme~\cite{Richman:1995wm}, it has
\begin{flalign}
\langle \phi\left(\vec{p}\right)|J_\mu^W\left(0\right)|D_s\left(p^\prime\right)\rangle&=\varepsilon_\nu^*\epsilon_{\mu\nu\alpha\beta} p^{\prime}_{\alpha} p_{\beta}\frac{2V}{m+M} +\left(M+m\right)\varepsilon^*_{\mu}A_1\nonumber\\
&+\frac{\varepsilon^*\cdot q}{M+m}\left(p+p^\prime\right)_\mu A_2-2m\frac{\varepsilon^*\cdot q}{q^2}q_\mu\left(A_0-A_3\right),
\end{flalign}
where form factors $V$ and $A_i~(i=0,1,2,3)$ are introduced, rather than $F_i~(i=0,1,2,3)$
in this work. $A_3\left(q^2\right)$ is not an independent form factor, since
\begin{flalign}
    A_3\left(q^2\right)=\frac{M+m}{2m}A_1\left(q^2\right)-\frac{M-m}{2m}A_2\left(q^2\right).
\end{flalign}
We also have the kinematic constraint $A_0\left(0\right) = A_3\left(0\right)$.
The form factors $F_i~(i=0,1,2,3)$
introduced in this work can be related to the conventional
form factors $V$ and $A_i~(i=0,1,2)$ by the following way
\begin{flalign}\label{eq:F_V_matching}
V&=\frac{\left(m+M\right)}{2mM}F_0,\nonumber\\
A_1&=\frac{F_1}{M+m},\nonumber\\
A_2&=\frac{M+m}{2mM^2}\left(MF_2+mF_3\right),\nonumber\\
A_0&= \frac{F_1}{2m}+\frac{m^2-M^2+q^2}{4m^2M}F_2+\frac{m^2-M^2-q^2}{4mM^2}F_3.
\end{flalign}
It is easy to obtain the ratios
\begin{flalign}
r_{V}&\equiv \frac{V\left(0\right)}{A_1\left(0\right)}=\frac{\left(m+M\right)^2}{2mM}\frac{F_0\left(0\right)}{F_1\left(0\right)},\nonumber\\
r_{2}&\equiv \frac{A_2\left(0\right)}{A_1\left(0\right)}=\frac{\left(m+M\right)^2}{2mM^2}\left[M\frac{F_2\left(0\right)}{F_1\left(0\right)}+m\frac{F_3\left(0\right)}{F_1\left(0\right)}\right],\nonumber\\
r_0&\equiv \frac{A_0\left(0\right)}{A_1\left(0\right)}=\frac{M+m}{2m}\left[1+\frac{m^2-M^2}{2mM}\frac{F_2\left(0\right)}{F_1\left(0\right)}+\frac{m^2-M^2}{2M^2}\frac{F_3\left(0\right)}{F_1\left(0\right)}\right].
\end{flalign}

Although these traditional form factors $V,A_0,A_1,A_2$ are obtained through a matching procedure as Eq.~(\ref{eq:F_V_matching}), they can basically be computed straightforwardly using the traditional parametrization. The two approaches are completely equivalent. We prefer the former in this work since it leads to a simple 
combination of scalar functions, as shown in Eq.~(\ref{eq:F_scalar}).

\subsection{Hadronic function}
\label{method:B}
The hadronic function $H_{\mu\nu}\left(\vec{x},t\right)$ can be extracted from the three-point function
\begin{flalign}
C_{\mu\nu}(\vec{x}, t,t_s)=\langle\mathcal{O}_{\phi_{\nu}}\left(t\right)J_{\mu}^{W}\left(0\right)\mathcal{O}_{D_s}^{\dagger}\left( -t_{s}\right)\rangle,
\end{flalign}
where the interpolating operators for the mesons are $\mathcal{O}_{\phi_{\nu}}\left(t\right)=\bar{s}\left(t\right)\gamma_\nu s\left(t\right)$ and $\mathcal{O}_{D_s}^{\dagger}\left( -t_{s}\right)=-\bar{c}\left(-t_s\right)\gamma_5s\left(-t_s\right)$.
In this work, we only consider the connected contributions. Then, it has the following quark contractions
\begin{flalign}
C_{\mu\nu}(\vec{x}, t,t_s) 
 =\langle \mathrm{Tr}[\gamma_5\gamma_5S_{-s}^\dagger(t,-t_s)\gamma_5\gamma_\nu S_s(t,0)\gamma_\mu(1-\gamma_5)S_c(0,-t_s)]\rangle,
\end{flalign}
where $S$ denotes the quark propagator. Wall-source propagator is used for $S^\dagger_{-s}\left(t,-t_s\right)$ and $S_c\left(0,-t_s\right)$, and point-source propagator is used for $S_s\left(t,0\right)$. To increase statistics economically, we average over $N_{\text{src}}$ wall-source propagators in the temporal direction and $N_{\text{src}}\times N_s$ point-source propagators in the temporal and spatial directions in total. In our calculation, it is found that increasing the number of point-source propagators effectively improves the precision. Compared with increasing the number of gauge configurations, this strategy avoids calculating two additional wall-source propagators and is therefore far more economical. Then, the hadronic function $H_{\mu\nu}\left(\vec{x},t\right)$ is determined directly
through
\begin{flalign}
H_{\mu\nu}\left(\vec{x},t\right)=\frac{2M}{Z_{D_s}}e^{Mt_s}C_{\mu\nu}\left(\vec{x},t;t_s\right).
\label{ap2}
\end{flalign}
For the computation of the three-point function $C_{\mu\nu}(\vec{x}, t,t_s) $, all propagators are produced on a large number of time slices to increase the statistics based on the invariance of time translation. 

In our calculations, $M$, $m$, $E$, $Z_{D_s}$, and $Z_\phi$ are extracted
from the two-point function,
\begin{flalign}
 C^{(2)}(\vec{p},t)=\sum_{\vec{x}}e^{-i\vec{p}\cdot\vec{x}}\langle \mathcal{O}_H\left(\vec{x},t\right)\mathcal{O}^\dagger_H\left(0\right)\rangle
\end{flalign}
by a single-state fit at large $t$. The fit function for $D_s$ is
\begin{flalign}
C^{(2)}\left(\vec{p},t\right)=\frac{Z_{D_s}^2}{2E_{D_s}}\left(e^{-E_{D_s}t}+e^{-E_{D_s}\left(T-t\right)}\right),
\label{Ds}
\end{flalign}
where $E_{D_s}$ is the energy of the $D_s$ particle, and for the ground state, $E_{D_s}=M$. The fit function for the $\phi$ meson is
\begin{flalign}
C^{(2)}\left(\vec{p},t\right)=\left(1+\frac{|\vec{p}|^2}{3m^2}\right)\frac{Z_\phi^2}{2E}\left[e^{-Et}+e^{-E\left(T-t\right)}\right],
\label{phi}
\end{flalign}
where we average the three gamma matrix results to get this form.

\section{Simulation Results}
\label{sec:res}
\subsection{Lattice set up}
\label{res:A}
We employ seven $(2+1)$-flavor Wilson-clover gauge
ensembles generated by the CLQCD collaboration~\cite{CLQCD:2023sdb,CLQCD:2024yyn}, the
parameters of which are shown in Table~\ref{lattice}. The dynamical
ensembles use tadpole-improved tree-level
Symanzik gauge action and tadpole-improved tree-level
clover fermions. The valence strange and charm quark masses are tuned using the “fictitious” meson $\eta_s$ and the $D_s$ meson masses. It was found that tuning the valence
strange quark mass by using $m_{\eta_s} = 689.89(49)~\text{MeV}$~\cite{Borsanyi:2020mff} on each ensemble can significantly suppress the effect of unphysical strange quark mass in the sea. In addition, the heavy quark improved normalization factor $Z_V^{c}$ and $Z_V^s$ can also suppress the discretization error, thus allowing a more reliable continuum extrapolation. Both normalization factors are fixed by the vector current conservation condition: $Z_V^c$ for the $\bar{c}\gamma_{\mu}c$ current with $\eta_c$ state and $Z_V^s$ for the $\bar{s}\gamma_{\mu}s$ current with $\eta_s$ state. As far as the flavor-changed current is concerned, like $\bar{c}\Gamma s$, the corresponding vector current normalization factor $Z_V^{cs}=\sqrt{Z_V^cZ_V^s}$ and the axial vector normalization factor $Z_A^{cs}=Z_V^{cs}{Z_A}/{Z_V}$. The numerical tests of these improvements on decay constants of charmed mesons have been presented in Ref.~\cite{CLQCD:2024yyn}. In this work, we perform all calculations using the same quark masses and renormalization constants as in Ref.~\cite{CLQCD:2024yyn}. In recent years, plenty of studies have been carried out based on these configurations~\cite{Meng:2024gpd,Yan:2024yuq,Meng:2024nyo,Yan:2024gwp,Chen:2024rgi,LatticeParton:2024zko, Wang:2025hew}. Since the ensembles have several different lattice spacings and pion masses, they are expected to provide a fully controlled continuous extrapolation.

 \renewcommand{\tablename}{Table}
        \begin{table}[htbp]
	\centering  
 \caption{Parameters of gauge ensembles used in this work. From top to bottom, we list the ensemble name, the lattice spacing $a$, the bare quark mass $am_{s,l,}$, the valence bare
strange and charm quark mass $am^{\mathrm{V}}_{s,c}$, the lattice size $L$ and $L^3\times T$, the number of the measurements of the correlation function
 for each ensemble $N_{\mathrm{mea}}=N_{\mathrm{cfg}}\times N_{\mathrm{src}}\times N_{\mathrm{s}}$ ($N_{\mathrm{cfg}}$ is the number of configurations; $N_{\mathrm{src}}$ and $N_s$ are the number of the propagators in the temporal and spatial directions, respectively), the pion mass $m_\pi$, the range of the time separation $t$ between the initial hadron and the current, the strange and charm quark vector normalization constant $Z_V^{s,c}$, and the ratio between axial vector normalization constant $Z_A$ and $Z_V$. Here, $L^3\times T$, and $t$ are given in the lattice units.}
 \addvspace{5pt}
 \scalebox{0.7}{
	\begin{tabular}{cccccccc}
		\hline\hline\noalign{\smallskip}	
		&C24P29&C32P23&C32P29&F32P30&F48P21&G36P29&H48P32\\
  \noalign{\smallskip}\hline\noalign{\smallskip} 
  $a~(\mathrm{fm})$&\multicolumn{3}{c}{$0.10524(05)(62)$} &\multicolumn{2}{c}{$0.07753(03)(45)$}&$0.06887(12)(41)$&$0.05199(08)(31)$\\
  $am_l$&$-0.2770$&$-0.2790$&$-0.2770$&$-0.2295$&$-0.2320$&$-0.2150$&$ -0.1850$\\
   $am_s$&$-0.2400$&$-0.2400$&$-0.2400$&$-0.2050$&$-0.2050$&$ -0.1926$&$-0.1700$\\
$am_s^{\mathrm{V}}$&$-0.2356(1)$&$-0.2337(1)$&$-0.2358(1)$&$-0.2038(1)$&$-0.2025(1)$&$ -0.1928(1)$&$-0.1701(1)$\\
  $am_c^{\mathrm{V}}$&$0.4159(07)$&$0.4190(07)$&$0.4150(06)$&$0.1974(05)$&$0.1997(04)$&$0.1433(12)$&$0.0551(07) $\\
    $L~(\text{fm})$ & $2.53$ & $3.37$ & $3.37$ & $2.48$ & $3.72$ & $2.48$ & $2.50$\\
  $L^3\times T$&$24^3\times 72$&$32^3\times 64$&$32^3\times 64$&$32^3\times 96$&$48^3\times 96$&$36^3\times 108$&$48^3\times 144$\\
$N_{\mathrm{mea}}$&$450\times 72\times 2$&$333\times 64\times 3$&$397\times 64\times 2$&$360\times 96\times 2$&$241\times 48\times 4$&$300\times 54\times 2$&$300\times 72\times 2$\\
  $m_\pi~(\mathrm{MeV})$&$292.3(1.0)$&$227.9(1.2)$&$293.1(0.8)$&$300.4(1.2)$&$207.5(1.1)$&$297.2(0.9)$&$316.6(1.0)$\\
  $t$&$2-17$&$2-20$&$2-20$&$4-22$&$4-26$&$2-32$&$8-30$\\
  $Z^s_V$&$0.85184(06)$&$0.85350(04)$&$0.85167(04)$&$ 0.86900(03)$&$ 0.86880(02)$&$ 0.87473(05)$&$0.88780(01)$\\
$Z^c_V$&$1.57353(18)$&$1.57644(12)$&$1.57163(14)$&$1.30566(07)$&$1.30673(04)$&$1.23990(13)$&$ 1.12882(11)$\\
  $Z_A/Z_V$&$1.07244(70)$&$1.07375(40)$&$1.07648(63)$&$1.05549(54)$&$1.05434(88)$&$ 1.04500(22)$&$1.03802(28)$\\
	\noalign{\smallskip}\hline\noalign{\smallskip}
	\end{tabular}}
 \label{lattice}
\end{table}

\subsection{Mass spectra, dispersion relations, and decay constants}
\label{res:B}

The energy levels of the particles $D_s$ and $\phi$ are extracted from the two-point functions, which are calculated by the point-source propagators. A single-state correlated fit with the formulas Eq.~(\ref{Ds}) and Eq.~(\ref{phi}) is utilized, and the numerical fitting results of the spectra are summarized in Table~\ref{result}. The effective energy levels of the particles $D_s$ and $\phi$ are shown in Fig.~\ref{2ptC24P29}-\ref{2ptH48P32} in Appendix~\ref{sec:ap2} for all the ensembles, and the horizontal bands therein denote the fitted center values and statistical errors. The dispersion relations of $D_s$ and $\phi$ particles are also checked utilizing their energy levels given in Eq.~(\ref{dispersion}),
\begin{flalign}
4\sinh^2\frac{E_h}{2}=4\sinh^2\frac{m_h}{2}+\mathcal{Z}_{\mathrm{latt}}^h\cdot 4\sum_i\sin^2\frac{{p}_i}{2},
\label{dispersion}
\end{flalign}
where the symbol $h$ denotes the particle $D_s$ or $\phi$. It is
found that the discrete dispersion relation describes the energies and momenta well. There is a nice linear behavior between $4\sinh^2({E_h}/{2})$ and $4\sum_i\sin^2({p_i}/{2})$ as illustrated in Fig.~\ref{2ptC24P29}-\ref{2ptH48P32} in Appendix~\ref{sec:ap2}. Since the physical mass and decay constant of the $D_s$ meson have already been presented in Ref.~\cite{CLQCD:2024yyn}, we only calculate the mass and decay constant of the $\phi$ meson in this work. The decay constant $f_\phi$ is given by
\begin{flalign}
f_\phi=\frac{Z_V^sZ_\phi}{m}.
\end{flalign}
The lattice results of all ensembles are listed in Table~\ref{result}.

 \renewcommand{\tablename}{Table}
        \begin{table}[htbp]
	\centering  
  \caption{Mass spectra, overlap function $Z_H$ and coefficients $\mathcal{Z}_{\mathrm{latt}}^h$ of the $D_s$ and $\phi$ particles.}
  \addvspace{5pt}
	\scalebox{0.75}{\begin{tabular}{cccccccc}
		\hline\hline\noalign{\smallskip}	
		&C24P29&C32P23&C32P29&F32P30&F48P21&G36P29&H48P32\\
  \noalign{\smallskip}\hline\noalign{\smallskip}
$aE_{D_s}~(|\vec{n}|^2=0)$&$1.04895(26)$&$1.04921(26)$&$1.04914(27)$&$0.77337(19)$&$0.77287(21)$&$0.68670(20)$&$0.51826(15)$\\
$aE_{D_s}~(|\vec{n}|^2=1)$&$1.07663(29)$&$1.06498(29)$&$1.06482(30)$&$0.79614(22)$&$0.78307(23)$&$0.70709(25)$&$0.53406(19)$\\
$aE_{D_s}~(|\vec{n}|^2=2)$&$1.10346(35)$&$1.08045(32)$&$1.08021(34)$&$0.81823(27)$&$0.79314(27)$&$0.72693(33)$&$0.54956(27)$\\
$aE_{D_s}~(|\vec{n}|^2=3)$&$1.12955(44)$&$1.09565(37)$&$1.09532(38)$&$0.83969(35)$&$0.80310(31)$&$0.74635(48)$&$0.56482(44)$\\
$aE_{D_s}~(|\vec{n}|^2=4)$&$1.15326(56)$&$1.10992(42)$&$1.10969(45)$&$0.85978(47)$&$0.81274(38)$&$0.76404(67)$&$0.57961(39)$\\
$\mathcal{Z}_{\text{latt}}^{D_s}$&$1.0402(48)$&$1.0389(72)$&$1.0346(75)$&$1.0324(45)$&$1.0276(92)$&$1.0168(62)$&$1.0334(58)$\\
$a^2Z_{D_s}$&$0.21942(42)$&$0.21768(50)$&$0.22088(49)$&$0.13947(26)$&$0.13604(39)$&$0.11502(30)$&$0.07221(21)$\\
\noalign{\smallskip}\hline\noalign{\smallskip}
$aE_{\phi}~(|\vec{n}|^2=0)$&$0.51803(83)$&$0.51320(83)$&$0.51937(81)$&$0.39477(61)$&$0.38776(66)$&$0.35470(82)$&$0.27361(59)$\\
$aE_{\phi}~(|\vec{n}|^2=1)$&$0.5781(10)$&$0.54897(93)$&$0.55431(94)$&$0.44002(68)$&$0.40932(76)$&$0.39297(92)$&$0.30358(66)$\\
$aE_{\phi}~(|\vec{n}|^2=2)$&$0.6322(16)$&$0.5822(11)$&$0.5869(10)$&$0.4808(10)$&$0.43078(80)$&$0.4320(11)$&$0.33185(88)$\\
$aE_{\phi}~(|\vec{n}|^2=3)$&$0.6821(28)$&$0.6140(15)$&$0.6175(14)$&$0.5218(13)$&$0.4507(10)$&$0.4674(19)$&$0.3583(13)$\\
$aE_{\phi}~(|\vec{n}|^2=4)$&$0.7315(42)$&$0.6443(23)$&$0.6454(22)$&$0.5554(24)$&$0.4696(12)$&$0.5020(16)$&$0.3827(19)$\\
$\mathcal{Z}_{\text{latt}}^{\phi}$&$1.027(12)$&$1.042(13)$&$1.021(13)$&$1.0324(93)$&$1.061(15)$ &$1.058(11)$ &$1.057(13)$\\
$a^2Z_{\phi}$&$0.08001(57)$&$0.07853(59)$&$0.08134(60)$&$0.04512(27)$&$0.04312(31)$&$0.03580(34)$&$0.02068(19)$\\
$f_\phi~(\text{GeV})$&$0.2463(23)$&$0.2445(24)$&$0.2497(24)$&$0.2524(21)$&$0.2455(23)$&$0.2525(29)$&$0.2543(29)$\\
 \noalign{\smallskip}\hline\noalign{\smallskip}
	\end{tabular}}
 \label{result}
\end{table}

\renewcommand{\thesubfigure}{(\roman{subfigure})}
\renewcommand{\figurename}{Figure}
\begin{figure}[htp!]
\centering  
\subfigure[$\phi$ mass at $m_\pi=0.135~\text{GeV}$.]{
\includegraphics[width=7cm]{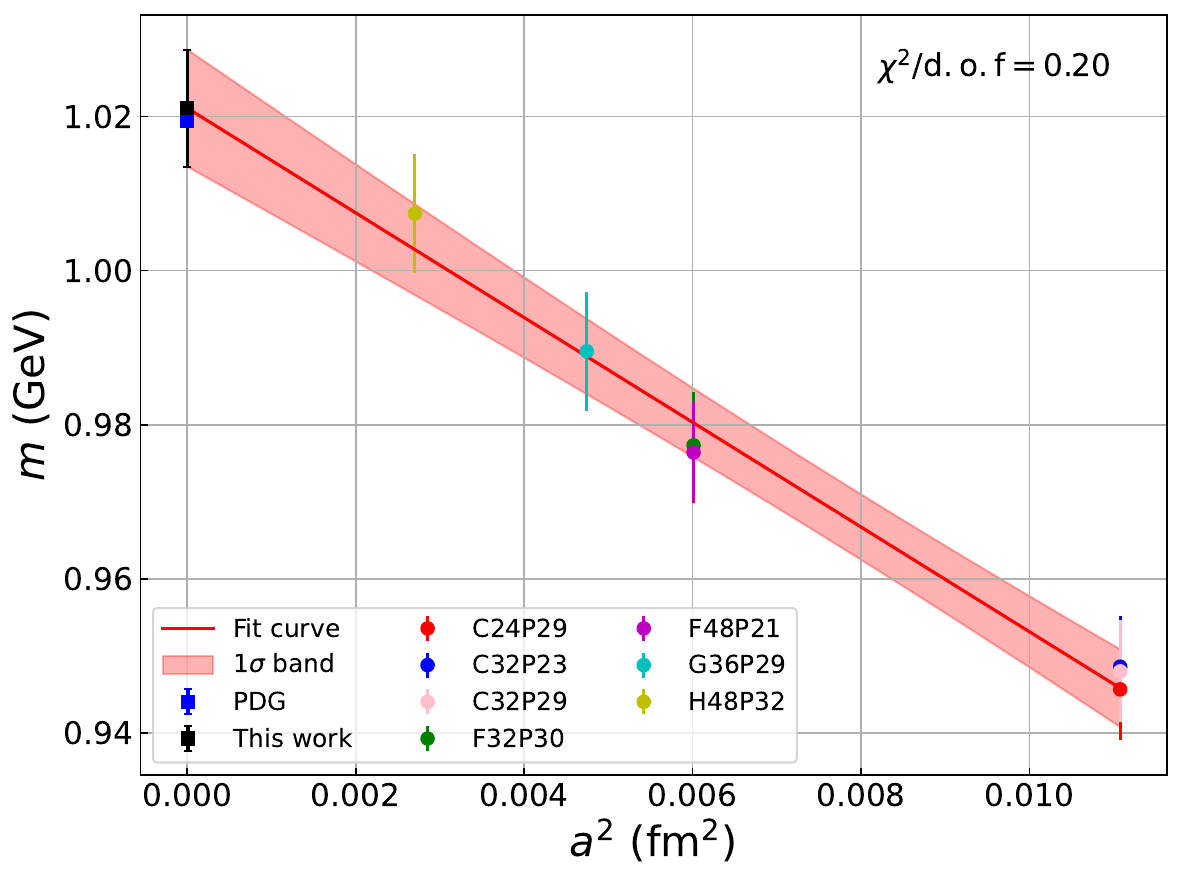}}
\subfigure[$\phi$ mass at $a=0.0~\text{fm}$.]{
\includegraphics[width=7cm]{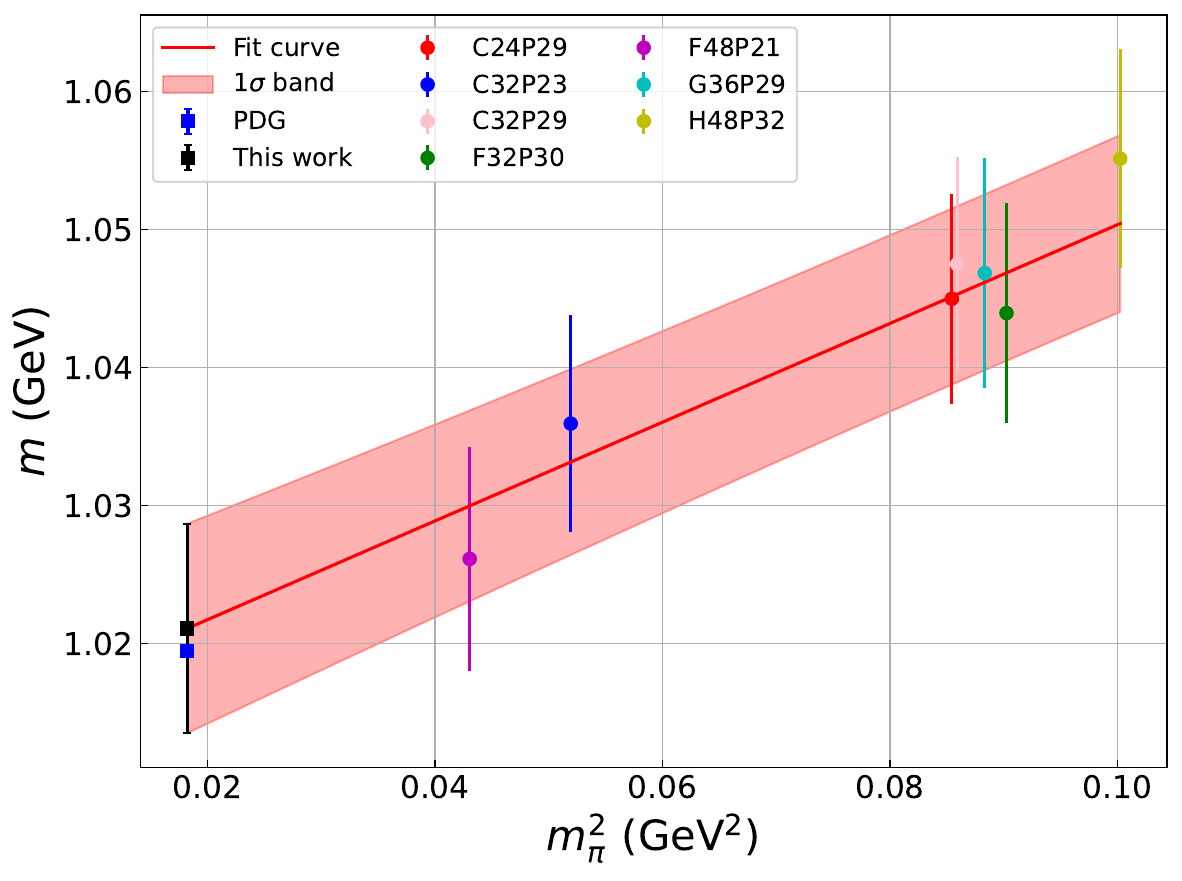}}
\subfigure[$\phi$ decay constant at $m_\pi=0.135~\text{GeV}$.]{
\includegraphics[width=7cm]{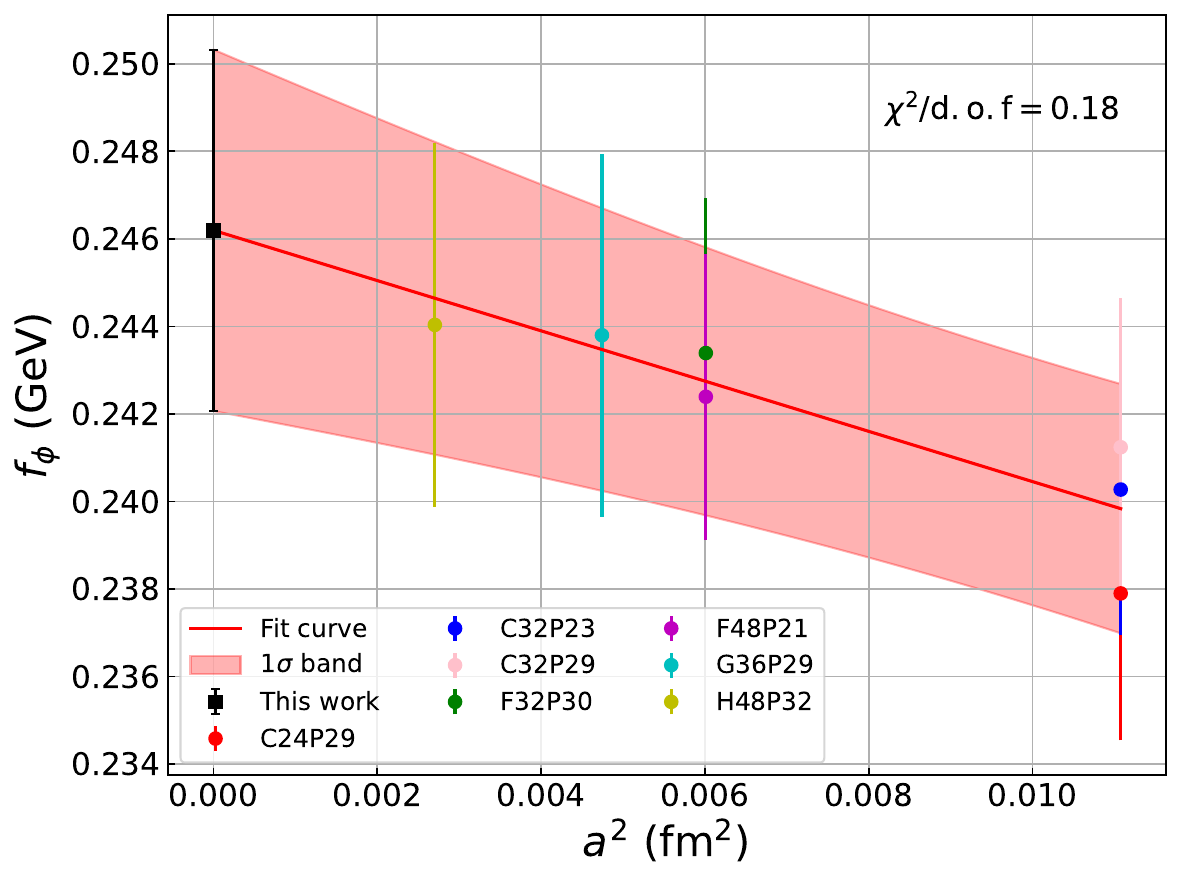}}
\subfigure[$\phi$ decay constant at $a=0.0~\text{fm}$.]{
\includegraphics[width=7cm]{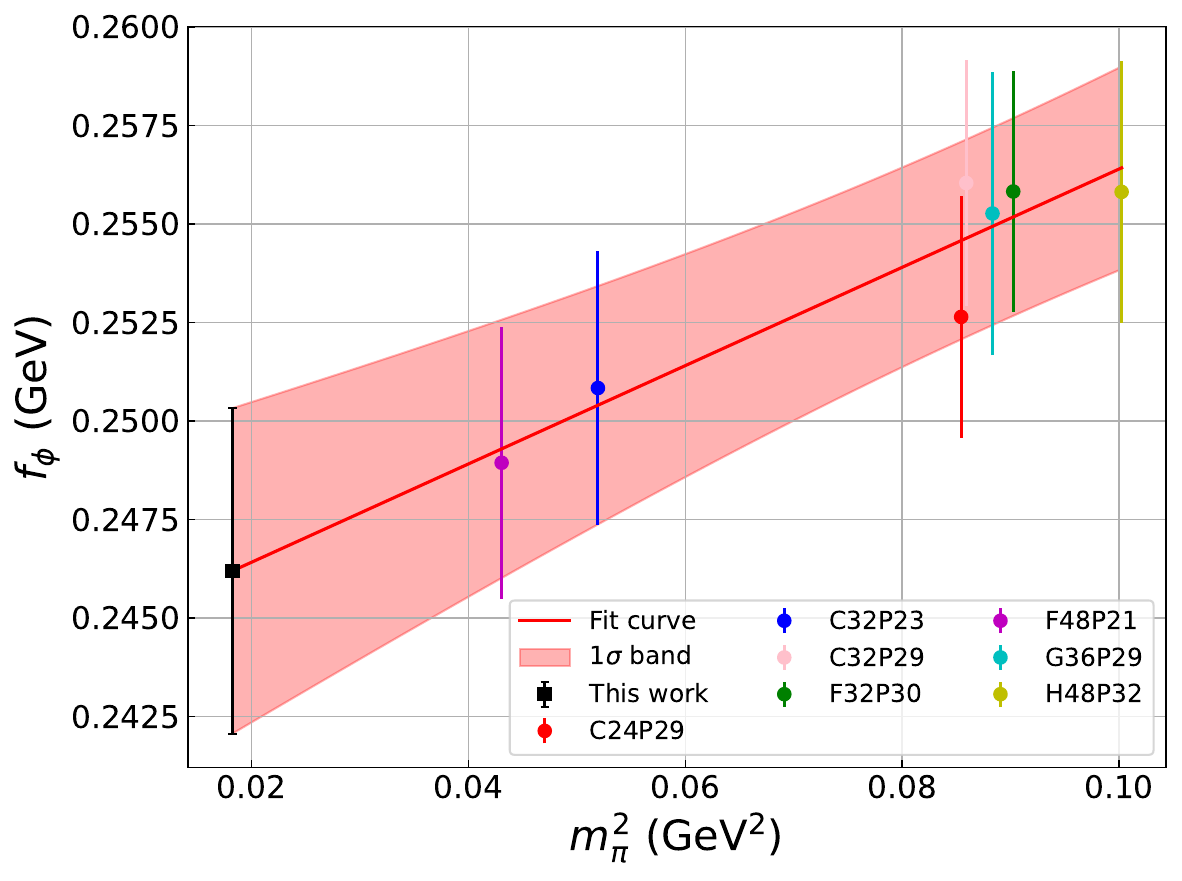}}
\caption{The lattice results of $\phi$ mass $m$ and decay constant $f_\phi$ as a function of the lattice spacing $a$ and pion mass $m_\pi$.}
\label{massanddecay}
\end{figure}
To obtain the mass and decay constant of the $\phi$ meson at the physical point, a continuum extrapolation is performed that includes both the lattice spacing $a$ and the pion mass $m_{\pi}$. The extrapolation function is of the form
\begin{flalign}
    m/f_\phi=c+d a^2+f\left(m_\pi^2-m^2_{\pi, \text{phys}}\right),
    \label{extra}
\end{flalign}
where $m/f_\phi$ denotes mass or decay constant; $c$, $d$ and $f$ are parameters to be determined by fitting, $m_{\pi,\text{phys}}=0.135~\text{GeV}$ is the physical pion mass. The lattice spacing and pion mass dependences are shown in Fig.~\ref{massanddecay}. 
The physical $\phi$ mass after extrapolation is $m=1.0211(76)~\text{GeV}$, which is consistent with the PDG~\cite{ParticleDataGroup:2024cfk} result. The physical $\phi$ decay constant after extrapolation is $f_\phi=0.2462(41)~\text{GeV}$. The value is consistent with the $\chi$QCD~\cite{Chen:2020qma} and HPQCD~\cite{Donald:2013pea} results, which are $f_\phi^{\chi\text{QCD}}=0.241(9)~\text{GeV}$ and $f_\phi^{\text{HPQCD}}=0.241(18)~\text{GeV}$, respectively. 
It is worth noting that only one or two lattice spacings are employed in the two studies mentioned above.

Although the physical $\phi$ particle is unstable and can decay into a pair of kaons, we neglect its decay in this work. First, at the heavier pion masses employed in our lattice ensembles, for example, $m_{\pi}=210\sim 320~\text{MeV}$, the $\phi$ particle is lighter than two kaons as shown in Table~\ref{Kmass}, so it remains stable. Second, its width is extremely small, making the stable-particle approximation reasonable. Finally, the mass and decay constant obtained under this assumption agree with the values in the literature. These observations demonstrate that treating the particle as stable is justified. As a sustainable extension of this work, we would investigate the effect of its finite width in future studies.

\renewcommand{\tablename}{Table}
        \begin{table}[htbp]
	\centering  
  \caption{$\phi$ and $K$ masses on each ensemble.}\label{Kmass}
\addvspace{5pt}
	\scalebox{0.75}{\begin{tabular}{cccccccc}
		\hline\hline\noalign{\smallskip}	
		&C24P29&C32P23&C32P29&F32P30&F48P21&G36P29&H48P32\\
  \noalign{\smallskip}\hline\noalign{\smallskip}

$am_{\phi}$&$0.51803(83)$&$0.51320(83)$&$0.51937(81)$&$0.39477(61)$&$0.38776(66)$&$0.35470(82)$&$0.27361(59)$\\
 $am_{K}$&$0.28341(35)$&$0.27383(35)$&$0.28295(29)$&$0.20912(24)$&$0.20007(18)$&$0.18551(23)$&$0.14217(14)$\\
 $2m_{K}-m_{\phi}$&$0.0488(11)$&$0.0345(11)$&$0.0465(10)$&$0.02347(78)$&$0.01238(75)$&$0.01632(94)$&$0.01073(65)$\\
 \noalign{\smallskip}\hline\noalign{\smallskip}
	\end{tabular}}
\end{table}

\subsection{Form factors}
\label{res:C}

The lattice results of $V\left(q^2\right)$ and $A_i\left(q^2\right)~(i=0,1,2)$ as a function of the time separation $t$ are shown in Fig.~\ref{3ptC24P29}-\ref{3ptH48P32} of Appendix~\ref{sec:ap2} for a series of momenta $\vec{p}=2\pi\vec{n}/L,~|\vec{n}|^2=1,2,3,4,5,6$. We present all results of seven gauge ensembles. It shows that both $V\left(q^2\right)$ and $A_i\left(q^2\right)~(i=0,1,2)$ have obvious $t$ dependence as $t$ increases, indicating sizable excited-state effects associated with the initial and final states. With large time intervals utilized in this work, obvious plateaus in proper time regions are observed. It is therefore natural to perform a constant fit for these lattice data in a suitable time region. All results of $V\left(q^2\right)$ and $A_i\left(q^2\right)~(i=0,1,2)$ with different momenta $\vec{p}$ are obtained in this way and are denoted by the color bands in
the figures. Since the local current is adopted in our calculation, the renormalization constant $Z_V^{cs}$ and $Z_A^{cs}$ are multiplied to obtain form factors $V\left(q^2\right)$ and $A_i\left(q^2\right)~(i=0,1,2)$ for each ensemble. The numerical fitting results are listed in Table~\ref{3pt} in Appendix~\ref{sec:ap2}.

\renewcommand{\thesubfigure}{(\roman{subfigure})}
\renewcommand{\figurename}{Figure}
\begin{figure}[htp]
\centering  
\subfigure{
\includegraphics[width=7cm]{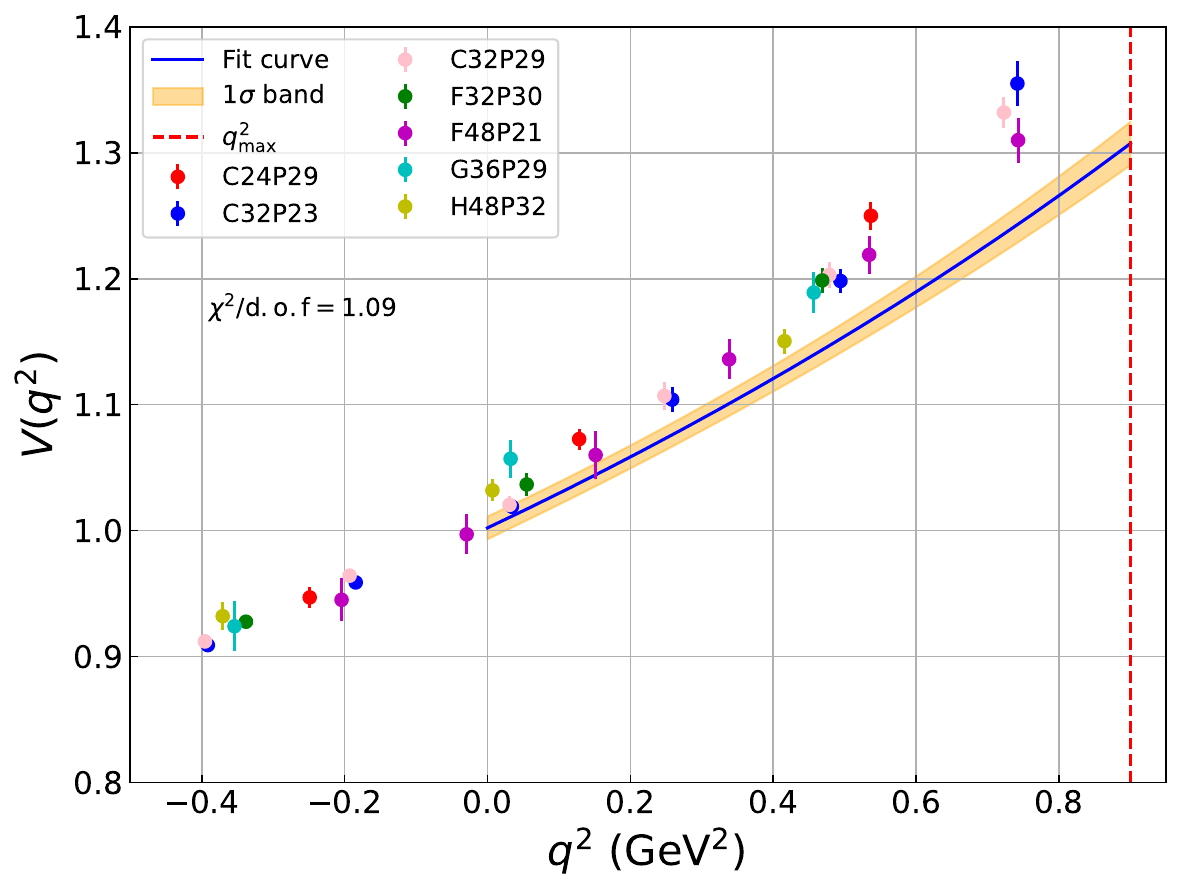}}
\subfigure{
\includegraphics[width=7cm]{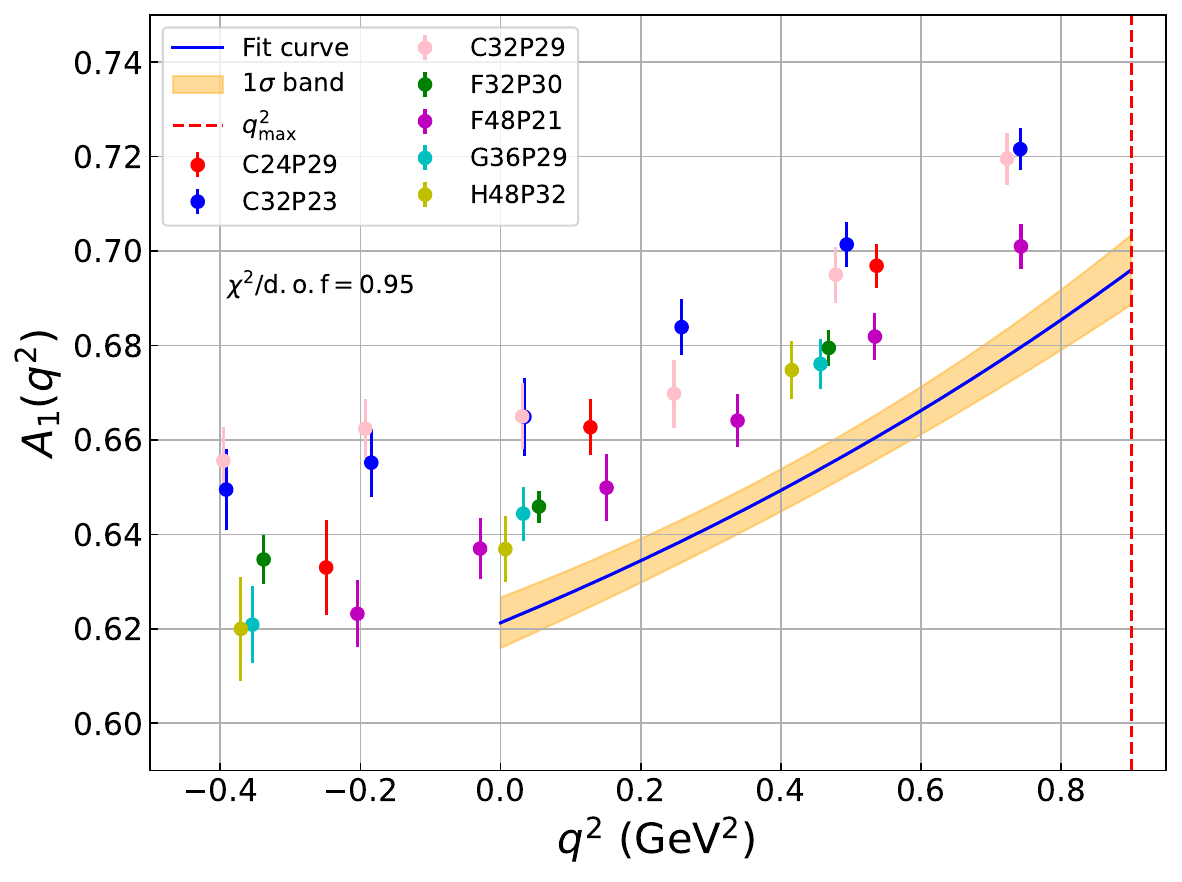}}
\subfigure{
\includegraphics[width=7cm]{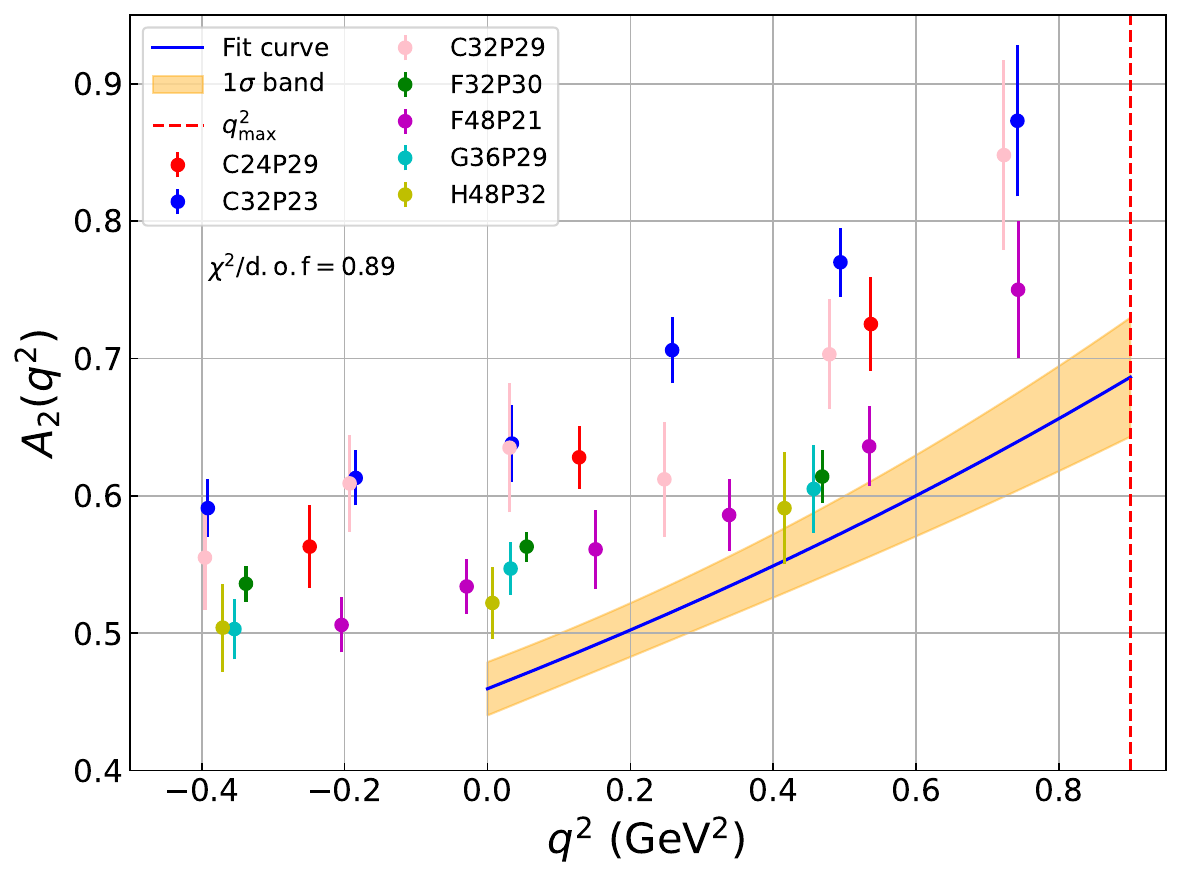}}
\subfigure{
\includegraphics[width=7cm]{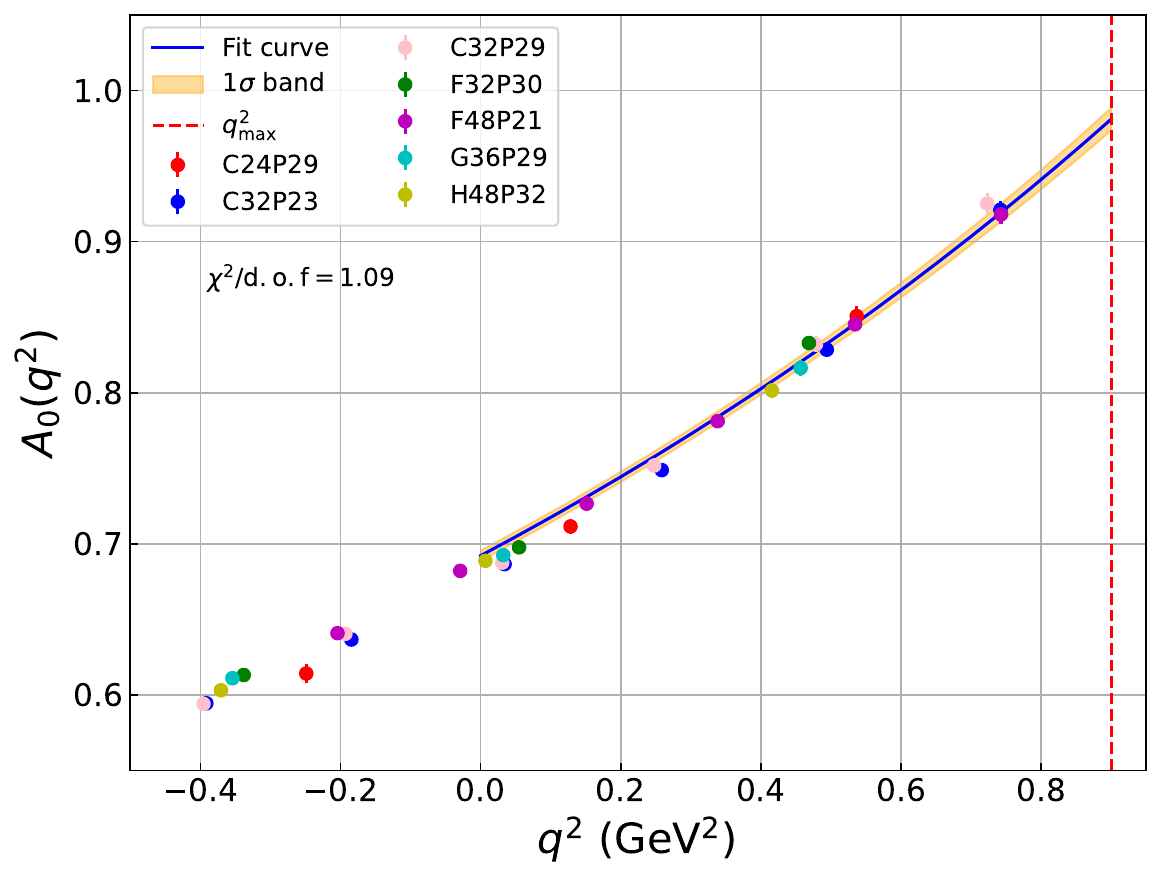}}
\caption{The lattice results of the $D_s\to\phi$ form factors and the extrapolation fitting. The shaded regions correspond to the final results at the continuum limit and physical pion mass.}
\label{extrapolation}
\end{figure}

To extrapolate these form factors to the continuum limit and physical pion mass globally, we use the ${z}$-expansion parameterization scheme. The fit functions are
\begin{flalign}
V\left(q^2,a,m_\pi\right)&=\frac{1}{1-q^2/m_{D_s^*}^2}\sum_{i=0}^2\left(c_i+d_ia^2\right)\left[1+f_i\left(m_\pi^2-m_{\pi,\text{phys}}^{2}\right)\right]z^{i},\nonumber\\
A_{0,1,2}\left(q^2,a,m_\pi\right)&=\frac{1}{1-q^2/m_{D_{s1}}^2}\sum_{i=0}^2\left(c_i+d_ia^2\right)\left[1+f_i\left(m_\pi^2-m_{\pi,\text{phys}}^{2}\right)\right]z^{i},
\label{expz}
\end{flalign}
with 
\begin{flalign}
z\left(q^2,t_0\right)=\frac{\sqrt{t_+-q^2}-\sqrt{t_+-t_0}}{\sqrt{t_+-q^2}+\sqrt{t_+-t_0}},
\label{z}
\end{flalign}
where $t_+=\left(m_{D_s}+m_\phi\right)^2$, $t_0=0$, $c_i$, $d_i$, and $f_i$ are parameters to be determined by fitting. Here, $m_{D_s}$, $m_{\phi}$, and pole masses $m_{D_s^*}$, $m_{D_{s1}}$ are fixed to their experiment values $m_{D_s}=1968.4~\text{MeV}$, $m_{\phi}=1019.5~\text{MeV}$, $m_{D_s^*}=2112.2~\text{MeV}$ and $m_{D_{s1}}=2459.5~\text{MeV}$~\cite{ParticleDataGroup:2024cfk}.

All results of the extrapolation and lattice data are shown in Fig.~\ref{extrapolation}, where the $q^2$-dependence of the form factors is presented. The solid lines with shaded regions denote the final physical results, whereas the colored data points with error bars are the numerical lattice results. Basically, $q^2$ in the phase space should be restricted in the region $\left(q^2_{\text{min}},q^2_{\text{max}}\right)=\left( 0, (m_{D_s}-m_\phi)^2\right)$. When the $\phi$ particle carries sufficiently high momentum, the $q^2$ can exceed the phase-space boundary, resulting in a negative value. These form factors with negative $q^2$ can be incorporated into the fit as more stringent constraints. In real calculations, the lattice data in the region $q^2\in (-0.4\mbox{ GeV}^2,q^2_{\text{max}})$ are utilized for the extrapolation. It is observed that the form factors can be described well by a continuum extrapolation that contains the $a^2$-order and the $m_{\pi}^2$-order terms. The fitting parameters and the covariance matrices are listed in the Appendix~\ref{sec:ap2}. 

\renewcommand{\thesubfigure}{(\roman{subfigure})}
\renewcommand{\figurename}{Figure}
\begin{figure}[htp]
\centering  
\subfigure[Form factors at $m_\pi=0.135~\text{GeV}$.]{
\includegraphics[width=7cm]{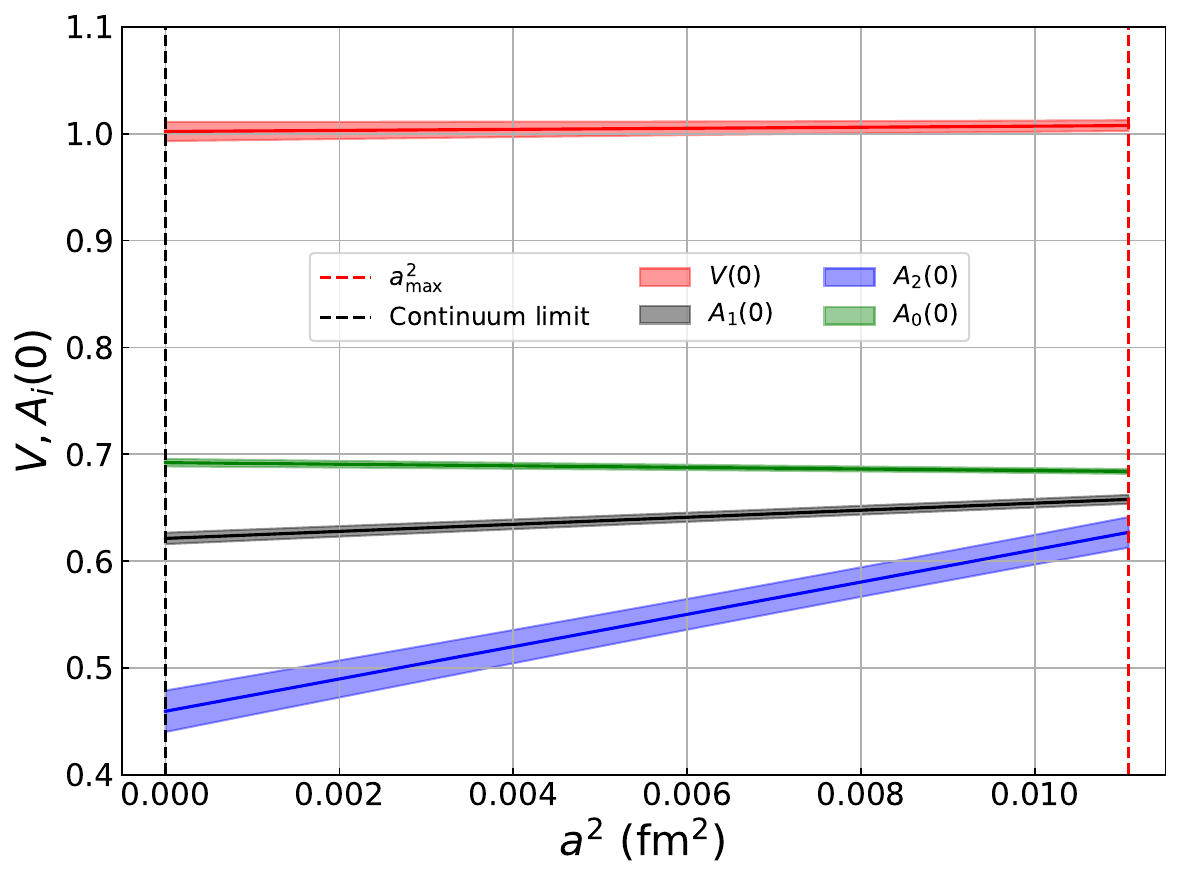}}
\subfigure[Form factors at $a=0.0~\text{fm}$.]{
\includegraphics[width=7cm]{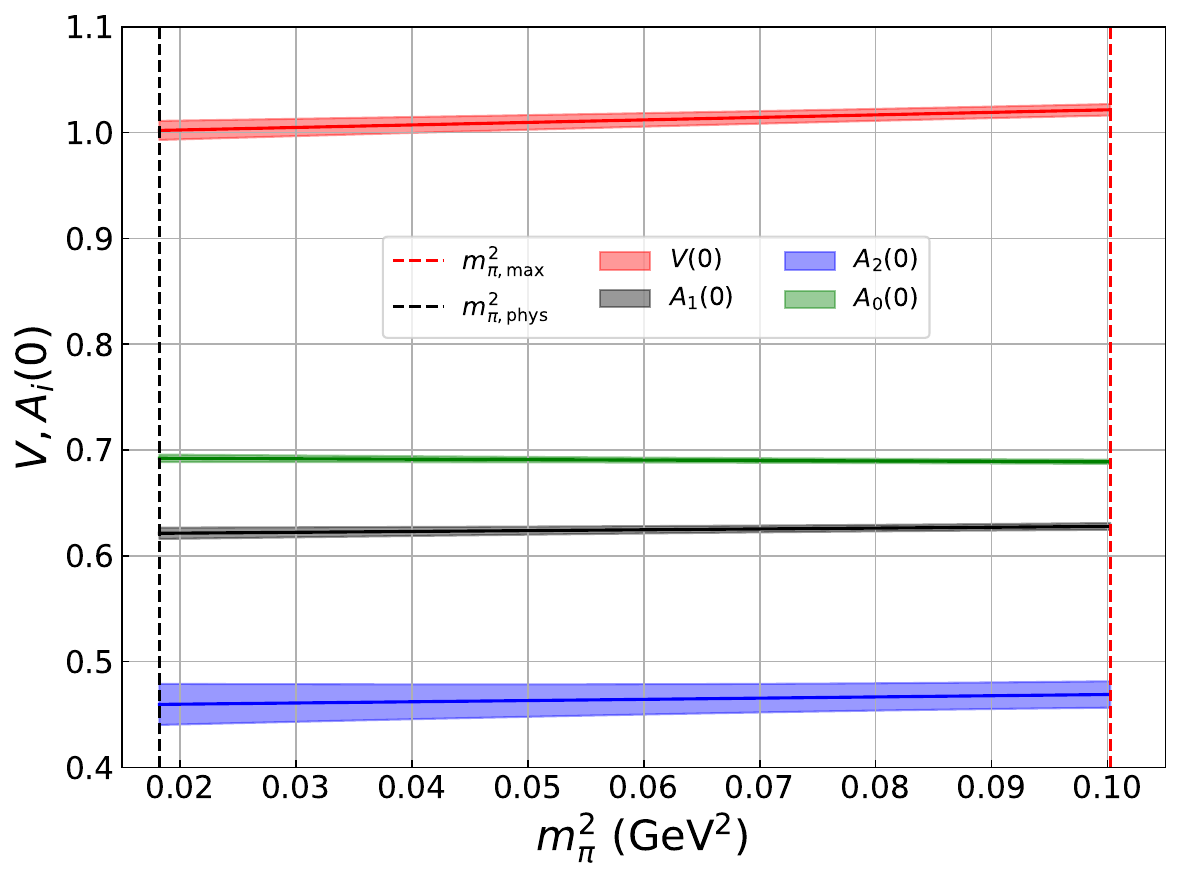}}
\caption{The lattice results of form factors at $q^2=0$ as a function of the lattice spacing $a$ and pion mass $m_\pi$.}
\label{aandpi}
\end{figure}

\renewcommand{\tablename}{Table}
        \begin{table}[htbp]
	\centering  
     \caption{Numerical results of the conventional form factors $V\left(0\right)$, $A_i\left(0\right)~(i=0,1,2)$, and $r_{V,A,0}$.}
 \label{limit}
 \addvspace{5pt}
	\scalebox{0.75}{\begin{tabular}{ccccccccc}
		\hline\hline\noalign{\smallskip}	
		&$V\left(0\right)$&$A_1\left(0\right)$&$A_2\left(0\right)$&$A_0\left(0\right)$&$A_3\left(0\right)-A_0\left(0\right)$&$r_V$&$r_2$&$r_0$\\
	\noalign{\smallskip}\hline\noalign{\smallskip}
This work &$1.002(9)$&$0.621(5)$&$0.460(19)$&$0.692(4)$&$0.004(12)$&$1.614(19)$&$0.741(31)$&$1.114(11)$\\
HPQCD~\cite{Donald:2013pea}&$1.059(124)$&$0.615(24)$&$0.457(78)$&$0.706(37)$&Enforced to $0$&$1.720(210)$&$0.740(120)$&$1.140(60)$\\
  \noalign{\smallskip}\hline
	\end{tabular}}

\end{table}

\begin{figure}[htp]
\centering  
\subfigure{
\includegraphics[width=10cm]{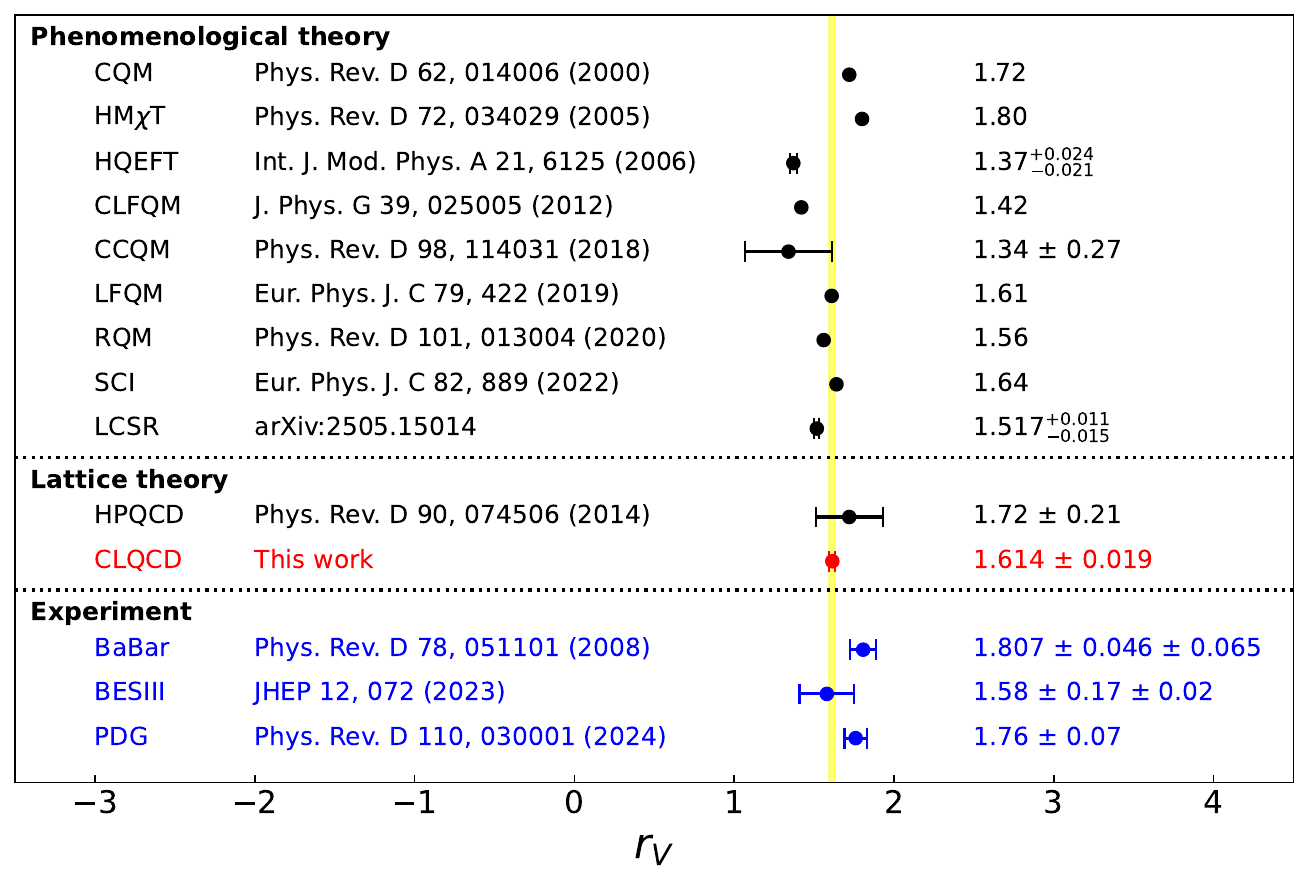}}
\subfigure{
\includegraphics[width=10cm]{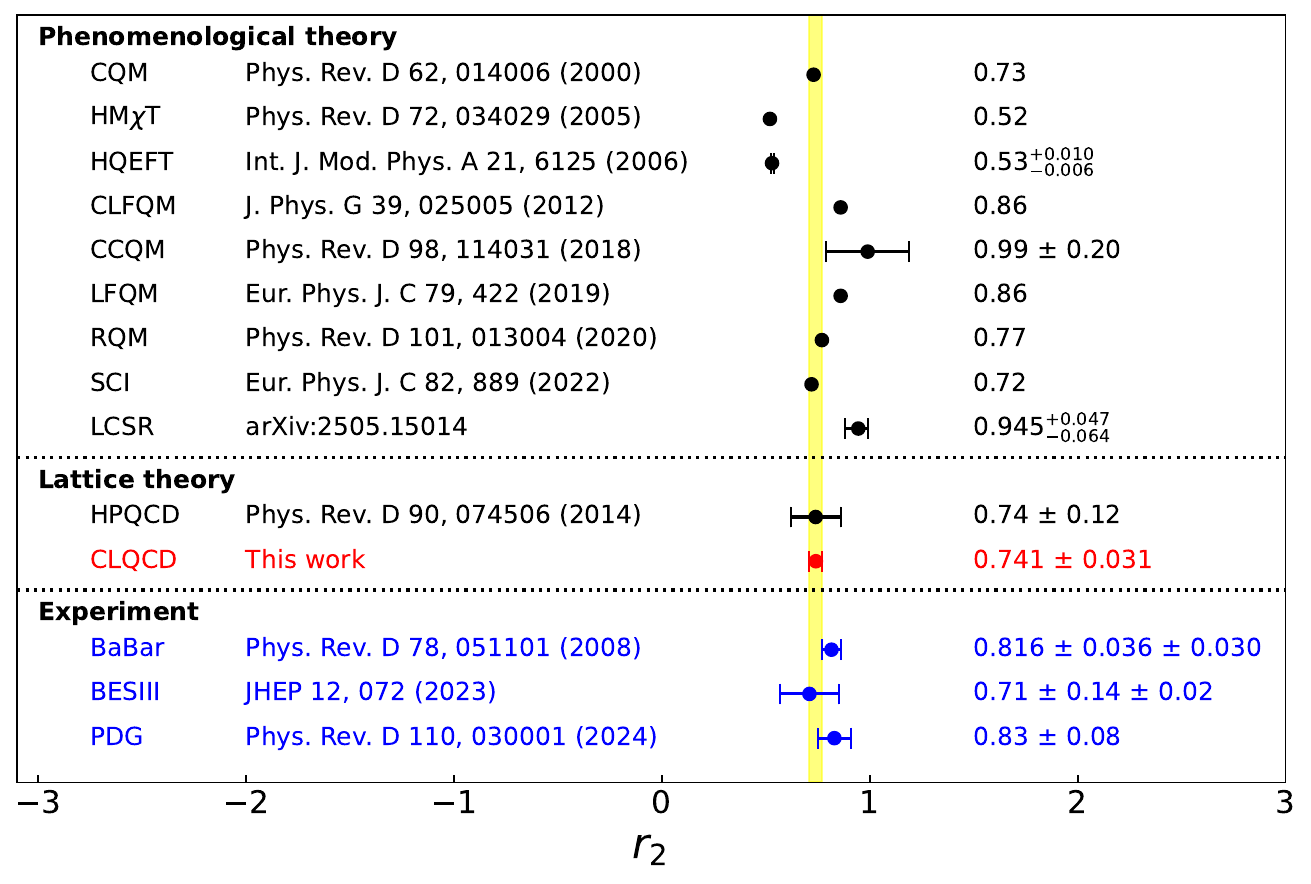}}
\caption{The comparison of the $r_V$ (top) and $r_2$ (bottom) calculated in this work, measured by experiments, and those given by different theoretical predictions.}
\label{rVr2}
\end{figure}

The numerical results of form factors at zero transfer momentum are shown in Fig.~\ref{aandpi}, where the dependences on the lattice spacing $a$ and pion mass $m_\pi$ are investigated separately.
The discretization effects of these form factors exhibit different behaviors, with the $a^2$-dependence of $A_2(0)$ being larger than others. None of the form factors show obvious dependence on the pion mass, which is probably because no light valence quarks are involved directly in the semileptonic decay $D_s\rightarrow \phi$. Compared with previous lattice calculations~\cite{Donald:2013pea}, we not only present the first results obtained at the physical point using four different lattice spacings and multiple pion masses, but also improve the precision of the form factors by up to an order of magnitude, thereby providing experiments with a much more precise theoretical benchmark. The calculated $A_3\left(0\right)-A_0\left(0\right)$ is consistent with $0$, satisfying the kinematic constraint. Detailed comparisons between our results and previous lattice/phenomenological theory/experiment results are listed in Table~\ref{limit} and Fig.~\ref{rVr2}. The current precision of the form factors has exceeded that of both the experimental measurements and the PDG. More stringent tests of the Standard Model require more precise measurements in future experiments.

\renewcommand{\thesubfigure}{(\roman{subfigure})}
\renewcommand{\figurename}{Figure}

\subsection{Differential decay width and branching fractions}
\label{res:D}

\begin{figure}[htp]
\centering  
\subfigure{
\includegraphics[width=7cm]{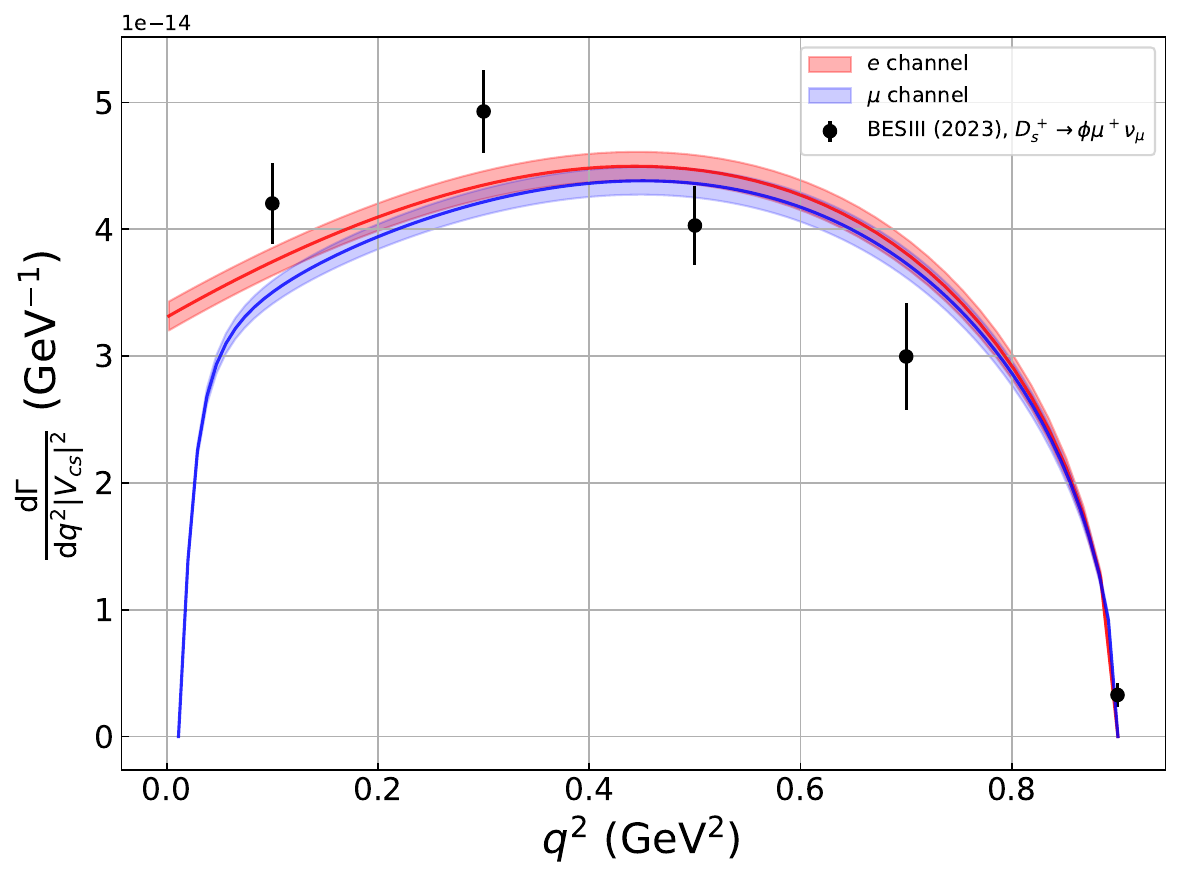}}
\subfigure{
\includegraphics[width=7cm]{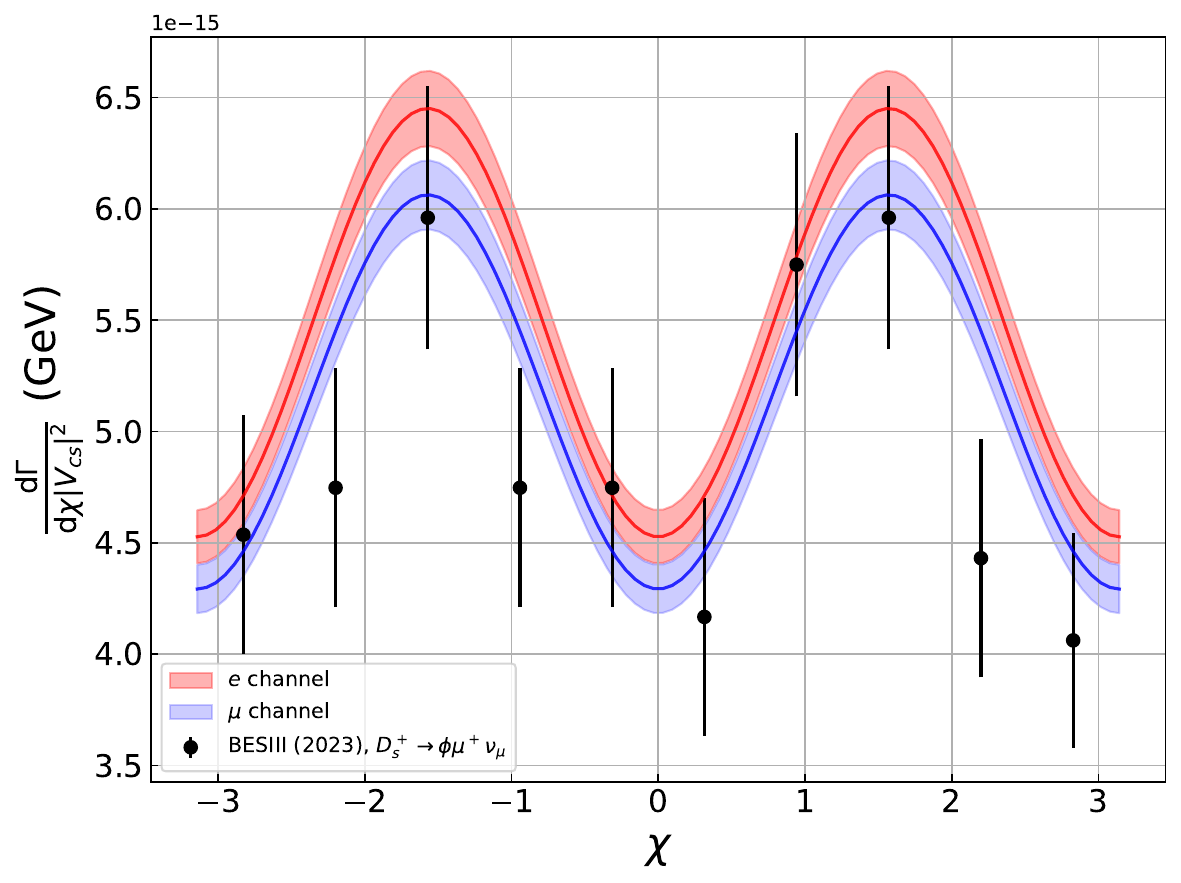}}
\subfigure{
\includegraphics[width=7cm]{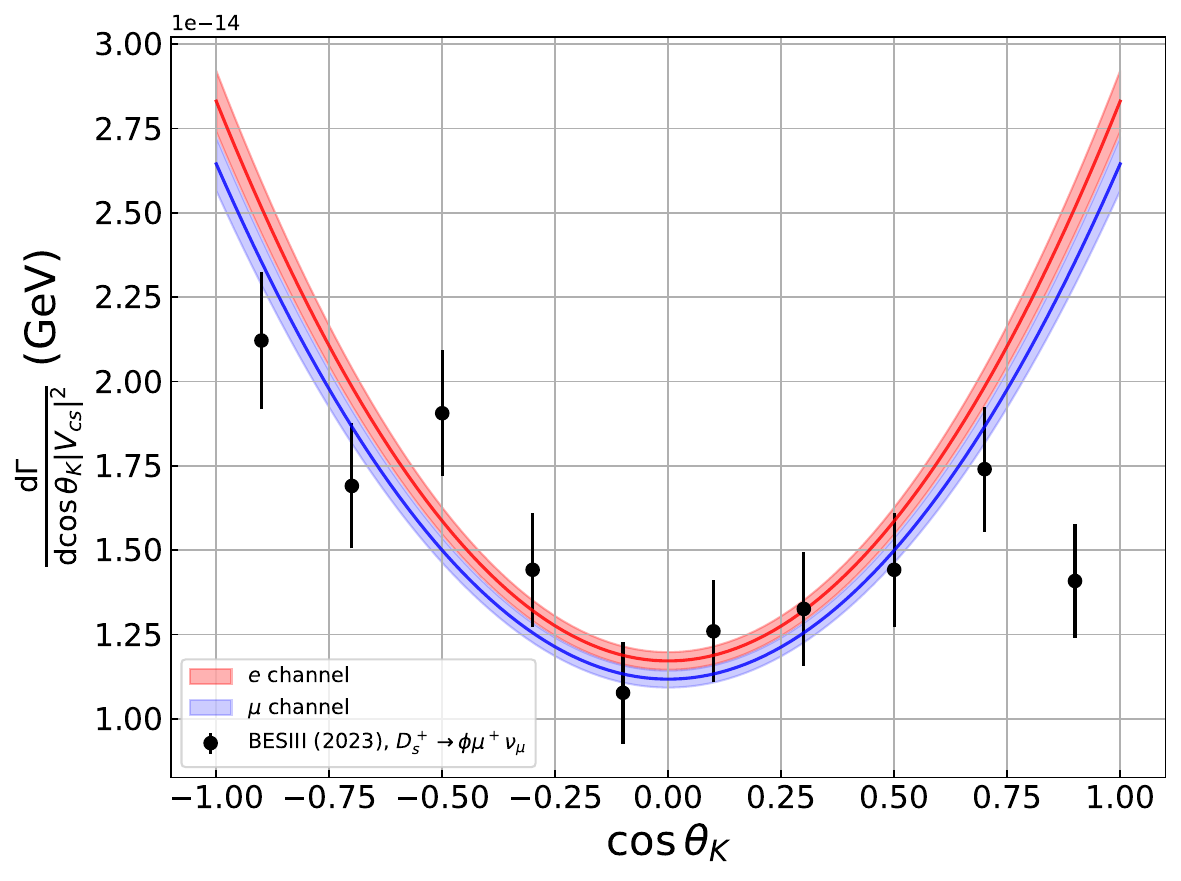}}
\subfigure{
\includegraphics[width=7cm]{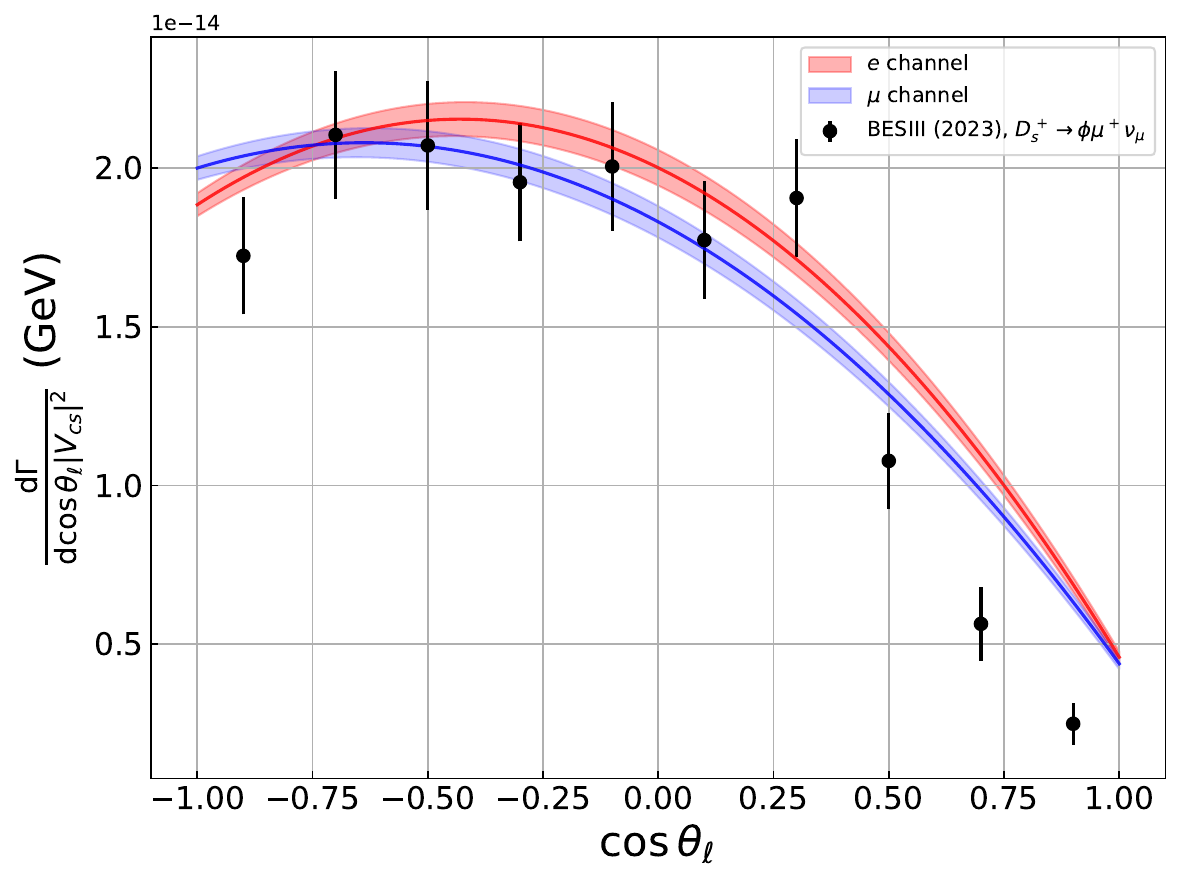}}
\caption{The differential decay width distributions of $e$ and $\mu$ channels.}
\label{width}
\end{figure}

As emphasized above, semileptonic decays of charmed hadrons into vector final states contain richer differential information. Plugging the form factors we have obtained before into Eq.~(\ref{eq:four_body_br}-\ref{eq:V_A_FFS}) yields the partial width as a function of $q^2$, $\cos\theta_l$, $\cos\theta_K$, and $\chi$, respectively. To facilitate direct comparison with experiments, one can integrate other variables and only reserve the one of interest, for example, the $q^2$-dependence. These individual differential distributions and the corresponding experimental measurements are shown in Fig.~\ref{width}. Since the calculation has considered the effect of the lepton mass, we can provide separate distributions for the electron and muon final states, as shown by the red and blue colors, respectively. 


\renewcommand{\thesubfigure}{(\roman{subfigure})}
\renewcommand{\figurename}{Figure}

The total decay width is computed directly after integrating the $q^2$ in the full phase space. Taking into account the PDG values of CKM matrix element $|V_{cs}|=0.975\pm 0.006$, $D_s$ lifetime $\tau_{D_s}=\left(5.012\pm 0.022\right)\times 10^{-13}~\text{s}$, and the Plank constant $\hbar$~\cite{ParticleDataGroup:2024cfk}, it then determines the branching fraction by the formula
$\mathcal{B}(D_s\to\phi\ell\nu_\ell)={\Gamma(D_s\to\phi\ell\nu_\ell)\times\tau_{D_s}}/{\hbar}$, where $\Gamma(D_s\to\phi\ell\nu_\ell)=\Gamma_{\text{latt}}\times |V_{cs}|^2$ is the total decay width. Finally, we obtain the total branching fraction of the corresponding decay channels
\begin{flalign}
\mathcal{B}(D_s\to\phi e\nu_e)&=2.493(66)_{\text{stat}}(31)_{|V_{cs}|}\times 10^{-2},
 \nonumber \\
\mathcal{B}(D_s\to\phi\mu\nu_\mu)&=2.351(60)_{\text{stat}}(29)_{|V_{cs}|}\times 10^{-2} ,
\label{eq:Br}
\end{flalign}
where the first errors are statistical, including various sources, such as the renormalization constants $Z_V^{c},Z_V^s$, uncertainty of the lattice spacing, pion mass error, continuum limit, physical pion mass extrapolation. The second errors are estimated from the uncertainty of the PDG value $|V_{cs}|$. The ratio of the $\mu$ channel branching fraction over the $e$ channel branching fraction $\mathcal{R}_{\mu/e}$ can also be determined directly, the value of which is collected in Table.~\ref{BF}.


 \renewcommand{\tablename}{Table}
        \begin{table}[htbp]
	\centering  
     \caption{The branching fractions predicted by this work and experiment results.}
 \label{BF}
 \addvspace{5pt}
	\begin{tabular}{cccccccc}
		\hline\hline\noalign{\smallskip}	
	$\mathcal{B}(D_s\to\phi \ell\nu_\ell)\times 10^{2}$	&{$\mu$ channel}&$e$ channel&$\mathcal{R}_{\mu/e}$\\
 \noalign{\smallskip}\hline\noalign{\smallskip}	
This work&$2.351(67)$&$2.493(73)$&$0.9432(13)$\\
BaBar~\cite{BaBar:2008gpr}&$-$&$2.61(17)$&$-$\\
CLEO~\cite{Hietala:2015jqa}&$-$&$2.14(19)$&$-$\\
BESIII (2018)~\cite{BESIII:2017ikf}&$1.94(54)$&$2.26(46)$&$0.86(29)$\\
BESIII (2023)~\cite{BESIII:2023opt}&$2.25(11)$&$-$&$-$\\
PDG~\cite{ParticleDataGroup:2024cfk}&$2.24(11)$&$2.34(12)$&$0.957(68)$\\
  \noalign{\smallskip}\hline
	\end{tabular}

\end{table}

\section{Discussion}
\label{sec:dis}

\subsection{Different parameterization schemes}
\label{dis:D}
In the above calculations, the physical results of the form factor are obtained by the $z$-expansion, which is widely used in the literature. To further investigate the systematic effects of the expansion, we also use three other parameterization schemes. They are the single pole form, modified pole form~\cite{Becirevic:1999kt}, and phase moment form~\cite{Yao:2019vty}. 

\begin{itemize}
    \item \textbf{Single pole form}: The fit function is
    \begin{flalign}
    &&
F\left(q^2,a,m_\pi\right)=\frac{1}{1-q^2/h^2}\left(c+d a^2\right)\left[1+f\left(m_\pi^2-m_{\pi,\text{phys}}^{2}\right)\right],
    &&
\end{flalign}
where $F$ denotes $V$ and $A_i~(i=0,1,2)$; $c,d,f,h$ are parameters to be determined by fitting.
    \item \textbf{Modified pole form}: The fit function is
    \begin{flalign}
    &&
F\left(q^2,a,m_\pi\right)=\frac{1}{\left(1-q^2/m_{\text{pole}}^2\right)\left(1-h q^2/m_{\text{pole}}^2\right)}\left(c+d a^2\right)\left[1+f\left(m_\pi^2-m_{\pi,\text{phys}}^{2}\right)\right],
    &&
\end{flalign}
where $F$ denotes $V$ and $A_i~(i=0,1,2)$; $c,d,f,h$ are parameters to be determined by fitting, and $m_{\text{pole}}$ is $m_{D_s^*}$ and $m_{D_{s1}}$ for vector and axial vector form factors, respectively.
\item \textbf{Phase moment  form}: The fit function is 
    \begin{flalign}\label{eq:phase_form}
    &&
F\left(q^2,a,m_\pi\right)=F\left(s_0\right)\left(1+d a^2\right)\left[1+f\left(m_\pi^2-m_{\pi,\text{phys}}^{2}\right)\right]\prod_{n=0}^\infty\exp\left(\frac{q^2-s_0}{s_{\text{th}}}\mathcal{A}^F_n\frac{q^{2n}}{s^n_{\text{th}}}\right),
    &&
\end{flalign}
where $F$ denotes $V$ and $A_i~(i=0,1,2)$; $F\left(s_0\right),d,f,\mathcal{A}^F_n$ are parameters to be determined by fitting, and $s_{\text{th}}\equiv\left(m_{D_s}+m_{\phi}\right)^2$. 
As stated in the Ref.~\cite{Yao:2019vty}, $\mathcal{A}^F_n$ are called the phase moments, which are related to phases of the form factors in the physical $D_s \phi$ scattering region. In the fitting, we set $s_0=0$ and take the truncation to the second order $n=1$. The phase moments $\mathcal{A}^F_n$ are extracted and listed in the Appendix~\ref{sec:ap2}.
\end{itemize}

The numerical results are shown in Table~\ref{comp}. The extrapolation results are well consistent with each other, which shows little parameterization scheme dependence. Since the pole and phase moment forms have more stringent constraints for the form factors, it naturally leads smaller fitting errors. From a more traditional and conservative perspective, we adopt the $z$-expansion scheme to obtain our nominal results in this work.

 \renewcommand{\tablename}{Table}
        \begin{table}[htbp]
	\centering  
     \caption{Comparisons of four parameterization schemes' results of form factors and branching fractions.}
 \label{comp}
 \addvspace{5pt}
	\begin{tabular}{ccccccc}
		\hline\hline\noalign{\smallskip}	
		&{$z$-expansion}&{Single pole}&{Modified pole}&{Phase moment}\\
	\noalign{\smallskip}\hline\noalign{\smallskip}
{$V\left(0\right)$}  &{$1.002(9)$} &{$1.002(9)$} &{$1.004(9)$}&{$0.998(9)$}\\		
{$A_1\left(0\right)$}&{$0.621(5)$} &{$0.624(4)$} &{$0.624(4)$}&{$0.622(4)$}\\		
{$A_2\left(0\right)$}&{$0.460(19)$}&{$0.470(19)$}&{$0.471(19)$}&{$0.464(18)$}\\		
{$A_0\left(0\right)$}&{$0.692(4)$} &{$0.688(3)$} &{$0.689(3)$}&{$0.688(3)$}\\	
$A_3\left(0\right)-A_0\left(0\right)$&$0.004(12)$&$0.008(11)$&$0.006(11)$&$0.008(11)$\\
{$r_V$}&{$1.614(19)$}&{$1.606(18)$} &{$1.609(18)$}&{$1.605(18)$}\\		
{$r_2$}&{$0.741(31)$}&{$0.753(31)$} &{$0.755(31)$}&{$0.746(29)$}\\	
{$r_0$}&{$1.114(11)$}&{$1.1026(85)$}&{$1.1042(86)$}&{$1.1061(86)$}\\	
$\mathcal{B}(D_s\to\phi \mu\nu_\mu)/ 10^{-2}$&$2.351(67)$&$2.367(54)$&$2.362(50)$&$2.363(55)$\\
$\mathcal{B}(D_s\to\phi e\nu_e)/10^{-2}$&$2.493(73)$&$2.511(59)$&$2.504(55)$&$2.505(60)$\\
 $\mathcal{R}_{\mu/e}$&$0.9432(13)$&$0.9427(12)$&$0.9431(12)$&$0.9432(11)$\\ 
  \noalign{\smallskip}\hline
	\end{tabular}
\end{table}

\subsection{Finite-volume effects}
\label{dis:A}
The physical volumes of these gauge ensembles are around $2.5\sim 3.7$ fm with pion mass ranging from $210\sim 320~\text{MeV}$. The ensembles C24P29 and C32P29 share the same lattice spacing and pion mass, and can be used to examine the finite-volume effects. The fitting formula of form factors $V\left(q^2\right),A_i\left(q^2\right)~(i=0,1,2)$ are given as
\begin{flalign}
    V&=\frac{1}{1-q^2/m_{D_s^*}^2}\left(a_0+a_1z+a_2z^2\right),\nonumber\\
    A_{0,1,2}&=\frac{1}{1-q^2/m_{D_{s1}}^2}\left(a_0+a_1z+a_2z^2\right).
\label{zexp}
\end{flalign}

\renewcommand{\tablename}{Table}
        \begin{table}[htbp]
	\centering  
     \caption{Numerical values of the form factors $V(0),A_0(0),A_1(0),A_2(0)$ and the corresponding $\chi^2/{\text{d.o.f}}$ values from the ensemble C24P29 and C32P29.}
 \label{comp2}
 \addvspace{5pt}
	\begin{tabular}{ccccccc}
		\hline\hline\noalign{\smallskip}	
		&{C24P29}& $\chi^2/\textrm{d.o.f}$  & {C32P29}& $\chi^2/\textrm{d.o.f}$  &{Combined} & $\chi^2/\textrm{d.o.f}$\\
	\noalign{\smallskip}\hline\noalign{\smallskip}
{$V\left(0\right)$}&{$1.0271(66)$}   & $0.03$ & {$1.0167(43)$}& $0.29$ &{$1.0202(36)$} & $0.63$ \\		
{$A_1\left(0\right)$}&{$0.6523(55)$} & $0.01$ &{$0.6639(37)$} & $0.24$ &{$0.6605(30)$} & $0.82$\\		
{$A_2\left(0\right)$}&{$0.605(18)$}  & $0.01$ &{$0.616(20)$}  & $0.53$ &{$0.609(14)$}  & $0.38$\\		
{$A_0\left(0\right)$}&{$0.6759(45)$} & $0.07$ &{$0.6835(29)$} & $0.26$ &{$0.6811(24)$} & $0.54$ \\	
  \noalign{\smallskip}\hline
	\end{tabular}
\end{table}

\begin{figure}[!htb]
\renewcommand{\figurename}{Figure}
      \centering
      \includegraphics[width=10cm]
      {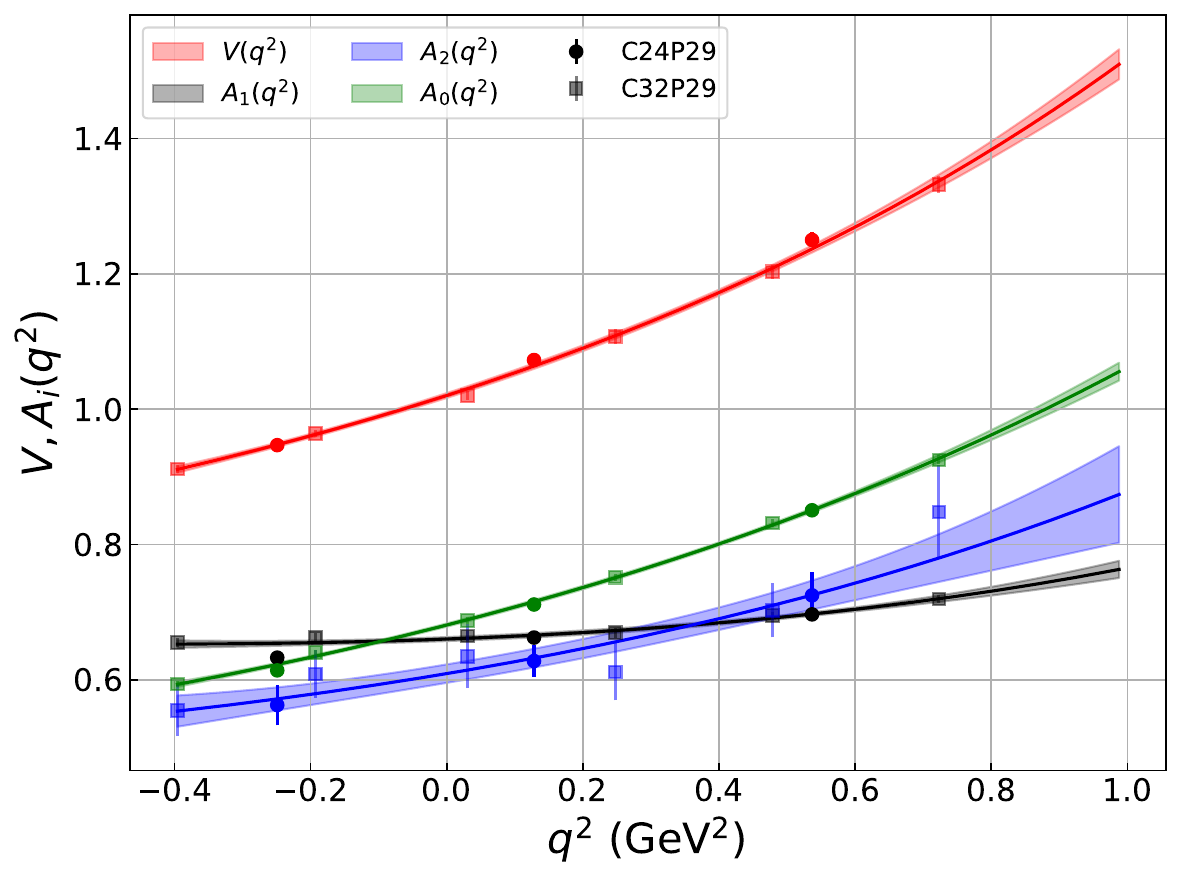}
      \caption{Form factor fitting results using C24P29 and C32P29 ensembles points with $z$-expansion from Eq. (\ref{zexp}).}
      \label{finiteV}
    \end{figure}

In Fig.~\ref{finiteV}, a joint fitting for C24P29 and C32P29 is presented. Points of the same color but different shapes are from different gauge ensembles. It is evident that all of them are well described by a single curve, and the $\chi^2/\textrm{d.o.f}$ are also quite reasonable, where the d.o.f is the number of degrees of freedom in each fit. Numerical results of the zero transfer momentum form factors on these two ensembles are summarized in Table.~\ref{comp2}. The values of joint fitting are well consistent with the other two individual fittings within $1\sigma$. It therefore provides reliable evidence that the finite-volume effects in our calculation are under control.    
\subsection{Comparisons with previous theory/experiment results}
\label{dis:B}

\begin{figure}[htp]
\centering  
\subfigure{
\includegraphics[width=10cm]{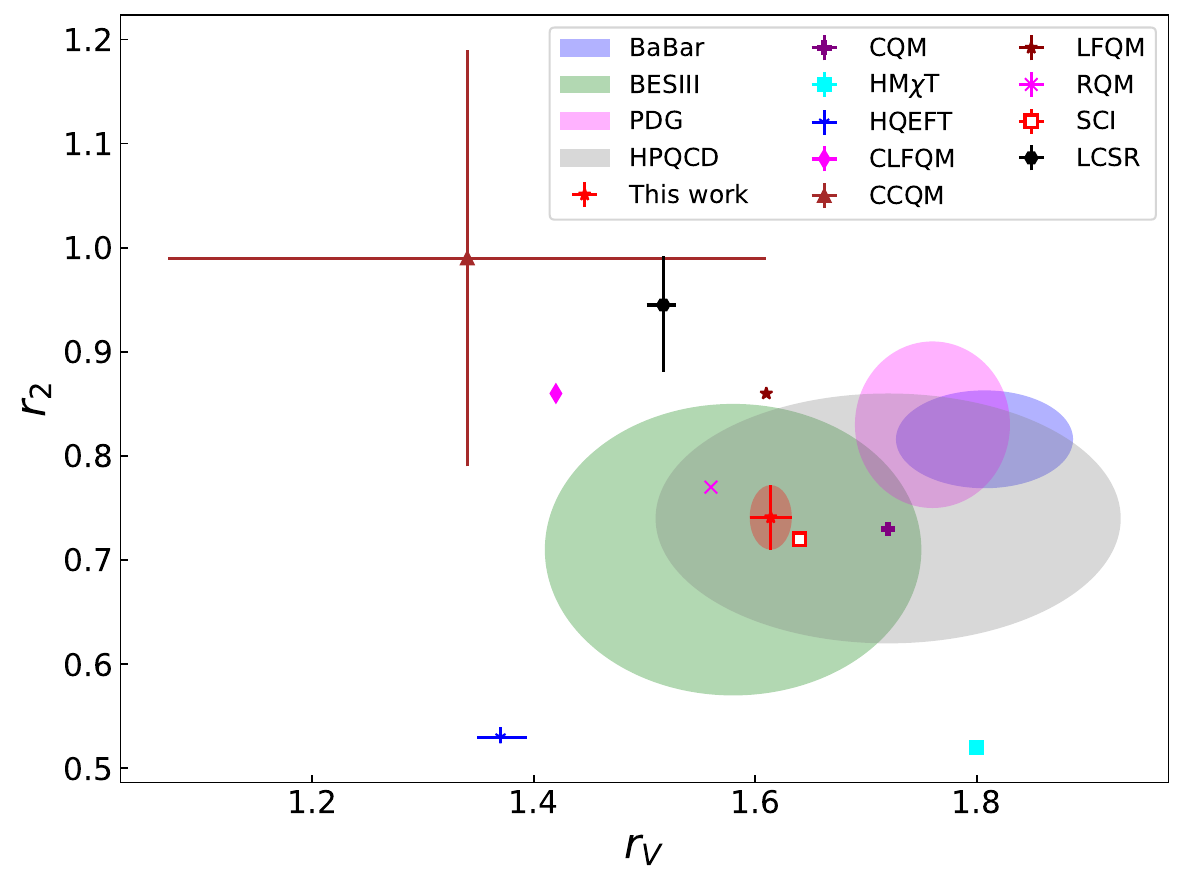}}
\caption{The comparison of the $r_V$, $r_2$, calculated in this work, and those given by different theoretical predictions and experimental measurements.}
\label{r2_rv}
\end{figure}

Our results of form factors reach the precision of $1\%-4\%$, which greatly improves the previous lattice QCD results and obtains the most precise determination to date. Compared with the previous lattice QCD calculations, we use multiple ensembles with different lattice spacings down to $0.05~\text{fm}$ and different pion masses to arrive at the physical point after the continuum limit and physical pion mass extrapolation. From the left figure of Fig.~\ref{r2_rv}, the $r_2$ is consistent with the PDG value within $1\sigma$, but $r_V$ has an inconsistency with the PDG. Since the PDG averages a result with large error and one with small error, its central value naturally converges toward the latter. A more precise experimental measurement in the future may clarify this discrepancy.

Our branch fraction results in Eq.~(\ref{eq:Br}) achieve a precision of $3\%$, in agreement with the latest BESIII experiment and providing the most accurate lattice QCD prediction available for future experimental tests. In the differential decay width distributions shown in Fig.~\ref{width}, we find that at high $q^2$ area and close to the border area of angular distributions, there is a visible difference between the lattice calculations and the BESIII data. These discrepancies remain to be clarified by future experiments with higher detection efficiency and larger statistical samples.

\renewcommand{\thesubfigure}{(\roman{subfigure})}
\renewcommand{\figurename}{Figure}

\renewcommand{\thesubfigure}{(\roman{subfigure})}
\renewcommand{\figurename}{Figure}
\begin{figure}[htp]
\centering  
\subfigure{
\includegraphics[width=10cm]{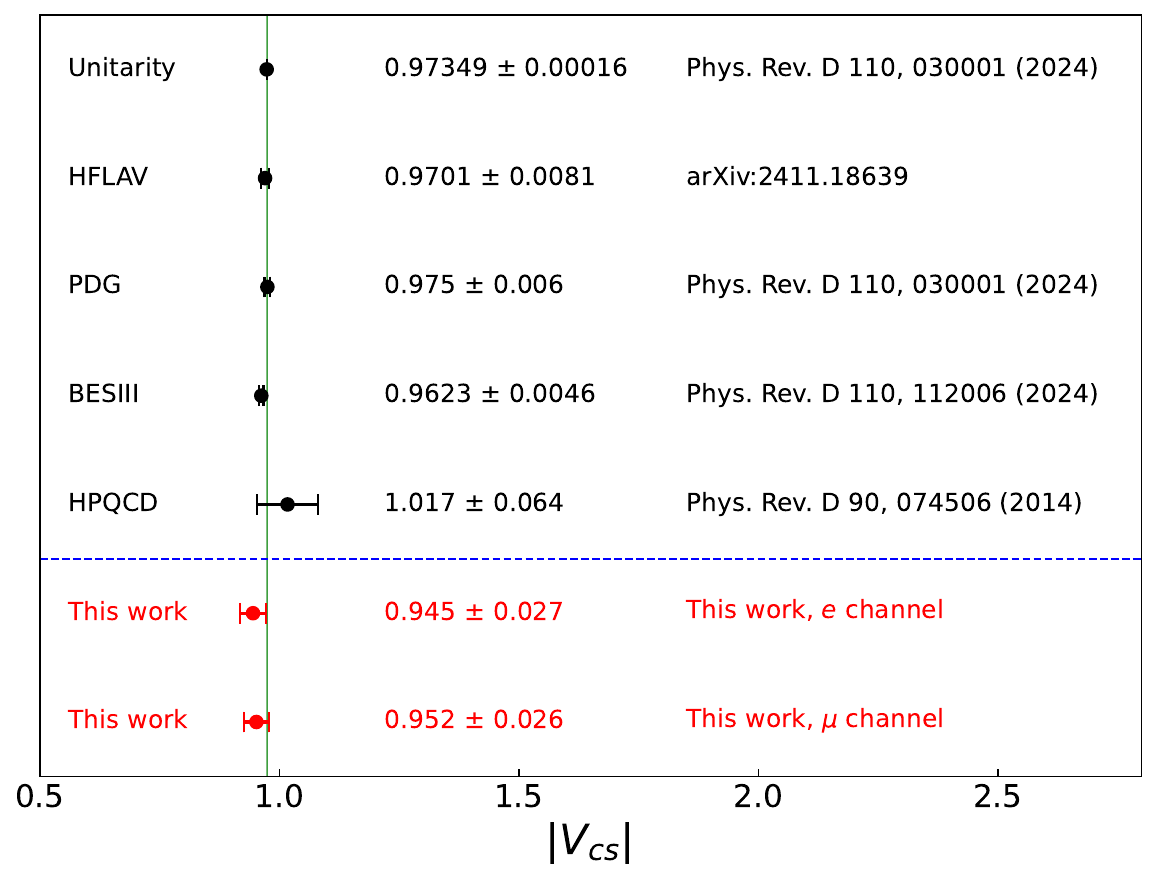}}
\caption{The $|V_{cs}|$ calculated in this work and those given by others.}
\label{Vcs}
\end{figure}

\subsection{CKM matrix element $|V_{cs}|$}
\label{dis:C}
We extract the CKM matrix element $|V_{cs}|$ by comparing the branching fractions to the PDG~\cite{ParticleDataGroup:2024cfk} results, which are $\mathcal{B}_{\text{PDG}}(D_s\to\phi \mu\nu_\mu)=2.24(11)\times 10^{-2}$ and $\mathcal{B}_{\text{PDG}}(D_s\to\phi e\nu_e)=2.34(12)\times 10^{-2}$. The lattice calculations and PDG results differ by a factor of $|V_{cs}|^2$, therefore we extract the $|V_{cs}|$ through 
\begin{flalign}
|V_{cs}|=\sqrt{\frac{1}{\Gamma_{\text{latt}}}\times\frac{\mathcal{B}_{\text{PDG}} \times\hbar}{\tau_{D_s}}}.
\end{flalign}
The $|V_{cs}|$ results are $0.952(12)_{\text{stat}}(23)_{\text{PDG}}$ and $0.945(12)_{\text{stat}}(24)_{\text{PDG}}$, which are extracted from the $\mu$ and $e$ channels, respectively. The first uncertainty comes from the lattice simulation, and the second comes from the PDG error. They are in agreement with the unitarity (global fit in the Standard Model)~\cite{ParticleDataGroup:2024cfk}, HFLAV~\cite{HeavyFlavorAveragingGroupHFLAV:2024ctg}, PDG~\cite{ParticleDataGroup:2024cfk}, BESIII~\cite{BESIII:2024slx}, and HPQCD~\cite{Donald:2013pea} values within $1\sigma$ (as shown in Fig.~\ref{Vcs}), which demonstrates consistency with the CKM unitarity within uncertainties. The error mainly comes from PDG error, and can still be improved from future experiments from BESIII or super $\tau$-charm factory.

\section{Conclusion}
\label{sec:conclu}

In this work, we present a systematic lattice calculation on $D_s\to\phi\ell\nu_\ell$ semileptonic decay. Seven (2+1)-flavor Wilson-clover gauge ensembles with different lattice spacings and pion masses are utilized. The calculations cover the full $q^2$ range, leading to a well-controlled accuracy. After the continuum limit and physical pion mass extrapolation, the form factor ratios are obtained to be $r_V=1.614(19)$ and $r_2=0.741(31)$ with $z$-expansion. For the $\phi$ meson, we obtain physical mass $m=1.0211(76)~\text{GeV}$ and decay constant $f_\phi=0.2462(41)~\text{GeV}$, which agrees with the previous $\chi$QCD and HPQCD results. The decay angular distributions are constructed using the helicity amplitudes, which are in good agreement with the BESIII experiment results. We finally obtain the branching fractions $\mathcal{B}(D_s\to\phi e\nu_e)=2.493(66)_{\text{stat}}(31)_{|V_{cs}|}\times 10^{-2}$ and $\mathcal{B}(D_s\to\phi\mu\nu_\mu)=2.351(60)_{\text{stat}}(29)_{|V_{cs}|}\times 10^{-2}$. The corresponding ratio
of the branching fractions between lepton $\mu$ and $e$ is $\mathcal{R}_{\mu/e}=0.9432(13)$, which improves the accuracy significantly and can be tested in future experiments. In addition, combining with the experimental data, the CKM matrix element $|V_{cs}|$ is extracted to be $0.952(12)_{\text{stat}}(23)_{\text{PDG}}$ and $0.945(12)_{\text{stat}}(24)_{\text{PDG}}$ for $\mu$ and $e$ channels, which is consistent with the experimental and CKM unitarity values.

The scalar function scheme employed in this work can be widely applied to other pseudoscalar to vector semileptonic decays, such as $D_s\to K^*$~\cite{BESIII:2018xre}, $D\to K^*$~\cite{BESIII:2024qnx}, and $D\to\rho$~\cite{BESIII:2024lxg} channels. For extensions of this work, future improvements could consider the $\phi$ decay width and estimate the contribution of disconnected diagrams.

\acknowledgments
We thank the CLQCD collaborations for providing us their gauge configurations with dynamical fermions~\cite{CLQCD:2023sdb,CLQCD:2024yyn}, which are generated on the Southern Nuclear Science Computing Center(SNSC), HPC Cluster of ITP-CAS, IHEP-CAS and CSNS-CAS, and the Siyuan-1 cluster supported by the Center for High Performance Computing at Shanghai Jiao Tong University. 
The calculations were performed using the Chroma software suite~\cite{Edwards:2004sx} with QUDA~\cite{Clark:2009wm,Babich:2011np,Clark:2016rdz} through HIP programming model~\cite{Bi:2020wpt}.
G.F. thanks Nakul Soni, De-Liang Yao, Jialei Shi, Tao Luo, Patrick Roudeau and Justine Serrano for helpful discussions. 
This work is supported in part by National Key Research and Development Program of China under Contract No. 2023YFA1606000.
The authors acknowledge support from NSFC under Grant No. 12293060, 12293063, 12293065, 12305094, 12192264, 12505099, 12075253. Y.M. also thanks the support from the Young Elite Scientists Sponsorship Program by Henan Association for Science and Technology with Grant No. 2025HYTP003. K.Z. thanks the support from the Cross Research Project of CCNU No. 30101250314. The numerical calculations are supported by the SongShan supercomputer at the National Supercomputing Center in Zhengzhou.

\clearpage
\appendix
\section{Scalar functions}
\label{sec:ap1}
The scalar functions can be extended to be
\begin{flalign}
\mathcal{I}_0&= \frac{2F_0}{m}\times\frac{Z_{\phi}e^{-Et}}{2E},\nonumber\\
\mathcal{I}_1&=\left(3F_1+\frac{E^2-m^2}{m^2}F_3\right)\times\frac{Z_{\phi}e^{-Et}}{2E},\nonumber\\
\mathcal{I}_2&=\left(-\frac{E}{m}F_2-\frac{E^2}{m^2}F_3\right)\times\frac{Z_{\phi}e^{-Et}}{2E},\nonumber\\
\mathcal{I}_3&=\left(\frac{M^2}{m^2}F_1-\frac{EM^2}{m^3}F_2-\frac{M^2}{m^2}F_3\right)\times\frac{Z_{\phi}e^{-Et}}{2E}.
\end{flalign}

\section{Fitting results}
\label{sec:ap2}

The fitting parameters and corresponding covariance matrices are listed in Table~\ref{para} and Eqs.~(\ref{cov1})-(\ref{cov4}). The numerical results of form factors at different transfer momenta are listed in Table~\ref{3pt}.

 \renewcommand{\tablename}{Table}
        \begin{table}[htbp]
	\centering  
     \caption{Fitting parameters in Eq. (\ref{expz}).}
    \addvspace{5pt}
	\begin{tabular}{ccccccc}
\hline\hline\noalign{\smallskip}
		&$V\left(q^2\right)$&$A_1\left(q^2\right)$&$A_2\left(q^2\right)$&$A_0\left(q^2\right)$\\
\noalign{\smallskip}\hline\noalign{\smallskip}
{$c_0$}&$1.0011(90)$ &$0.6217(54)$ &$0.458(20)$ &$0.6909(35)$\\
{$d_0$}&$0.50(59)$   &$3.30(37)$   &$15.1(1.6)$  &$-0.74(27)$\\
{$f_0$}&$0.240(94)$  &$0.13(11)$   &$0.28(42)$  &$-0.059(60)$\\
{$c_1$}&$-3.65(51)$  &$1.47(33)$   &$-4.6(1.4)$  &$-4.71(26)$\\
{$d_1$}&$-3.0(9.9)$   &$5.5(9.5)$    &$-0.3(10.0)$  &$-7.0(9.4)$\\
{$f_1$}&$-2.1(2.0)$   &$3.6(4.0)$    &$-10.9(2.4)$ &$0.39(84)$\\
{$c_2$}&$19.3(8.0)$   &$14.3(7.1)$   &$4.0(9.8)$   &$24.9(6.2)$\\
{$d_2$}&$0.4(10.0)$    &$0.2(10.0)$    &$0.1(10.0)$   &$0.4(10.0)$\\
{$f_2$}&$11.7(9.1)$   &$8.2(9.1)$    &$1.0(10.0)$   &$19.3(8.3)$\\
  \noalign{\smallskip}\hline
	\end{tabular}
 \label{para}
\end{table}

\newpage
\begin{adjustbox}{angle=90}
\begin{minipage}{\textheight}
\begin{flalign}
    \mathcal{C}_V=
\begin{pmatrix}
  0.00008172 & -0.00444039 & -0.00068620 &  0.00105618 & -0.00447844 &  0.00368754 & -0.00402920 &  0.00013857 &  0.00365604 \\
 -0.00444039 &  0.34662535 &  0.02301784 & -0.05621517 &  0.25851482 & -0.16805696 & -0.14414746 & -0.01770993 & -0.14870297 \\
 -0.00068620 &  0.02301784 &  0.00879097 & -0.01085557 &  0.03148669 & -0.03625461 & -0.02584266 & -0.00054831 & -0.11937720 \\
  0.00105618 & -0.05621517 & -0.01085557 &  0.26298768 & -1.07792927 &  0.92470733 &  0.70549134 &  0.00604313 & -0.04458816 \\
 -0.00447844 &  0.25851482 &  0.03148669 & -1.07792927 & 98.06103256 & -0.78333407 &  0.26841872 &  0.02629234 &  0.72529203 \\
  0.00368754 & -0.16805696 & -0.03625461 &  0.92470733 & -0.78333407 &  4.09730318 &  0.27820244 &  0.00139167 & -1.83082767 \\
 -0.00402920 & -0.14414746 & -0.02584266 &  0.70549134 &  0.26841872 &  0.27820244 & 63.44942376 & -0.35847817 & -23.40576455 \\
  0.00013857 & -0.01770993 & -0.00054831 &  0.00604313 &  0.02629234 &  0.00139167 & -0.35847817 & 99.99569885 & -0.23234159 \\
  0.00365604 & -0.14870297 & -0.11937720 & -0.04458816 &  0.72529203 & -1.83082767 & -23.40576455 & -0.23234159 & 83.49830448
\end{pmatrix}
\label{cov1}
\end{flalign}

\begin{flalign}
    \mathcal{C}_{A_1}=
\begin{pmatrix}
  0.00002916 & -0.00126454 & -0.00048075 &  0.00105408 & -0.00835085 & -0.01200967 &  0.00236950 &  0.00019269 &  0.00418906 \\
 -0.00126454 &  0.13462380 &  0.00725013 & -0.02334251 &  0.85232093 &  0.19722049 & -0.33152303 & -0.02286334 & -0.20850564 \\
 -0.00048075 &  0.00725013 &  0.01154454 & -0.02278872 &  0.03671618 &  0.28159452 & -0.06708070 & -0.00073909 & -0.12020621 \\
  0.00105408 & -0.02334251 & -0.02278872 &  0.10898830 & -0.76591184 & -1.16359667 &  1.06290642 &  0.00596025 & -0.15438566 \\
 -0.00835085 &  0.85232093 &  0.03671618 & -0.76591184 & 90.63630341 &  0.75302223 &  2.14494120 &  0.14498171 &  2.21574956 \\
 -0.01200967 &  0.19722049 &  0.28159452 & -1.16359667 &  0.75302223 & 15.81485093 & -7.56739725 & -0.04023243 &  5.43825798 \\
  0.00236950 & -0.33152303 & -0.06708070 &  1.06290642 &  2.14494120 & -7.56739725 & 50.01678429 & -0.46836782 & -25.20894251 \\
  0.00019269 & -0.02286334 & -0.00073909 &  0.00596025 &  0.14498171 & -0.04023243 & -0.46836782 & 99.99281585 & -0.25083139 \\
  0.00418906 & -0.20850564 & -0.12020621 & -0.15438566 &  2.21574956 &  5.43825798 & -25.20894251 & -0.25083139 & 83.61838715
\end{pmatrix}
\label{cov2}
\end{flalign}
\end{minipage}
\end{adjustbox}

\begin{adjustbox}{angle=90}
\begin{minipage}{\textheight}
\begin{flalign}
    \mathcal{C}_{A_2}=
\begin{pmatrix}
  0.00038085 & -0.02159334 & -0.00643142 &  0.00755884 & -0.00138358 &  0.00955081 & -0.00201340 &  0.00005167 &  0.00051084 \\
 -0.02159334 &  2.49015137 &  0.10905006 & -0.12633458 &  0.09242919 & -0.40674482 & -0.37292535 & -0.01134125 & -0.03441505 \\
 -0.00643142 &  0.10905006 &  0.17802191 & -0.18151298 &  0.02149573 & -0.21628314 & -0.05360504 & -0.00057683 & -0.05341426 \\
  0.00755884 & -0.12633458 & -0.18151298 &  1.92733979 & -0.84734565 &  2.11375134 &  1.31554025 &  0.01044933 &  0.10169343 \\
 -0.00138358 &  0.09242919 &  0.02149573 & -0.84734565 & 99.93406245 & -0.04374356 & -0.02073468 &  0.00073717 &  0.00437398 \\
  0.00955081 & -0.40674482 & -0.21628314 &  2.11375134 & -0.04374356 &  5.86386059 & -0.47618325 & -0.00447402 & -0.30099049 \\
 -0.00201340 & -0.37292535 & -0.05360504 &  1.31554025 & -0.02073468 & -0.47618325 & 96.99901521 & -0.02398347 & -0.62429493 \\
  0.00005167 & -0.01134125 & -0.00057683 &  0.01044933 &  0.00073717 & -0.00447402 & -0.02398347 & 99.99976916 & -0.00493715 \\
  0.00051084 & -0.03441505 & -0.05341426 &  0.10169343 &  0.00437398 & -0.30099049 & -0.62429493 & -0.00493715 & 99.85049495
\end{pmatrix}
\label{cov3}
\end{flalign}

\begin{flalign}
    \mathcal{C}_{A_0}=
\begin{pmatrix}
  0.00001257 & -0.00073957 & -0.00017487 &  0.00031558 & -0.00337074 &  0.00098619 & -0.00050960 &  0.00015037 &  0.00285996 \\
 -0.00073957 &  0.07323715 &  0.00616626 & -0.01379851 &  0.24323407 & -0.04527857 & -0.15031236 & -0.02058543 & -0.04126536 \\
 -0.00017487 &  0.00616626 &  0.00356590 & -0.00589155 &  0.03449121 & -0.01627688 & -0.00450439 & -0.00010779 & -0.09735622 \\
  0.00031558 & -0.01379851 & -0.00589155 &  0.06670735 & -0.85874460 &  0.19457960 &  0.56488665 &  0.00283191 & -0.33687532 \\
 -0.00337074 &  0.24323407 &  0.03449121 & -0.85874460 & 88.01703411 & -0.83639394 &  0.74606717 &  0.17248493 &  3.08818969 \\
  0.00098619 & -0.04527857 & -0.01627688 &  0.19457960 & -0.83639394 &  0.71287734 &  1.41951698 &  0.00695553 & -1.34991969 \\
 -0.00050960 & -0.15031236 & -0.00450439 &  0.56488665 &  0.74606717 &  1.41951698 & 38.17448876 & -0.52035122 & -38.67703835 \\
  0.00015037 & -0.02058543 & -0.00010779 &  0.00283191 &  0.17248493 &  0.00695553 & -0.52035122 & 99.98845123 & -0.31531639 \\
  0.00285996 & -0.04126536 & -0.09735622 & -0.33687532 &  3.08818969 & -1.34991969 & -38.67703835 & -0.31531639 & 68.32432985
\end{pmatrix}
\label{cov4}
\end{flalign}
\end{minipage}
\end{adjustbox}

\renewcommand{\tablename}{Table}
        \begin{table}[htbp]
	\centering  
 \caption{The numerical results of form factors at different transfer momenta for the ensembles of C24P29, C32P23, C32P29, F32P30, F48P21, G36P29, and
H48P32.}
 \addvspace{5pt}
	\scalebox{0.85}{{\begin{tabular}{cccccccc}
		\hline\hline\noalign{\smallskip}	
		&C24P29&C32P23&C32P29&F32P30&F48P21&G36P29&H48P32\\
  \noalign{\smallskip}\hline\noalign{\smallskip}
$V~(|\vec{n}|^2=1)$&$1.250(11)$&$1.355(18)$&$1.332(12)$&$1.1986(96)$&$1.310(18)$&$1.189(16)$&$1.1504(99)$\\
$V~(|\vec{n}|^2=2)$&$1.0726(83)$&$1.1982(92)$&$1.203(10)$&$1.0366(88)$&$1.219(15)$&$1.057(15)$&$1.0319(87)$\\
$V~(|\vec{n}|^2=3)$&$0.9469(81)$&$1.104(10)$&$1.107(11)$&$0.9276(53)$&$1.136(16)$&$0.924(20)$&$0.932(11)$\\
$V~(|\vec{n}|^2=4)$&$0.850(11)$&$1.0190(60)$&$1.0204(68)$&$0.8298(65)$&$1.060(19)$&$0.8453(86)$&$0.823(14)$\\
$V~(|\vec{n}|^2=5)$&$-$&$0.9588(49)$&$0.9642(55)$&$-$&$0.997(16)$&$-$&$-$\\
$V~(|\vec{n}|^2=6)$&$-$&$0.9090(50)$&$0.9119(58)$&$-$&$0.945(17)$&$-$&$-$\\
  \noalign{\smallskip}\hline\noalign{\smallskip}
$A_1~(|\vec{n}|^2=1)$&$0.6969(47)$&$0.7216(44)$&$0.7196(55)$&$0.6795(38)$&$0.7010(48)$&$0.6761(53)$&$0.6748(61)$\\
$A_1~(|\vec{n}|^2=2)$&$0.6627(59)$&$0.7014(47)$&$0.6950(59)$&$0.6459(34)$&$0.6819(49)$&$0.6444(57)$&$0.6369(70)$\\
$A_1~(|\vec{n}|^2=3)$&$0.633(10)$&$0.6839(59)$&$0.6698(72)$&$0.6347(51)$&$0.6641(56)$&$0.6209(81)$&$0.620(11)$\\
$A_1~(|\vec{n}|^2=4)$&$0.566(18)$&$0.6649(82)$&$0.6650(70)$&$0.6162(77)$&$0.6499(71)$&$0.606(13)$&$0.593(18)$\\
$A_1~(|\vec{n}|^2=5)$&$-$&$0.6552(72)$&$0.6624(63)$&$-$&$0.6370(64)$&$-$&$-$\\
$A_1~(|\vec{n}|^2=6)$&$-$&$0.6495(86)$&$0.6556(71)$&$-$&$0.6232(71)$&$-$&$-$\\
  \noalign{\smallskip}\hline\noalign{\smallskip}
$A_2~(|\vec{n}|^2=1)$&$0.725(34)$&$0.873(55)$&$0.848(69)$&$0.614(19)$&$0.750(50)$&$0.605(32)$&$0.591(41)$\\
$A_2~(|\vec{n}|^2=2)$&$0.628(23)$&$0.770(25)$&$0.703(40)$&$0.563(11)$&$0.636(29)$&$0.547(19)$&$0.522(26)$\\
$A_2~(|\vec{n}|^2=3)$&$0.563(30)$&$0.706(24)$&$0.612(42)$&$0.536(13)$&$0.586(26)$&$0.503(22)$&$0.504(32)$\\
$A_2~(|\vec{n}|^2=4)$&$0.400(42)$&$0.638(28)$&$0.635(47)$&$0.505(17)$&$0.561(29)$&$0.462(29)$&$0.430(42)$\\
$A_2~(|\vec{n}|^2=5)$&$-$&$0.613(20)$&$0.609(35)$&$-$&$0.534(20)$&$-$&$-$\\
$A_2~(|\vec{n}|^2=6)$&$-$&$0.591(21)$&$0.555(38)$&$-$&$0.506(20)$&$-$&$-$\\
 \noalign{\smallskip}\hline\noalign{\smallskip}
$A_0~(|\vec{n}|^2=1)$&$0.8507(64)$&$0.9211(58)$&$0.9252(69)$&$0.8329(40)$&$0.9180(63)$&$0.8164(50)$&$0.8015(48)$\\
$A_0~(|\vec{n}|^2=2)$&$0.7115(51)$&$0.8285(44)$&$0.8320(53)$&$0.6977(25)$&$0.8453(46)$&$0.6925(27)$&$0.6888(36)$\\
$A_0~(|\vec{n}|^2=3)$&$0.6142(65)$&$0.7488(44)$&$0.7518(51)$&$0.6132(28)$&$0.7812(42)$&$0.6111(25)$&$0.6031(37)$\\
$A_0~(|\vec{n}|^2=4)$&$0.5639(95)$&$0.6866(47)$&$0.6876(55)$&$0.5443(34)$&$0.7266(43)$&$0.5521(26)$&$0.5522(46)$\\
$A_0~(|\vec{n}|^2=5)$&$-$&$0.6367(37)$&$0.6405(44)$&$-$&$0.6821(34)$&$-$&$-$\\
$A_0~(|\vec{n}|^2=6)$&$-$&$0.5945(38)$&$0.5941(45)$&$-$&$0.6409(33)$&$-$&$-$\\
 \noalign{\smallskip}\hline\noalign{\smallskip}
	\end{tabular}}}
 \label{3pt}
\end{table}
The phase moments $\mathcal{A}^F_n$ are listed in Table. \ref{phase}, which are obtained by Eq.~(\ref{eq:phase_form}).

 \renewcommand{\tablename}{Table}
        \begin{table}[htbp]
	\centering  
     \caption{Phase moments $\mathcal{A}_n^F$ obtained by fitting.}
 \label{phase}
 \addvspace{5pt}
	\begin{tabular}{ccccccc}
		\hline\hline\noalign{\smallskip}	
		&{$\mathcal{A}_n^V$}&{$\mathcal{A}_n^{A_1}$}&{$\mathcal{A}_n^{A_2}$}&{$\mathcal{A}_n^{A_0}$}\\
	\noalign{\smallskip}\hline\noalign{\smallskip}
{$\mathcal{A}_0^{F}$}  &{$2.743(49)$} &{$0.713(66)$} &{$1.94 (24)$}&{$3.242(36)$}\\		
{$\mathcal{A}_1^{F}$}&{$6.8(1.1)$} &{$4.4(1.1)$} &{$16.6 (5.2)$}&{$5.06(75)$}\\		
  \noalign{\smallskip}\hline
	\end{tabular}
\end{table}

\renewcommand{\thesubfigure}{(\roman{subfigure})}
\renewcommand{\figurename}{Figure}
\begin{figure}[htp]
\centering  
\subfigure[$V\left(q^2\right)$ at C24P29.]{
\includegraphics[width=7cm]{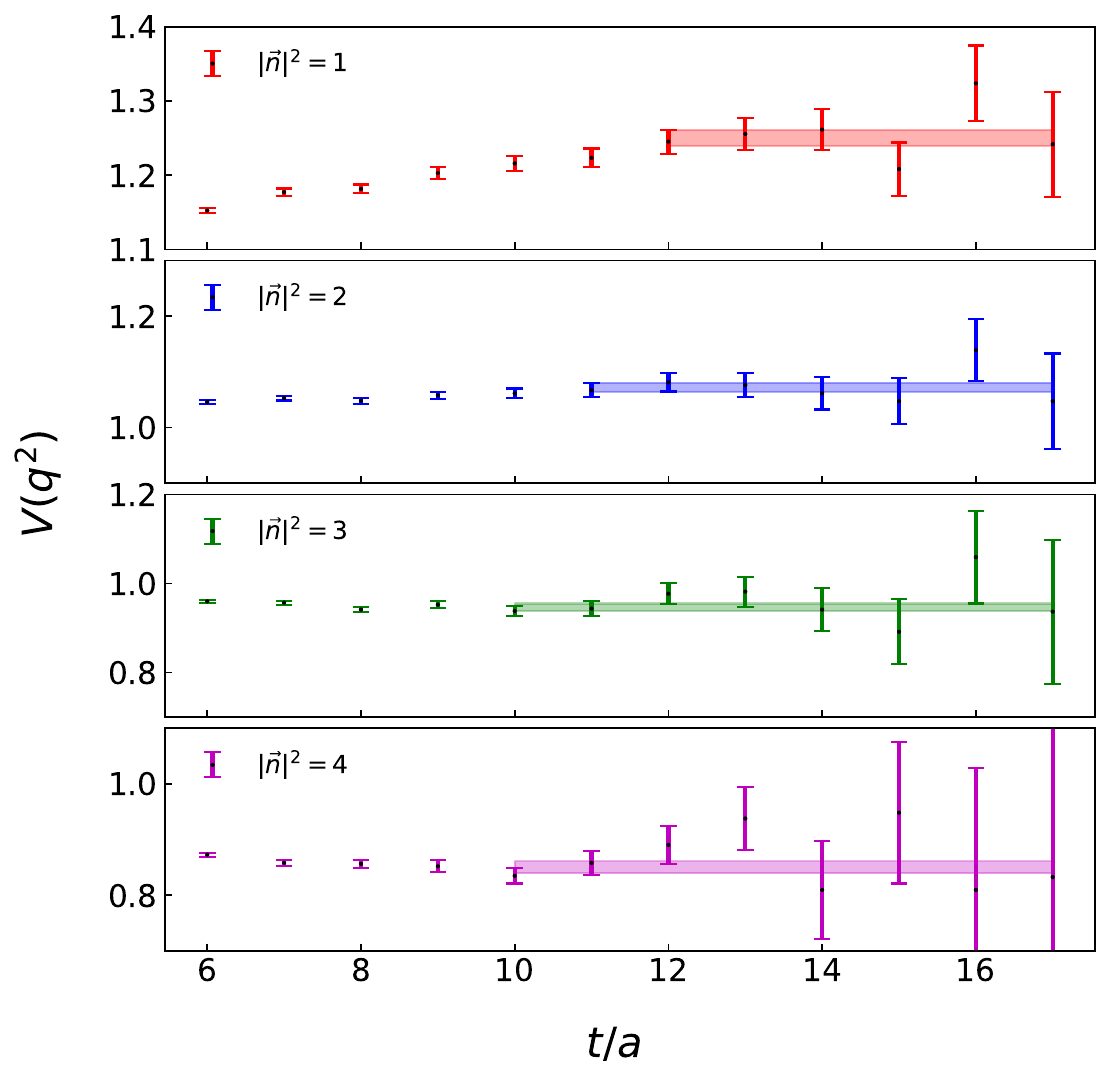}}
\subfigure[$A_0\left(q^2\right)$ at C24P29.]{
\includegraphics[width=7cm]{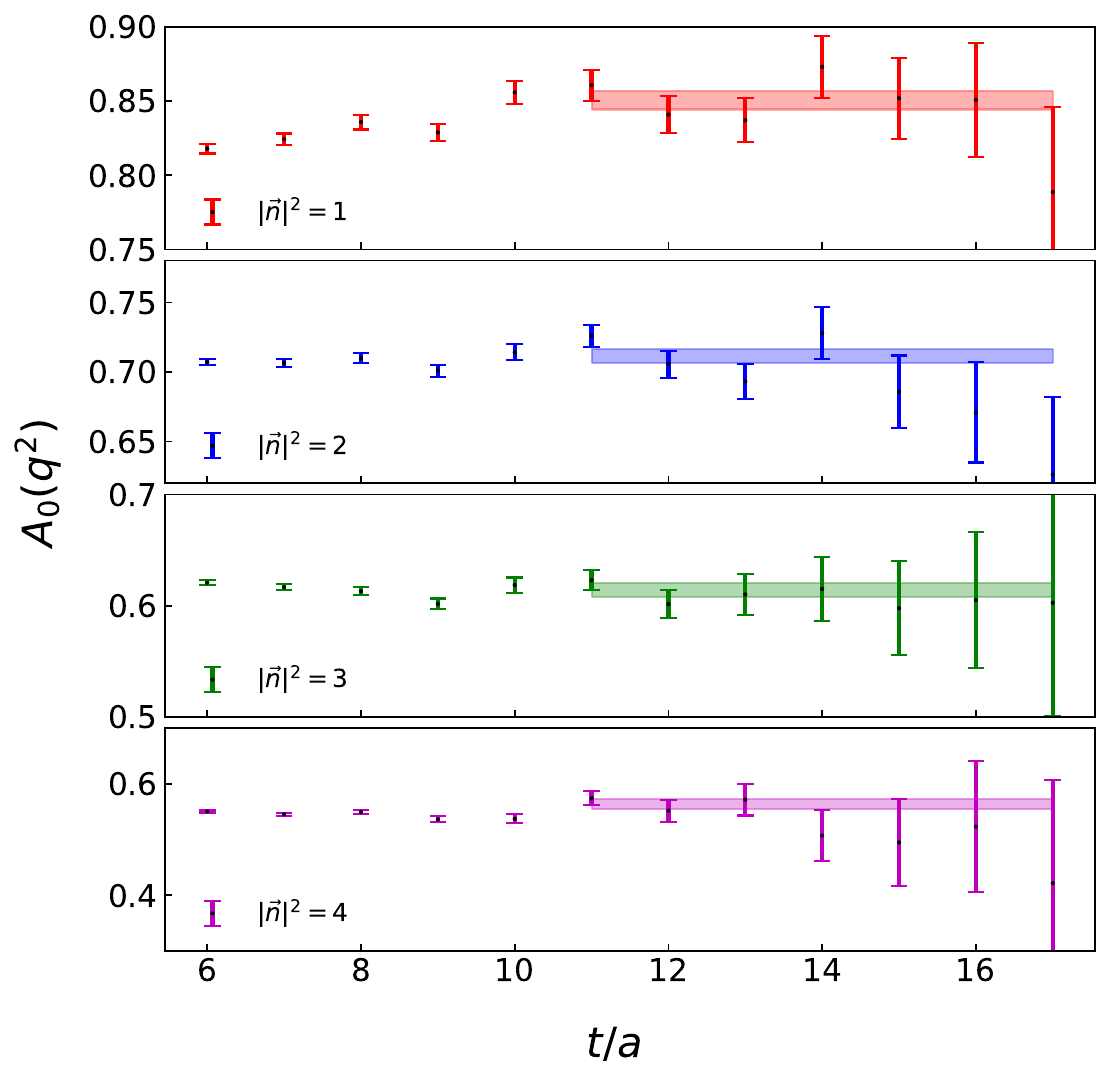}}
\subfigure[$A_1\left(q^2\right)$ at C24P29.]{
\includegraphics[width=7cm]{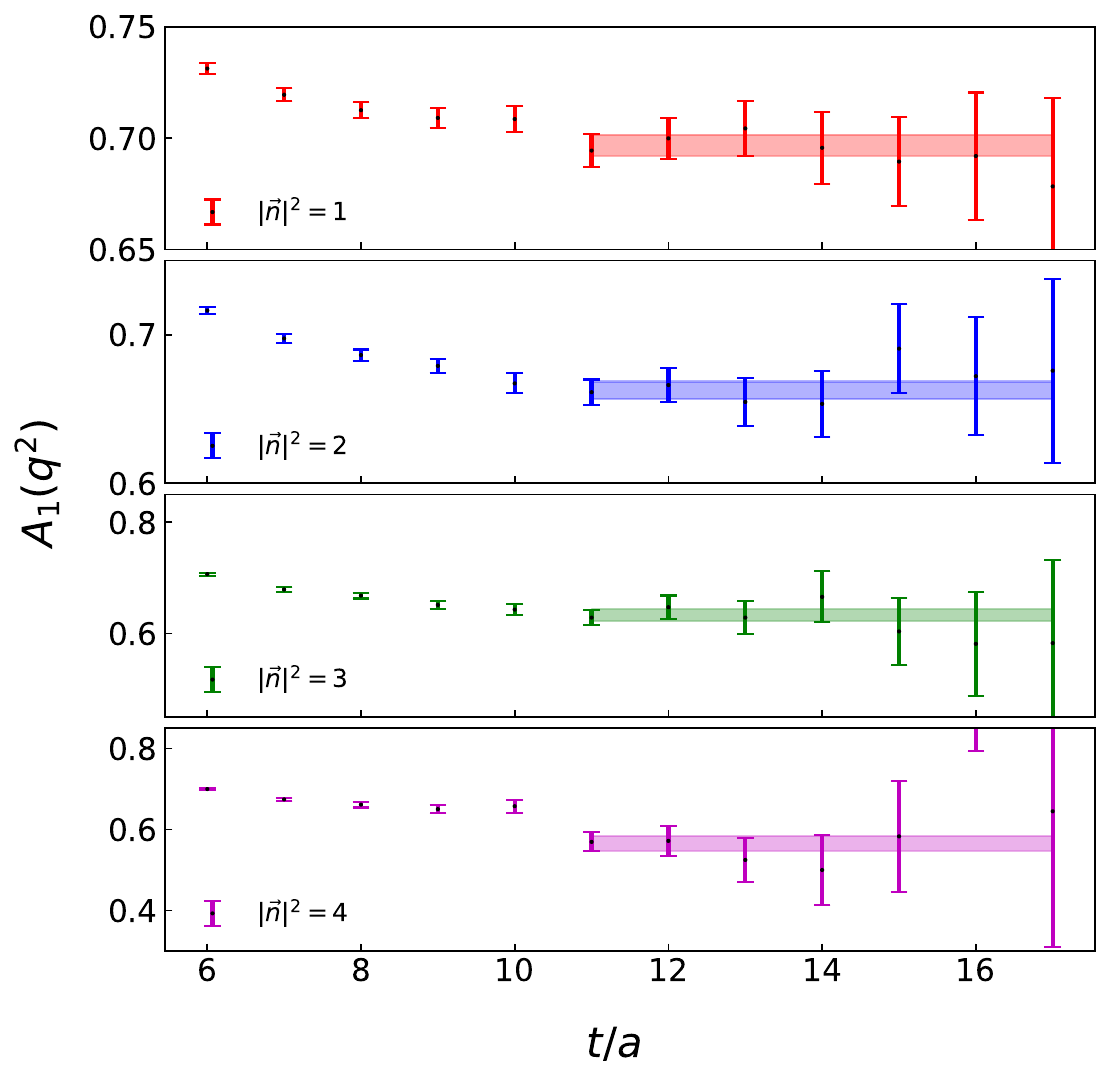}}
\subfigure[$A_2\left(q^2\right)$ at C24P29.]{
\includegraphics[width=7cm]{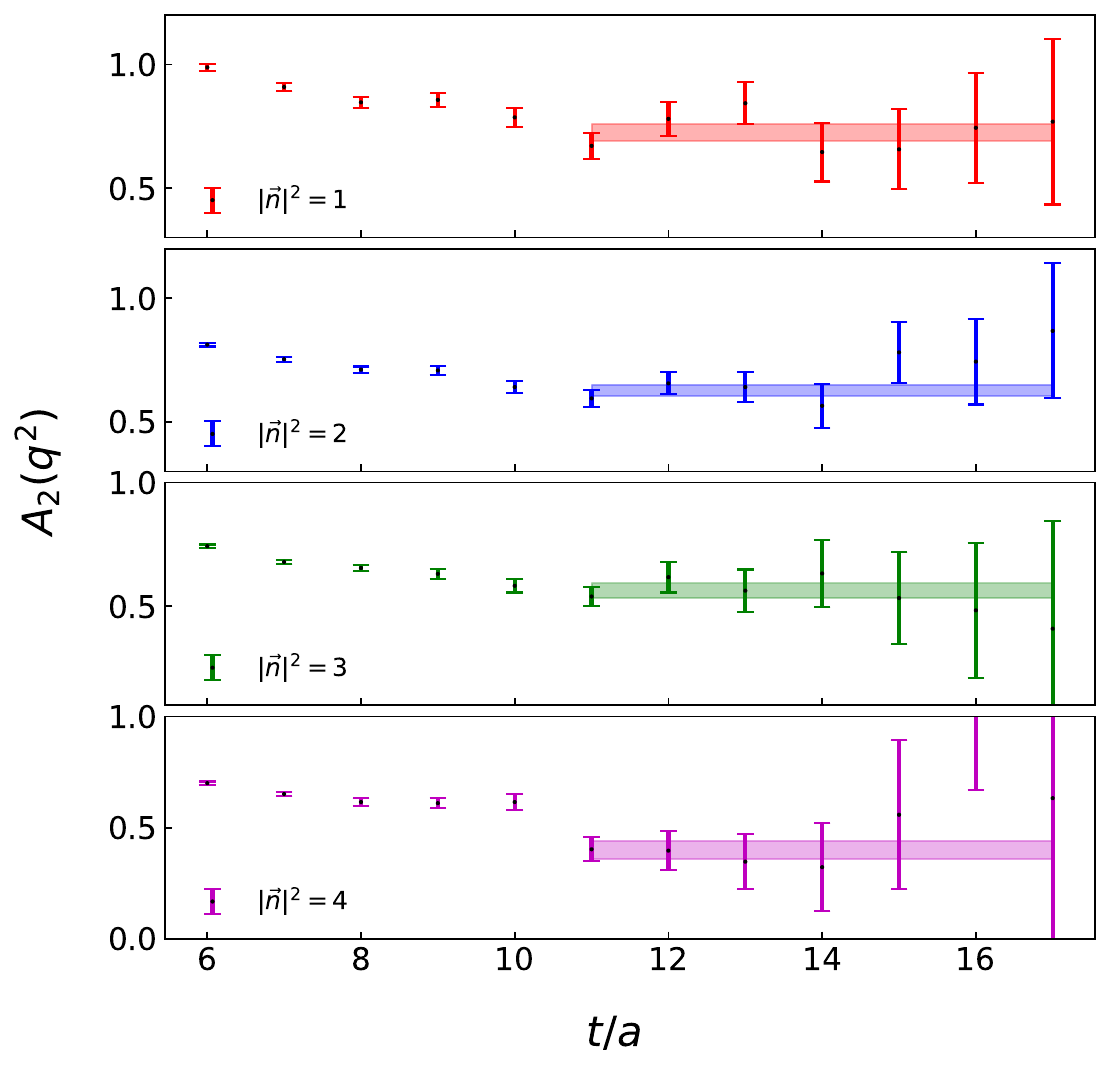}}
\caption{The form factors with different momentum $\vec{p}=2\pi\vec{n}/L,~|\vec{n}|^2=1,2,3,4$ at ensemble C24P29.}
\label{3ptC24P29}
\end{figure}

\renewcommand{\thesubfigure}{(\roman{subfigure})}
\renewcommand{\figurename}{Figure}
\begin{figure}[htp]
\centering  
\subfigure[$V\left(q^2\right)$ at C32P23.]{
\includegraphics[width=7cm]{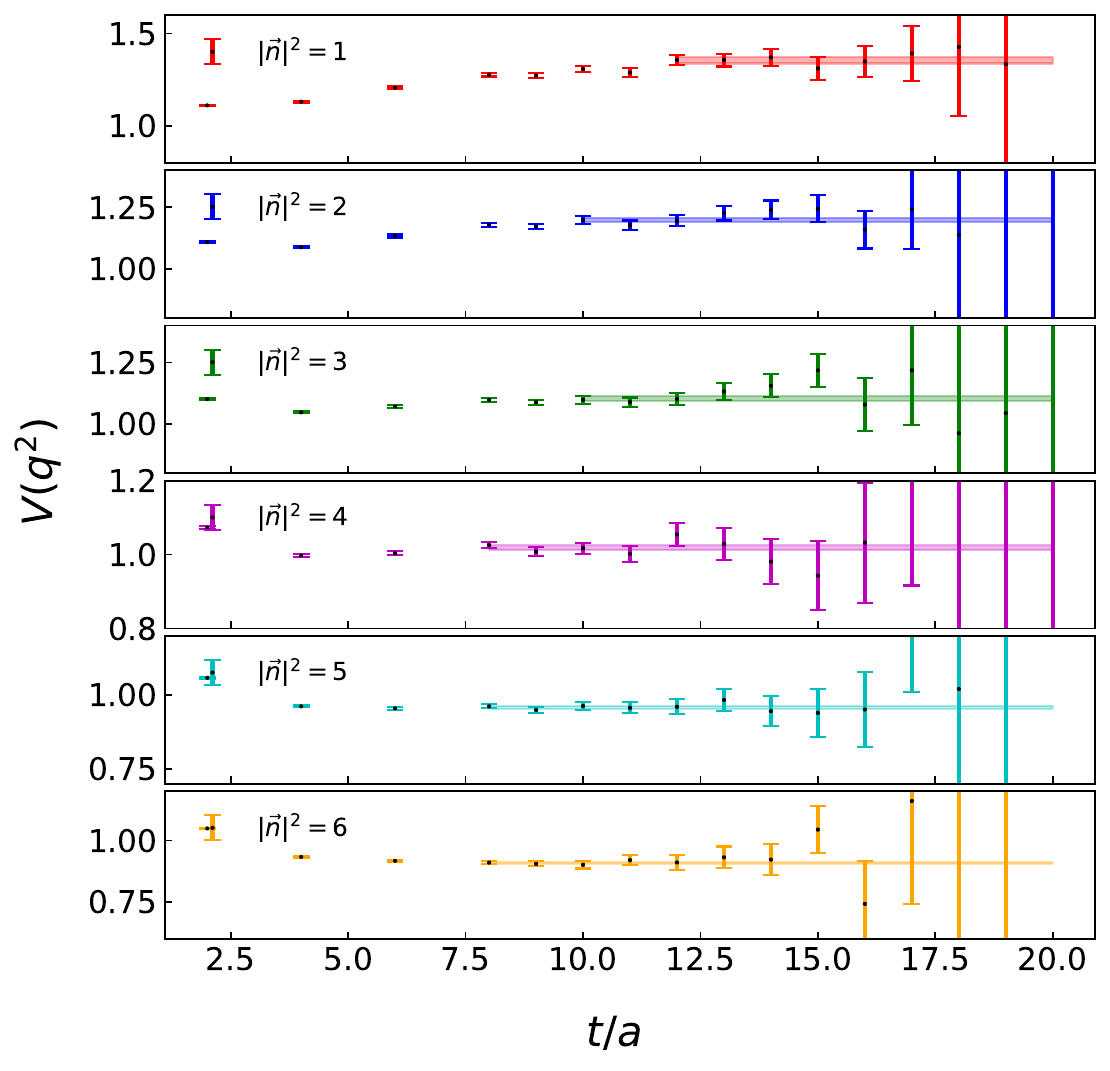}}
\subfigure[$A_0\left(q^2\right)$ at C32P23.]{
\includegraphics[width=7cm]{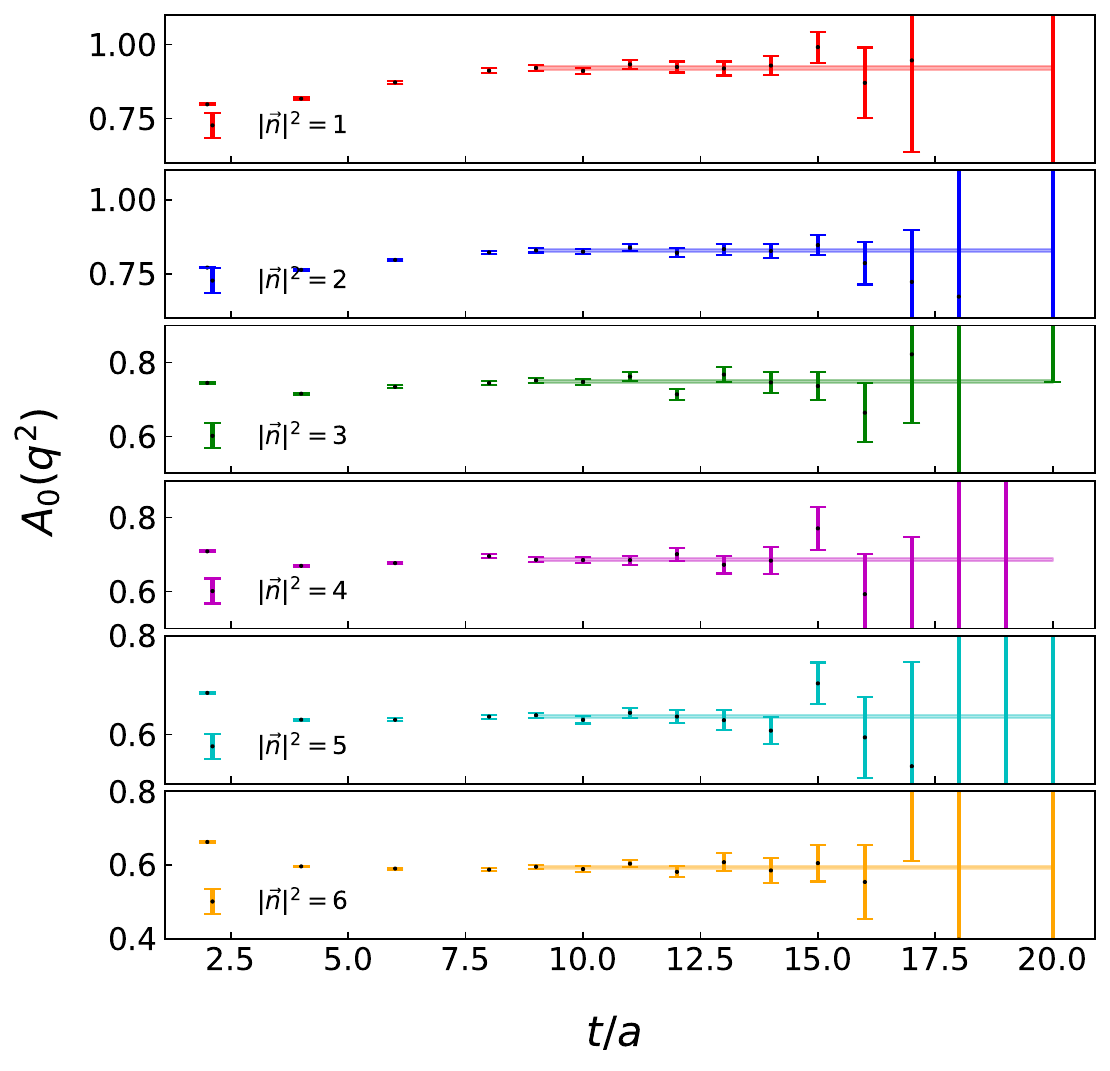}}
\subfigure[$A_1\left(q^2\right)$ at C32P23.]{
\includegraphics[width=7cm]{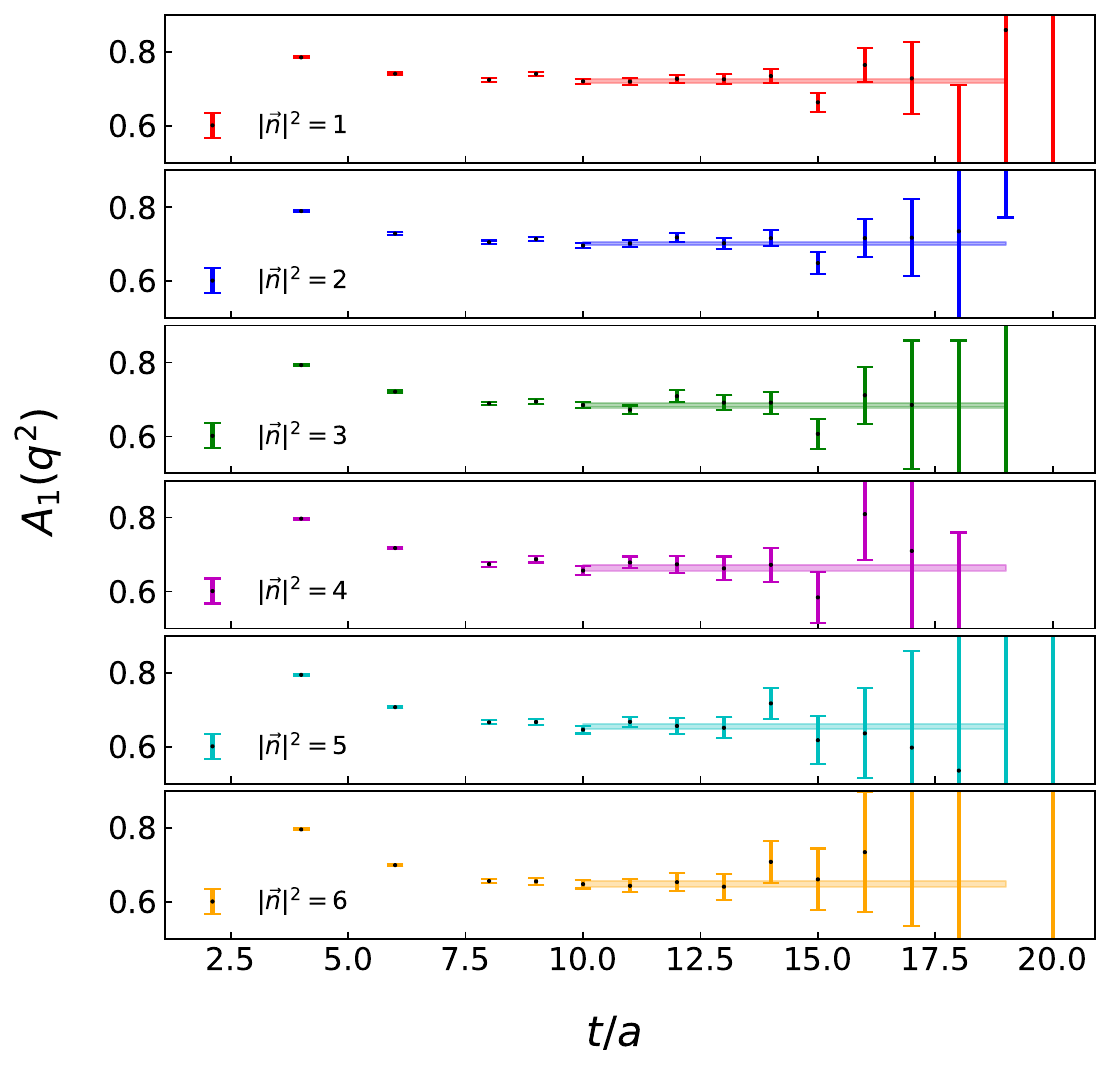}}
\subfigure[$A_2\left(q^2\right)$ at C32P23.]{
\includegraphics[width=7cm]{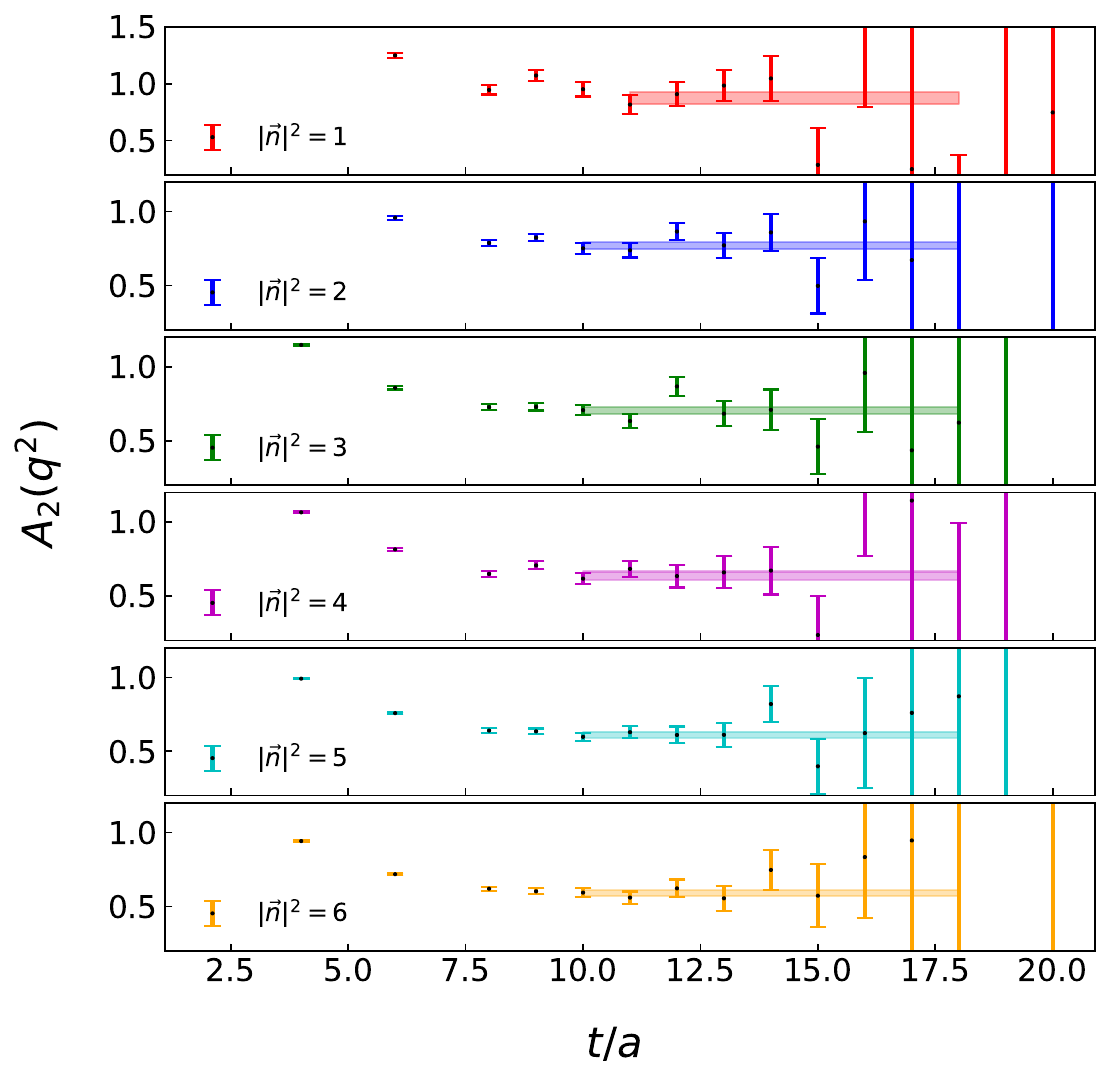}}
\caption{The form factors with different momentum $\vec{p}=2\pi\vec{n}/L,~|\vec{n}|^2=1,2,3,4,5,6$ at ensemble C32P23.}
\label{3ptC32P23}
\end{figure}

\renewcommand{\thesubfigure}{(\roman{subfigure})}
\renewcommand{\figurename}{Figure}
\begin{figure}[htp]
\centering  
\subfigure[$V\left(q^2\right)$ at C32P29.]{
\includegraphics[width=7cm]{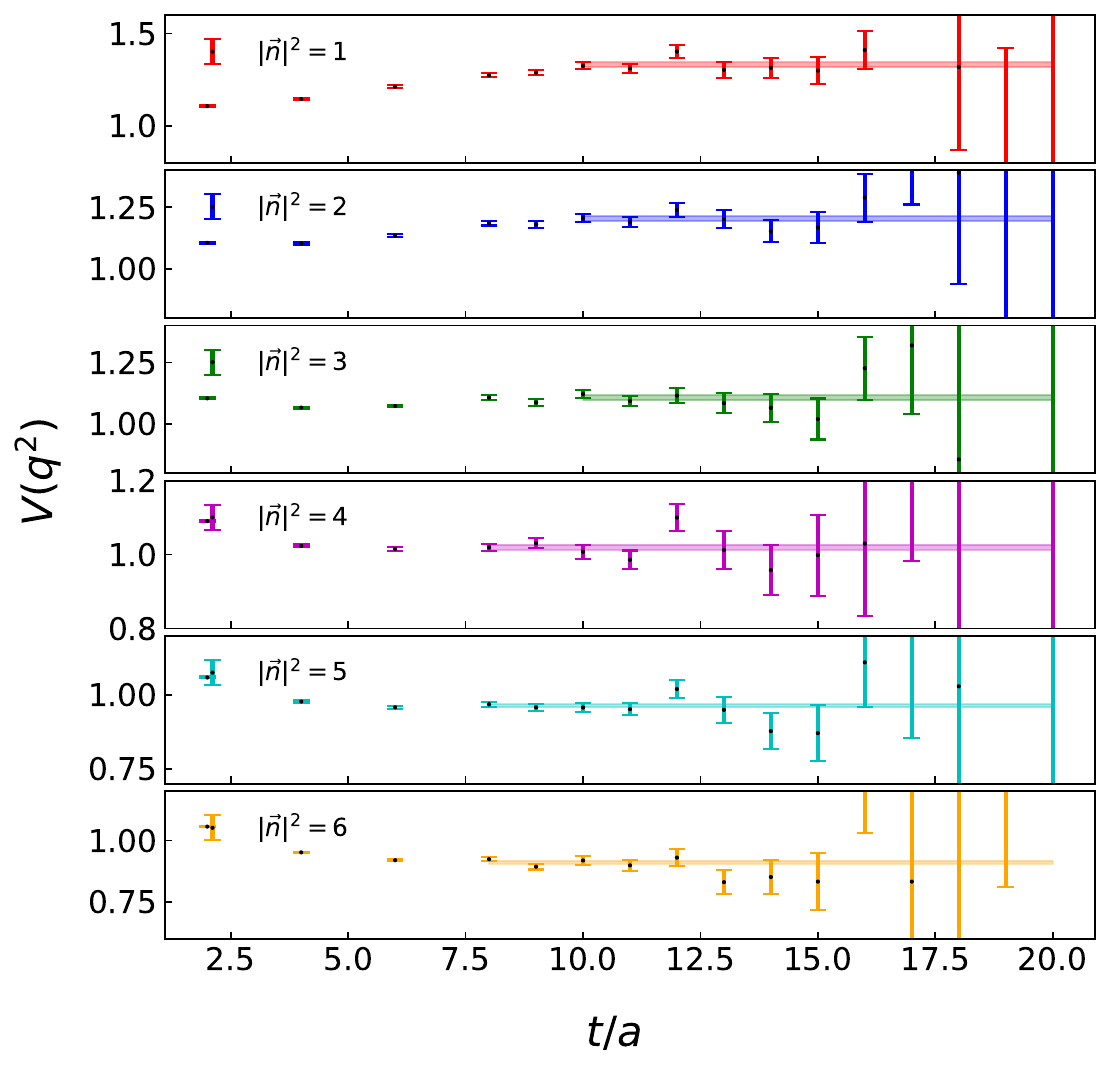}}
\subfigure[$A_0\left(q^2\right)$ at C32P29.]{
\includegraphics[width=7cm]{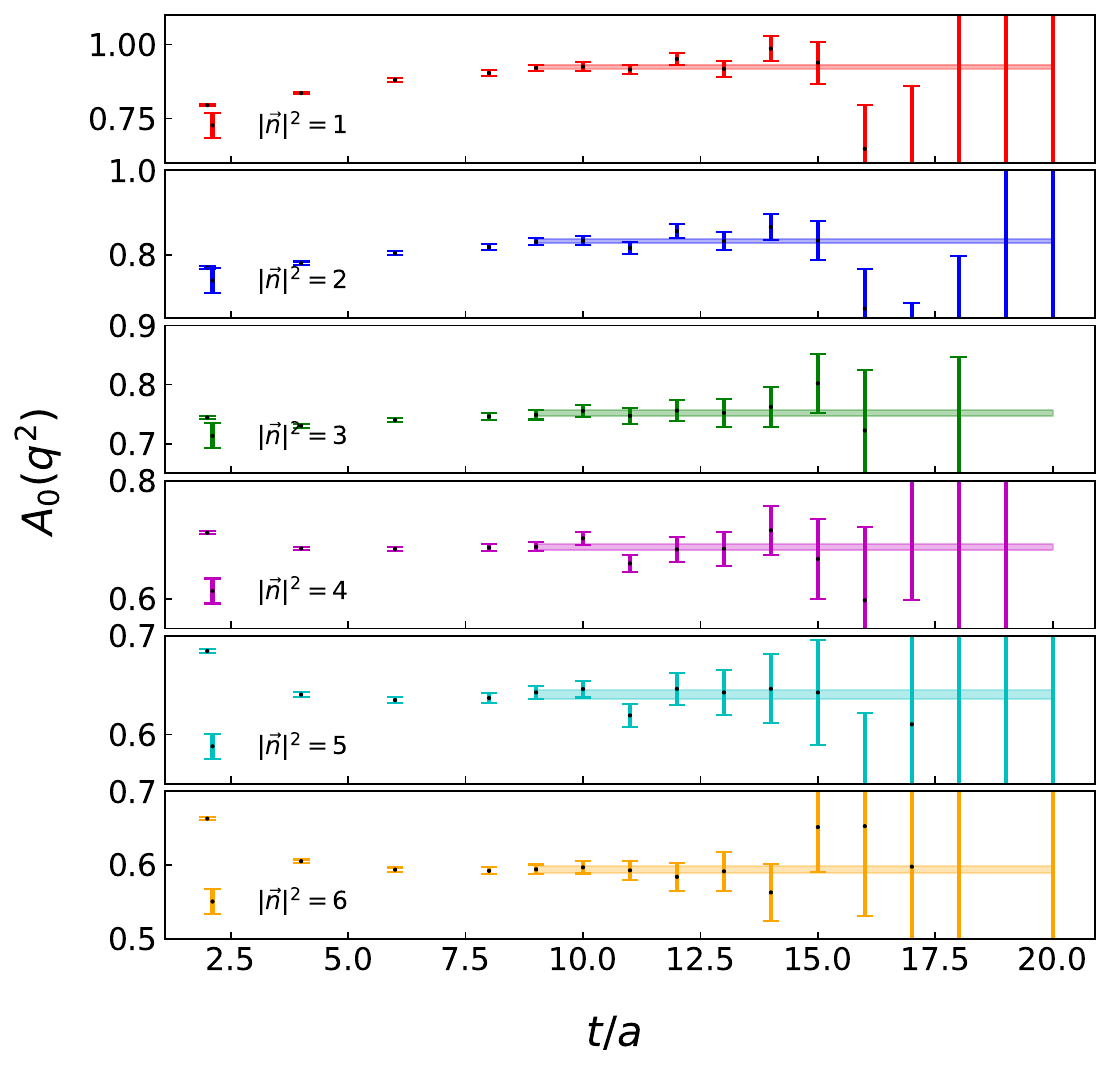}}
\subfigure[$A_1\left(q^2\right)$ at C32P29.]{
\includegraphics[width=7cm]{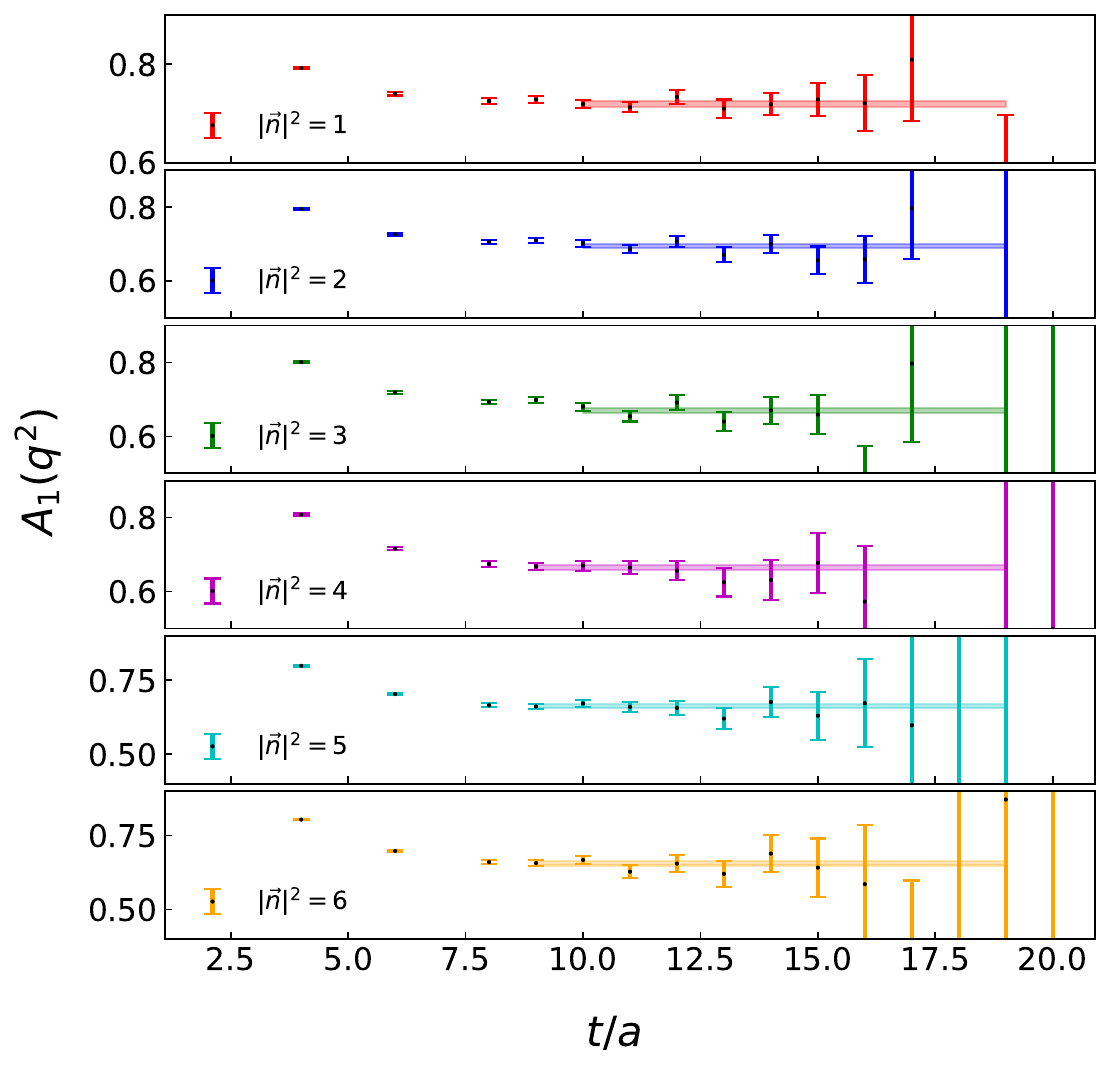}}
\subfigure[$A_2\left(q^2\right)$ at C32P29.]{
\includegraphics[width=7cm]{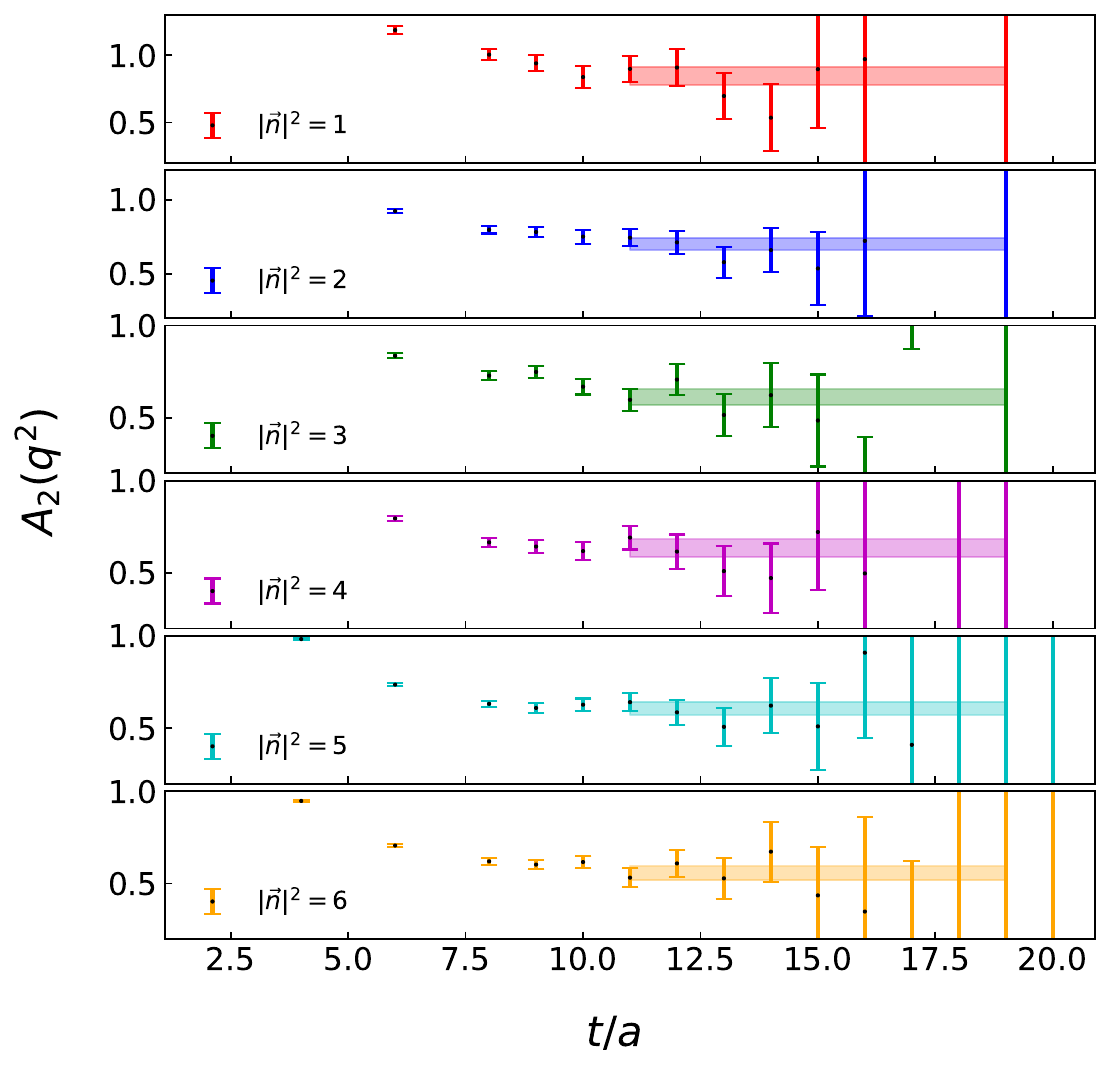}}
\caption{The form factors with different momentum $\vec{p}=2\pi\vec{n}/L,~|\vec{n}|^2=1,2,3,4$ at ensemble C32P29.}
\label{3ptC32P29}
\end{figure}

\renewcommand{\thesubfigure}{(\roman{subfigure})}
\renewcommand{\figurename}{Figure}
\begin{figure}[htp]
\centering  
\subfigure[$V\left(q^2\right)$ at F32P30.]{
\includegraphics[width=7cm]{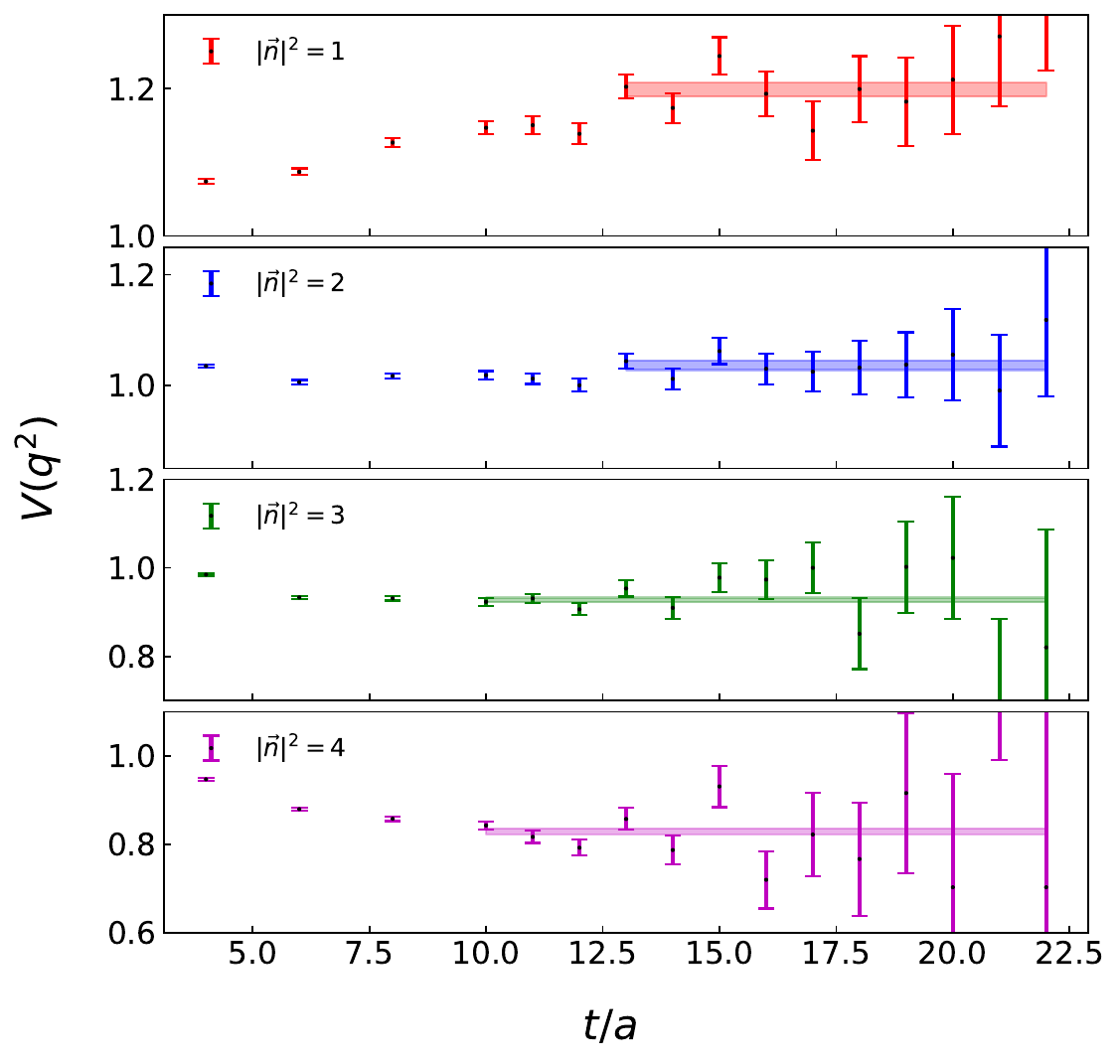}}
\subfigure[$A_0\left(q^2\right)$ at F32P30.]{
\includegraphics[width=7cm]{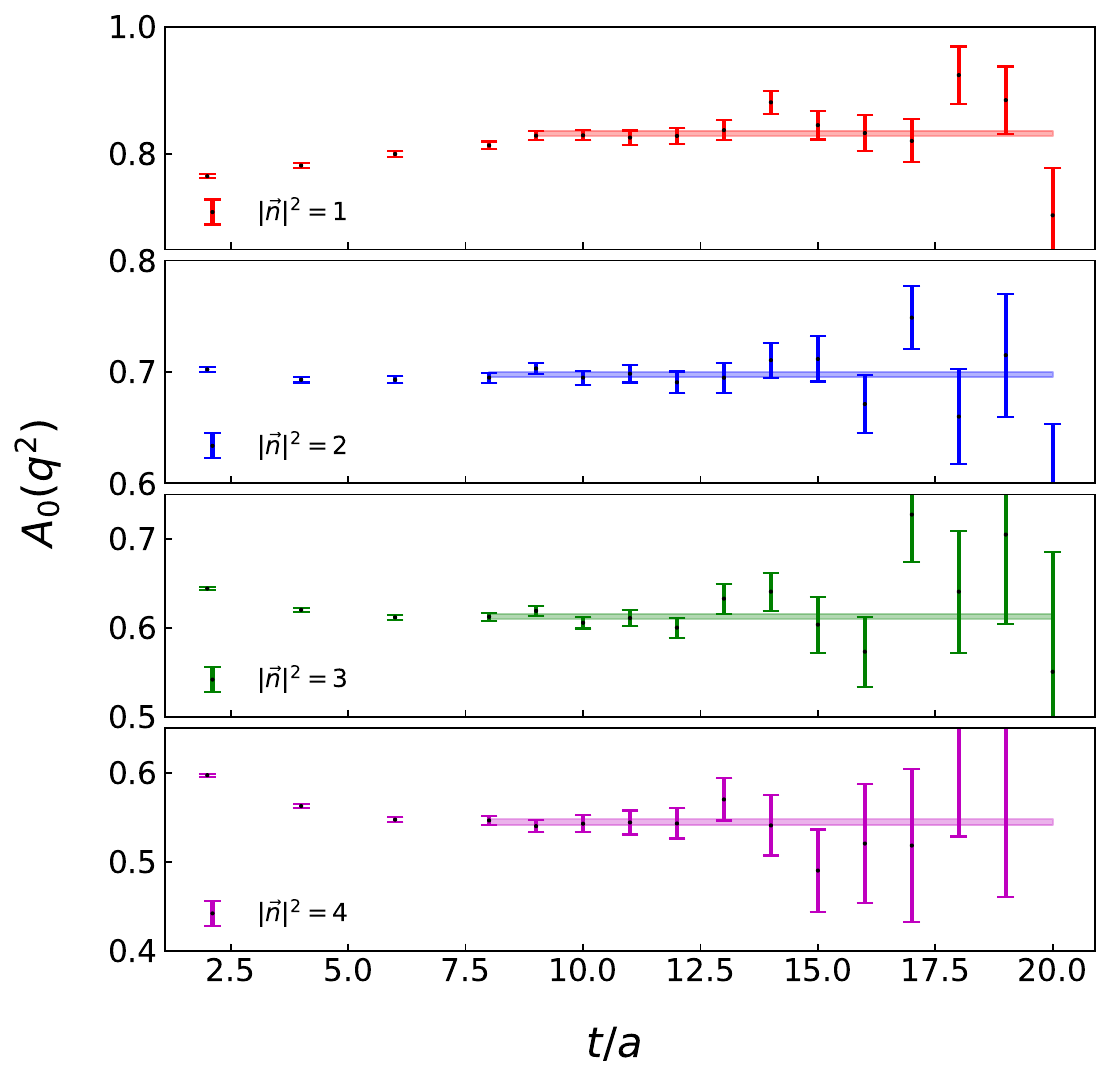}}
\subfigure[$A_1\left(q^2\right)$ at F32P30.]{
\includegraphics[width=7cm]{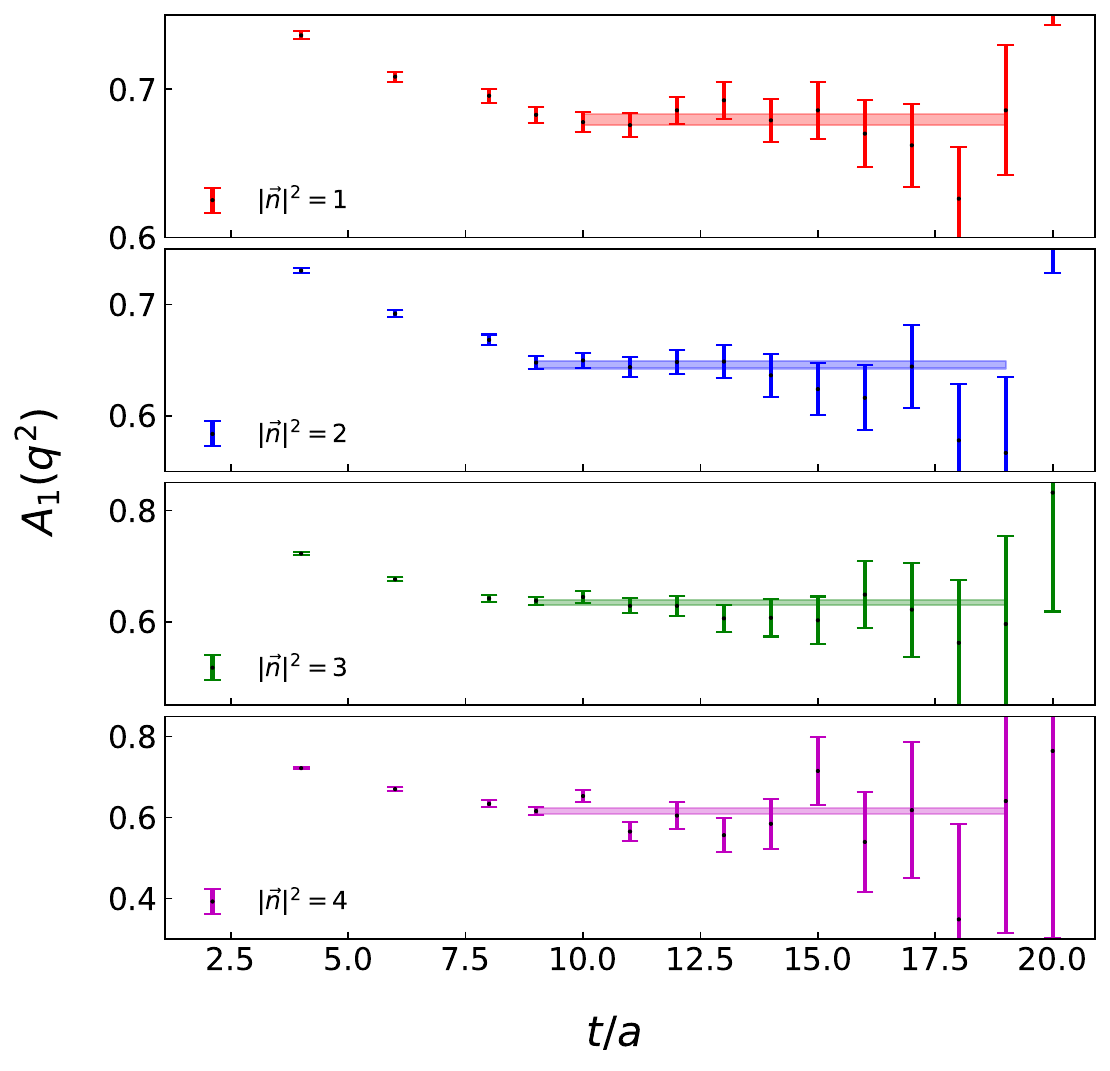}}
\subfigure[$A_2\left(q^2\right)$ at F32P30.]{
\includegraphics[width=7cm]{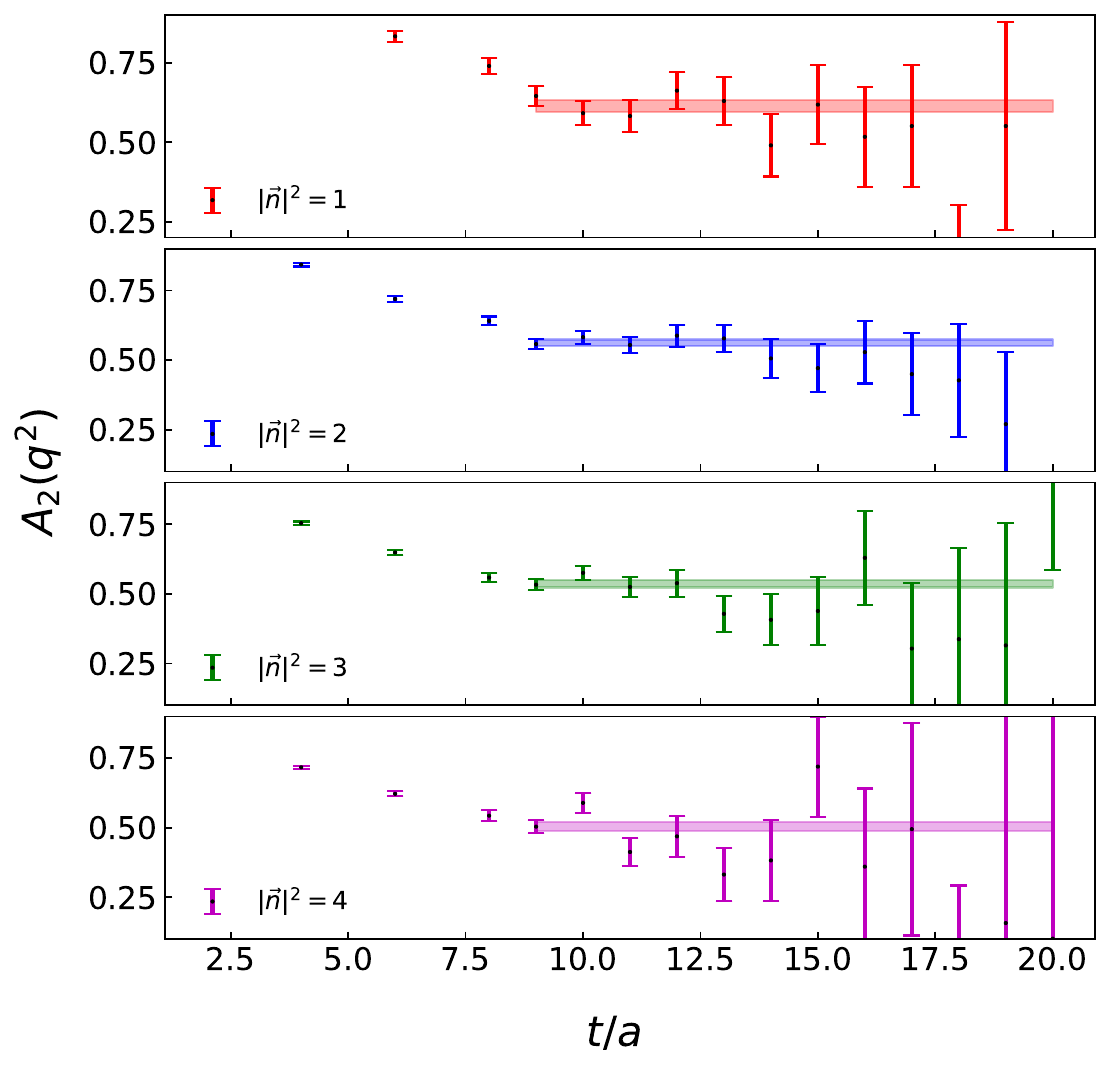}}
\caption{The form factors with different momentum $\vec{p}=2\pi\vec{n}/L,~|\vec{n}|^2=1,2,3,4$ at ensemble F32P30.}
\label{3ptF32P30}
\end{figure}

\renewcommand{\thesubfigure}{(\roman{subfigure})}
\renewcommand{\figurename}{Figure}
\begin{figure}[htp]
\centering  
\subfigure[$V\left(q^2\right)$ at F48P21.]{
\includegraphics[width=7cm]{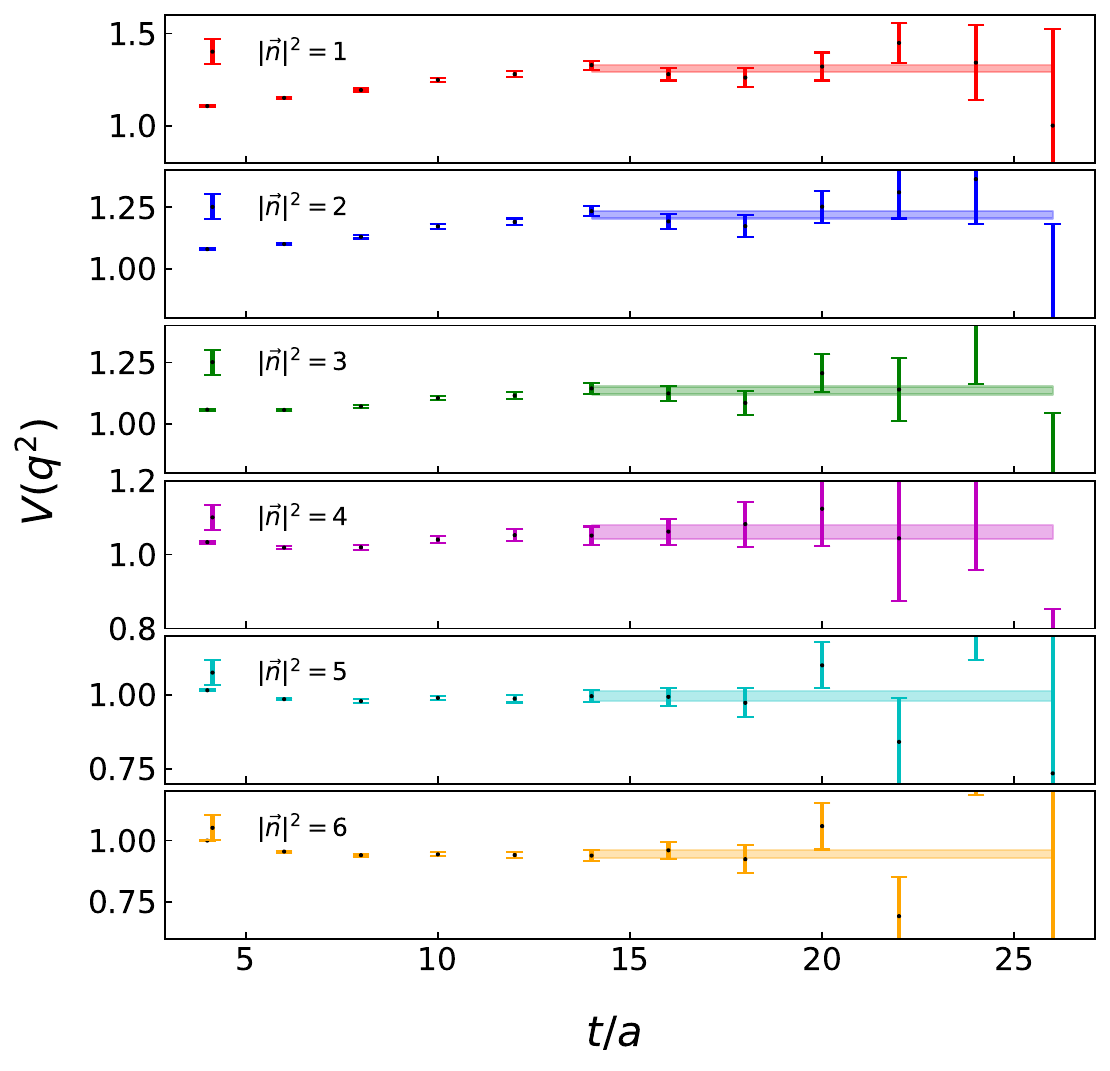}}
\subfigure[$A_0\left(q^2\right)$ at F48P21.]{
\includegraphics[width=7cm]{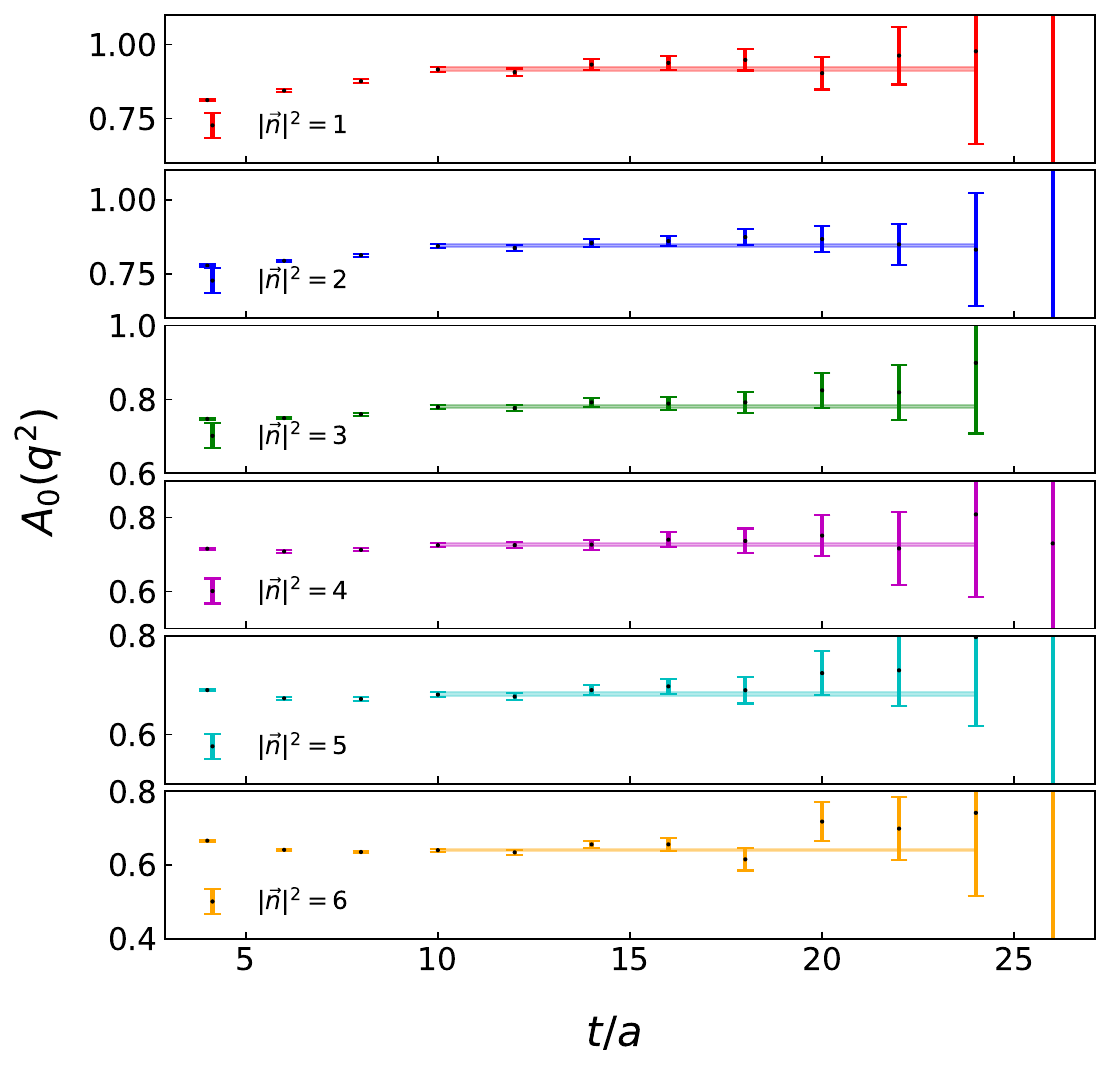}}
\subfigure[$A_1\left(q^2\right)$ at F48P21.]{
\includegraphics[width=7cm]{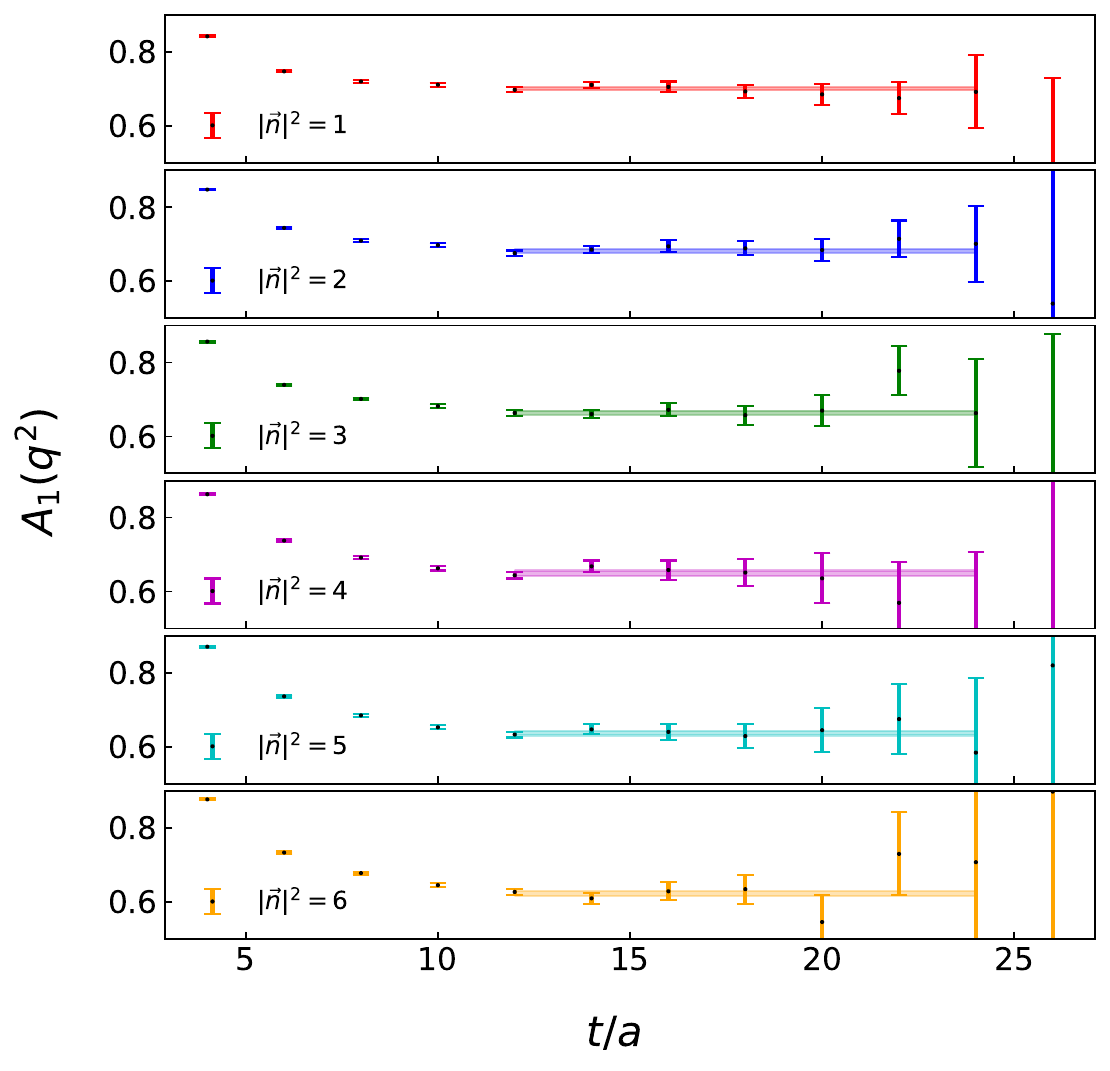}}
\subfigure[$A_2\left(q^2\right)$ at F48P21.]{
\includegraphics[width=7cm]{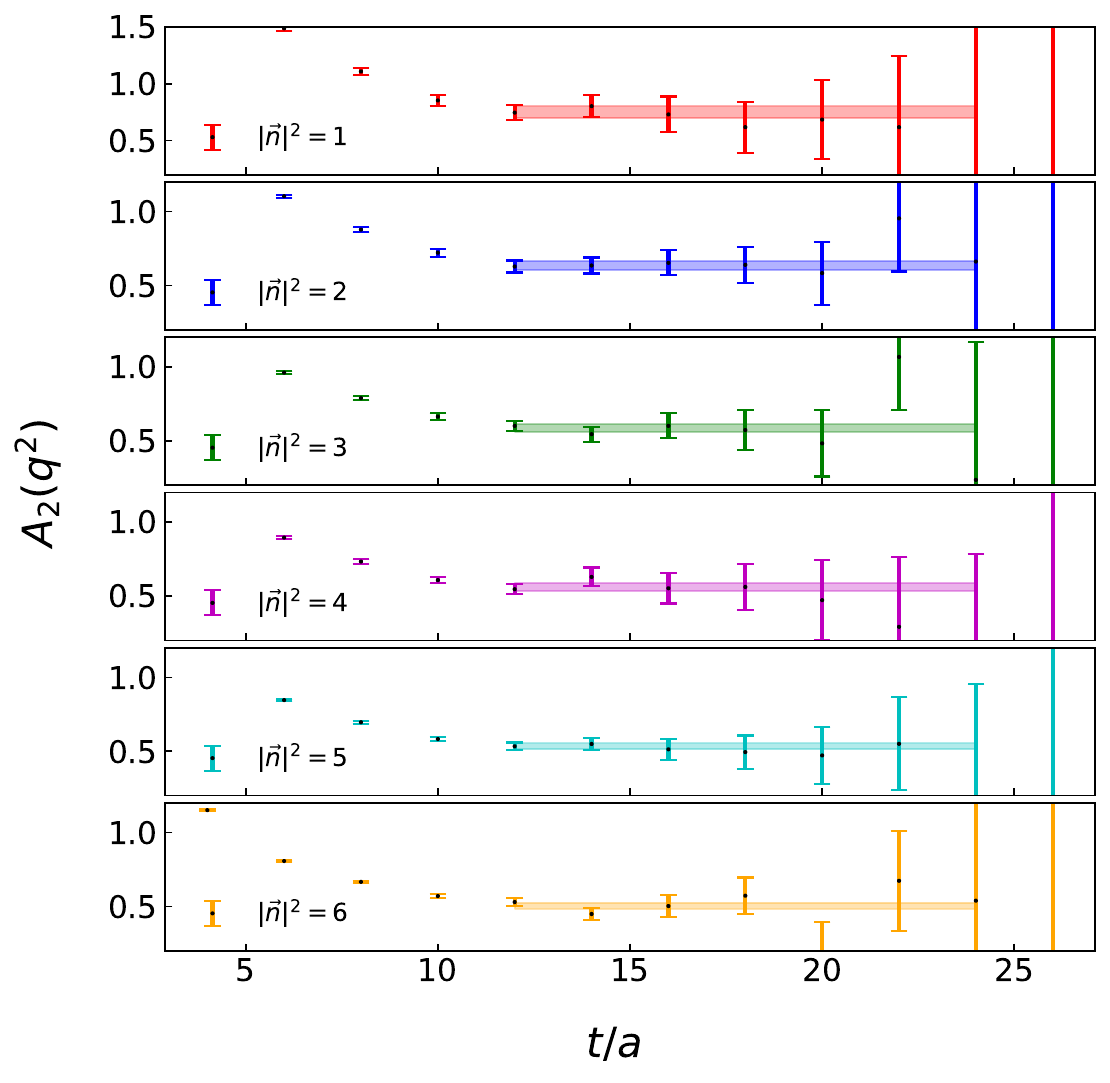}}
\caption{The form factors with different momentum $\vec{p}=2\pi\vec{n}/L,~|\vec{n}|^2=1,2,3,4,5,6$ at ensemble F48P21.}
\label{3ptF48P21}
\end{figure}

\renewcommand{\thesubfigure}{(\roman{subfigure})}
\renewcommand{\figurename}{Figure}
\begin{figure}[htp]
\centering  
\subfigure[$V\left(q^2\right)$ at G36P29.]{
\includegraphics[width=7cm]{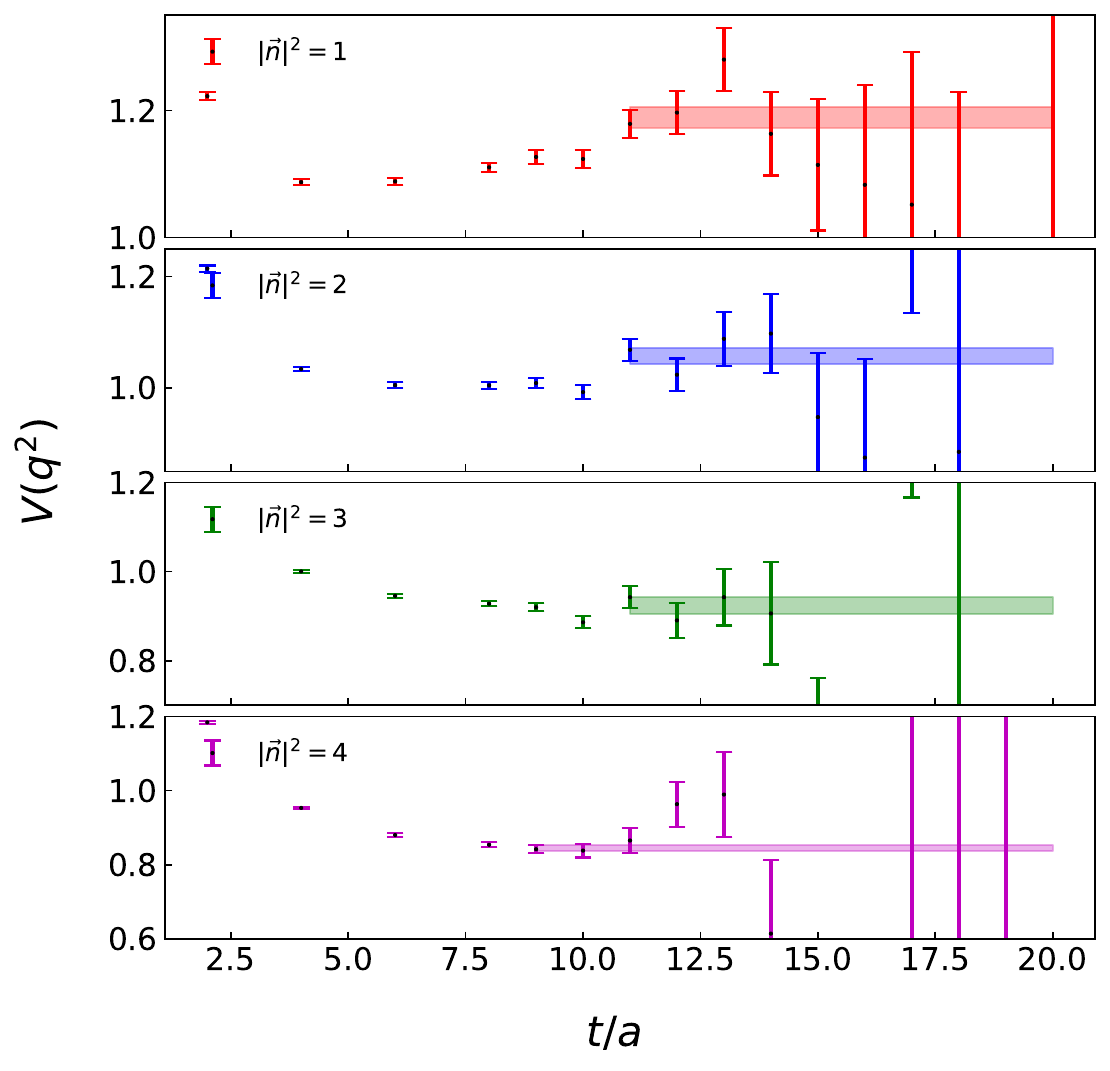}}
\subfigure[$A_0\left(q^2\right)$ at G36P29.]{
\includegraphics[width=7cm]{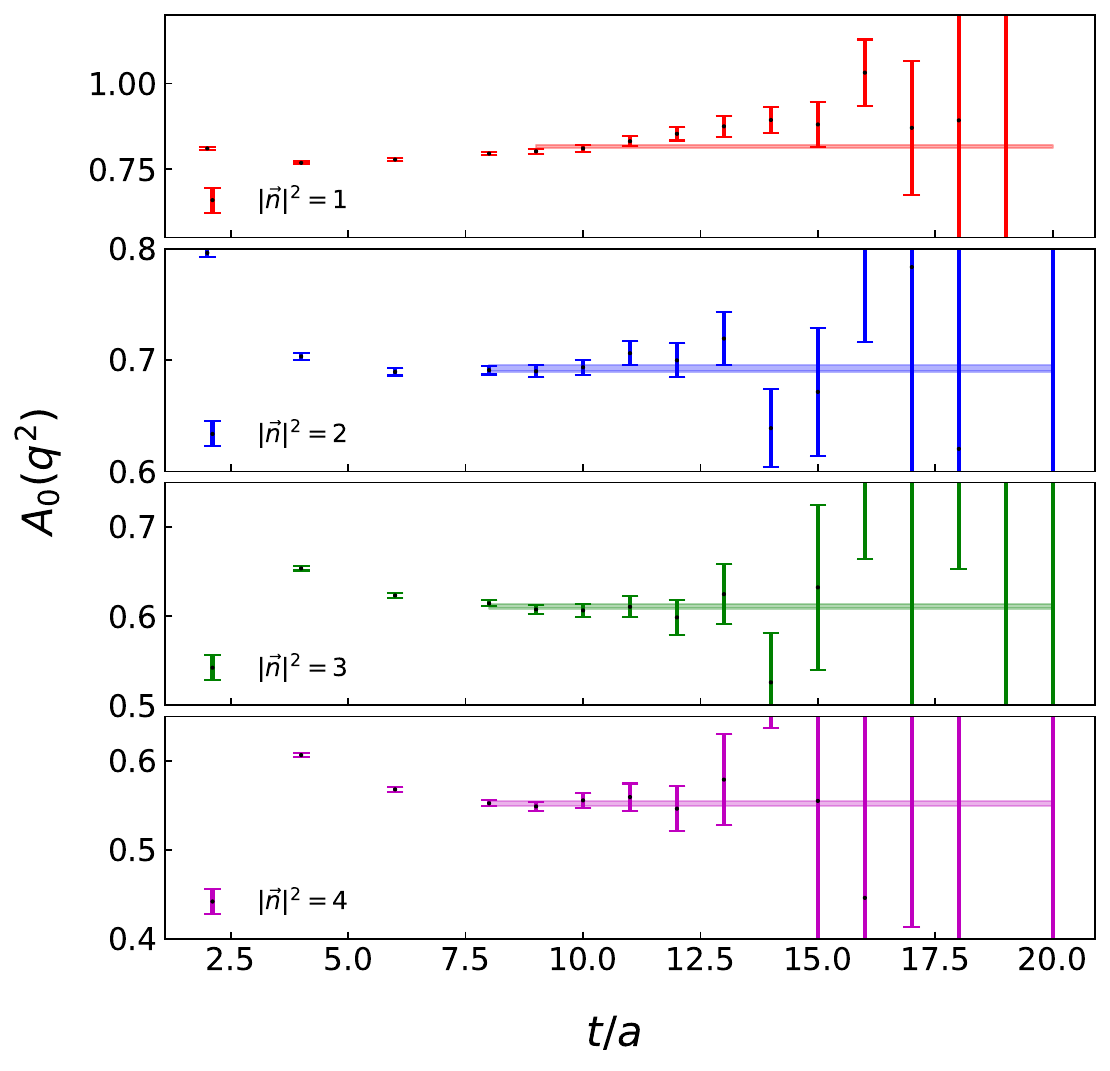}}
\subfigure[$A_1\left(q^2\right)$ at G36P29.]{
\includegraphics[width=7cm]{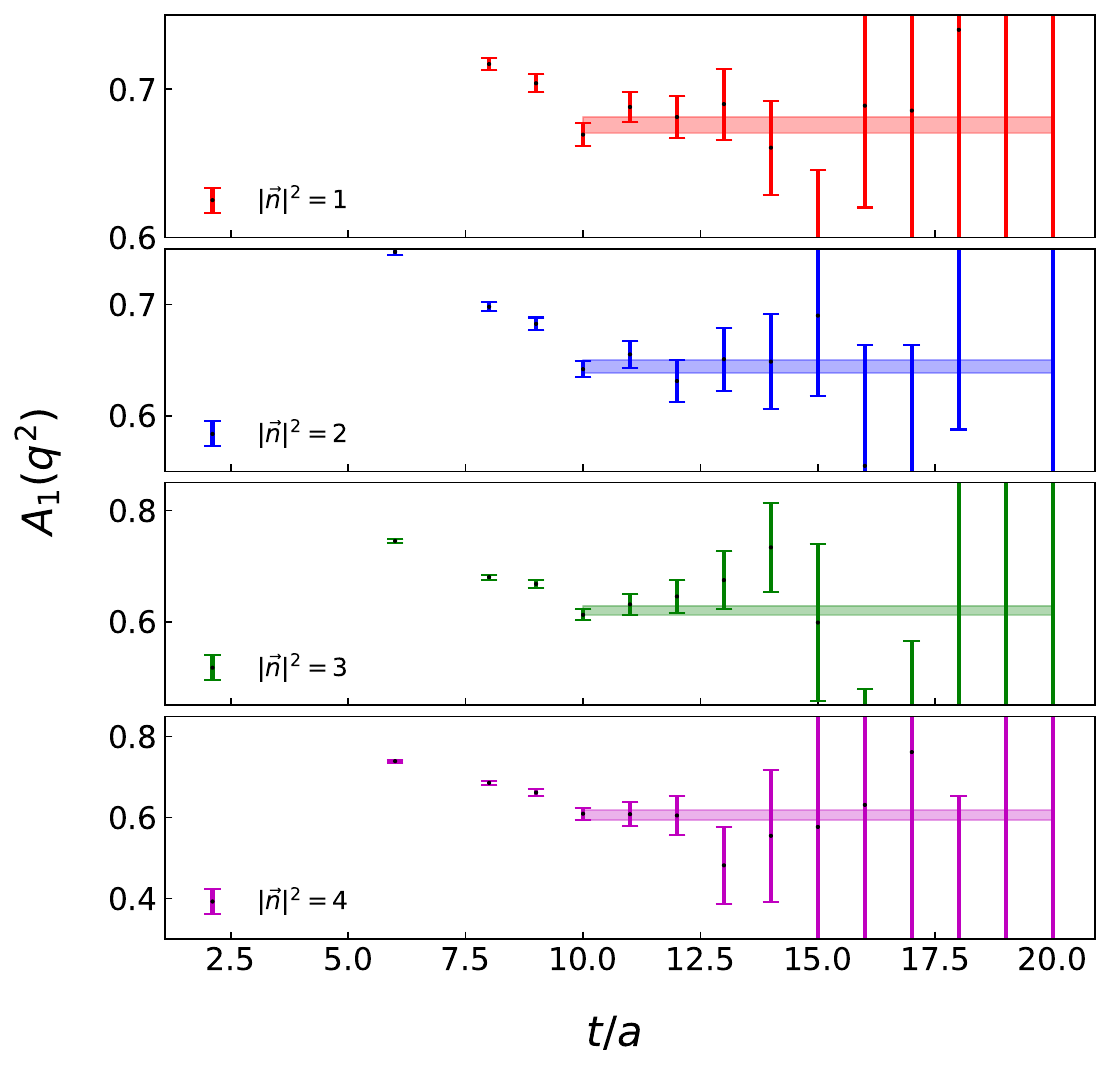}}
\subfigure[$A_2\left(q^2\right)$ at G36P29.]{
\includegraphics[width=7cm]{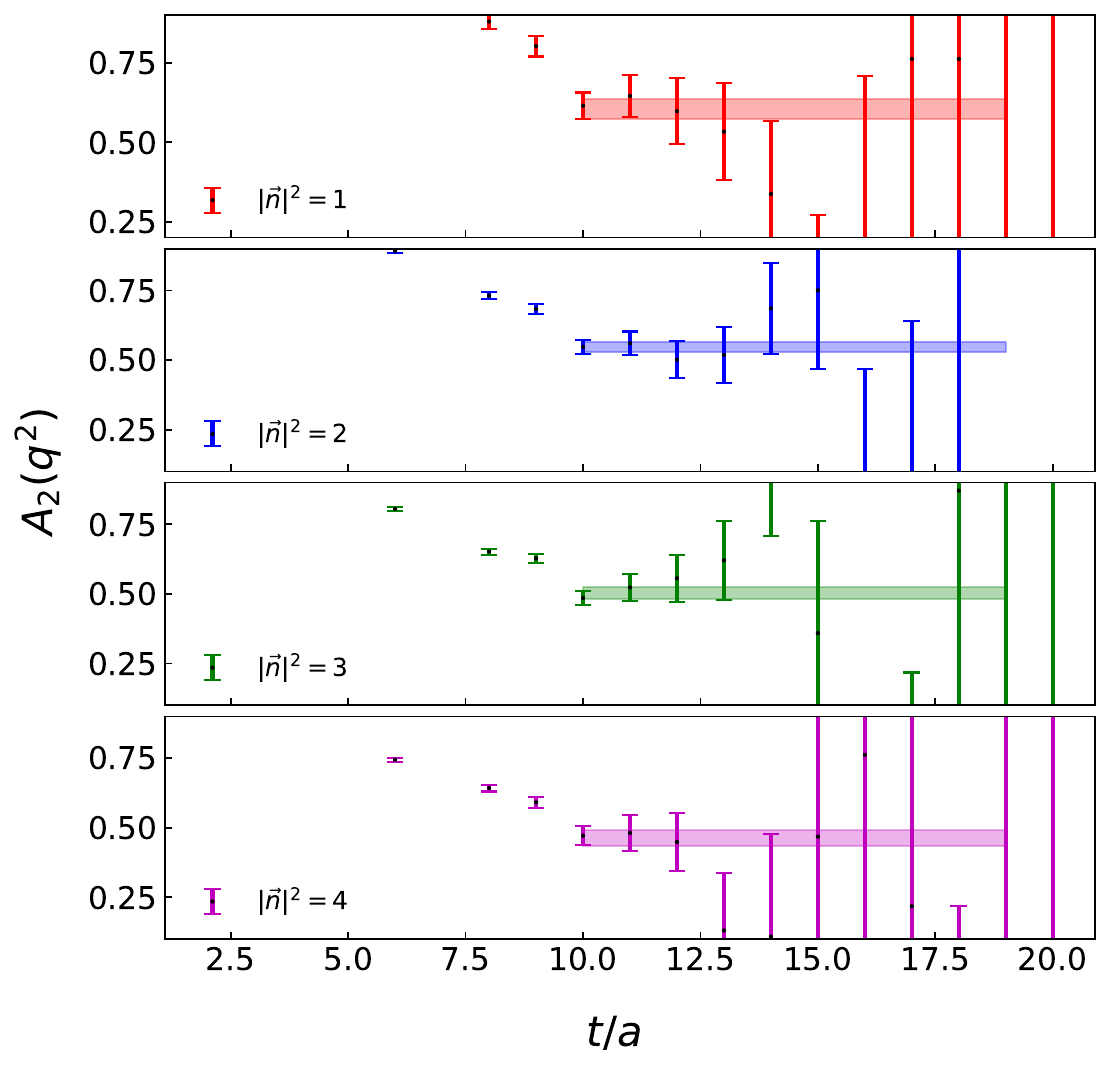}}
\caption{The form factors with different momentum $\vec{p}=2\pi\vec{n}/L,~|\vec{n}|^2=1,2,3,4$ at ensemble G36P29.}
\label{3ptG36P29}
\end{figure}

\renewcommand{\thesubfigure}{(\roman{subfigure})}
\renewcommand{\figurename}{Figure}
\begin{figure}[htp]
\centering  
\subfigure[$V\left(q^2\right)$ at H48P32.]{
\includegraphics[width=7cm]{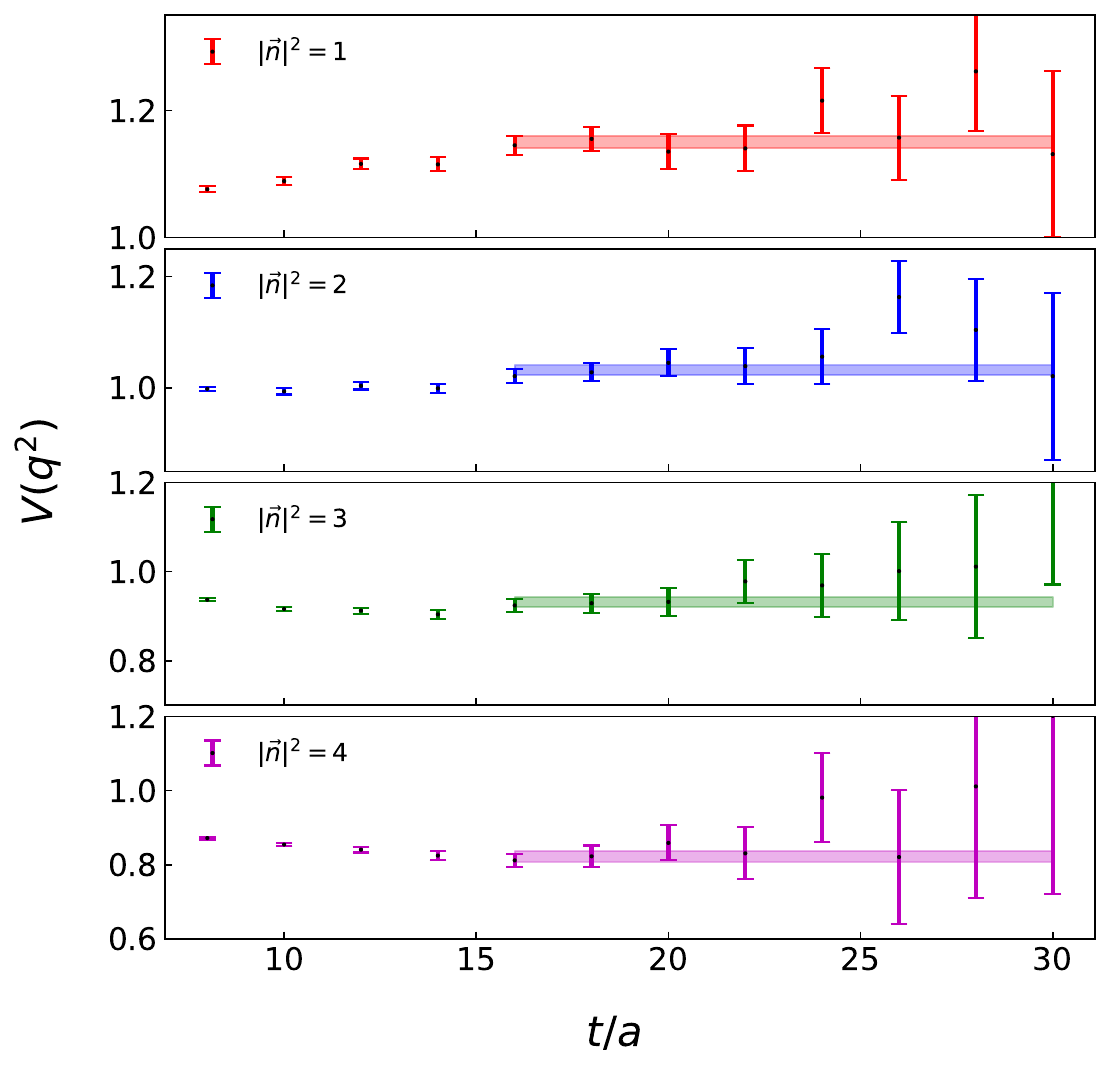}}
\subfigure[$A_0\left(q^2\right)$ at H48P32.]{
\includegraphics[width=7cm]{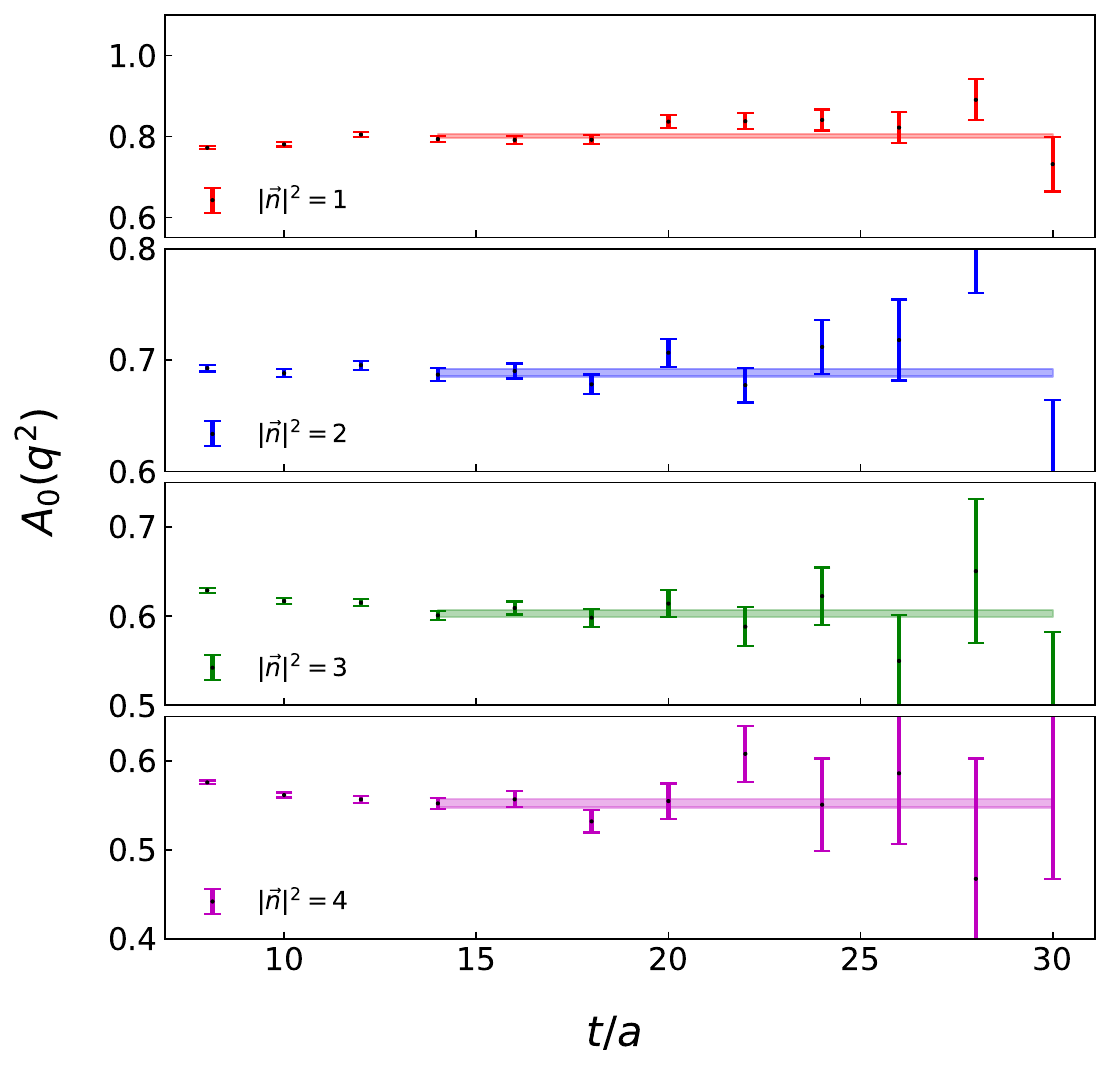}}
\subfigure[$A_1\left(q^2\right)$ at H48P32.]{
\includegraphics[width=7cm]{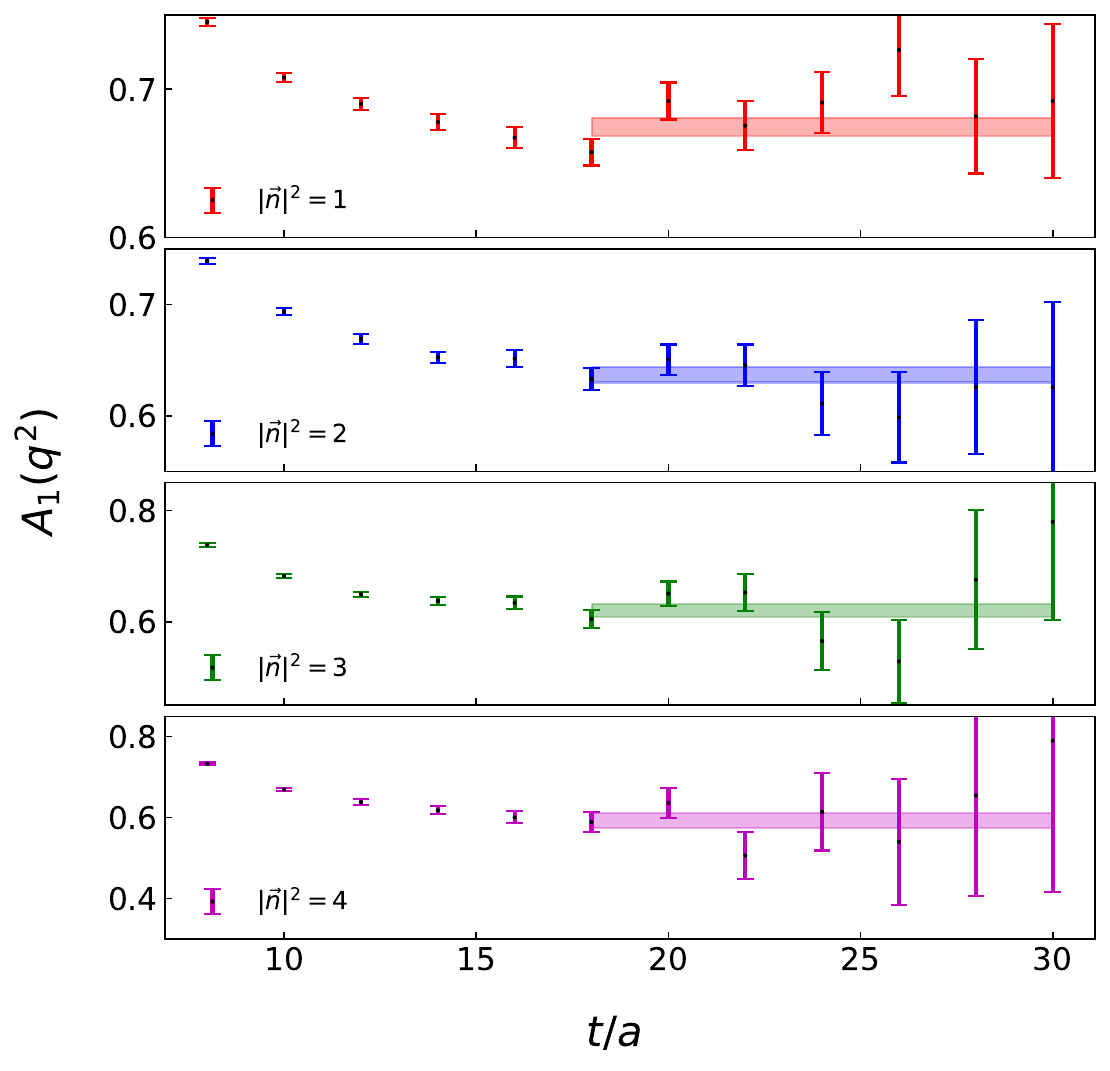}}
\subfigure[$A_2\left(q^2\right)$ at H48P32.]{
\includegraphics[width=7cm]{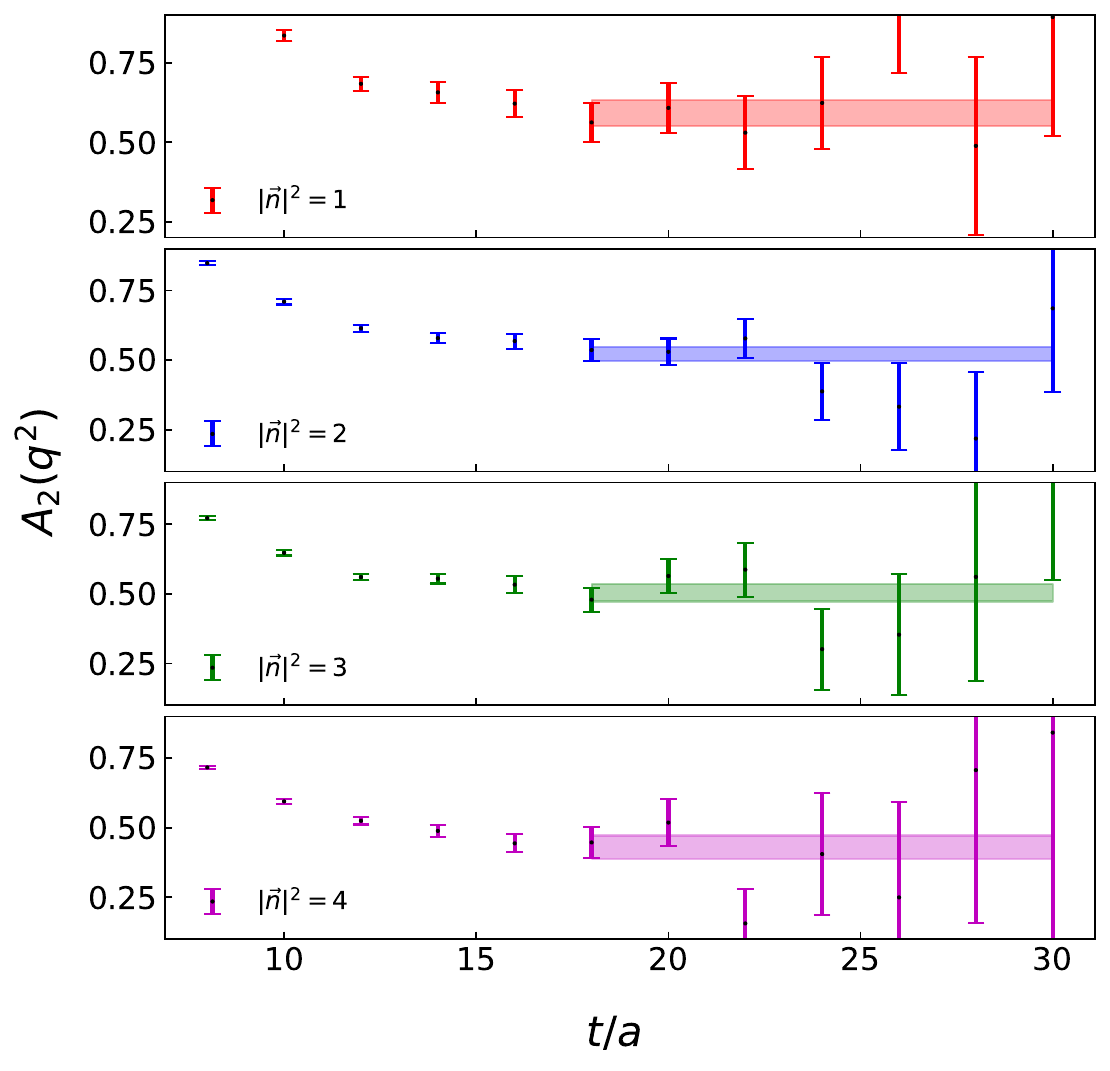}}
\caption{The form factors with different momentum $\vec{p}=2\pi\vec{n}/L,~|\vec{n}|^2=1,2,3,4$ at ensemble H48P32.}
\label{3ptH48P32}
\end{figure}

\renewcommand{\thesubfigure}{(\roman{subfigure})}
\renewcommand{\figurename}{Figure}
\begin{figure}[htp]
\centering  
\subfigure[For C24P29 $D_s$ meson.]{
\includegraphics[width=7cm]{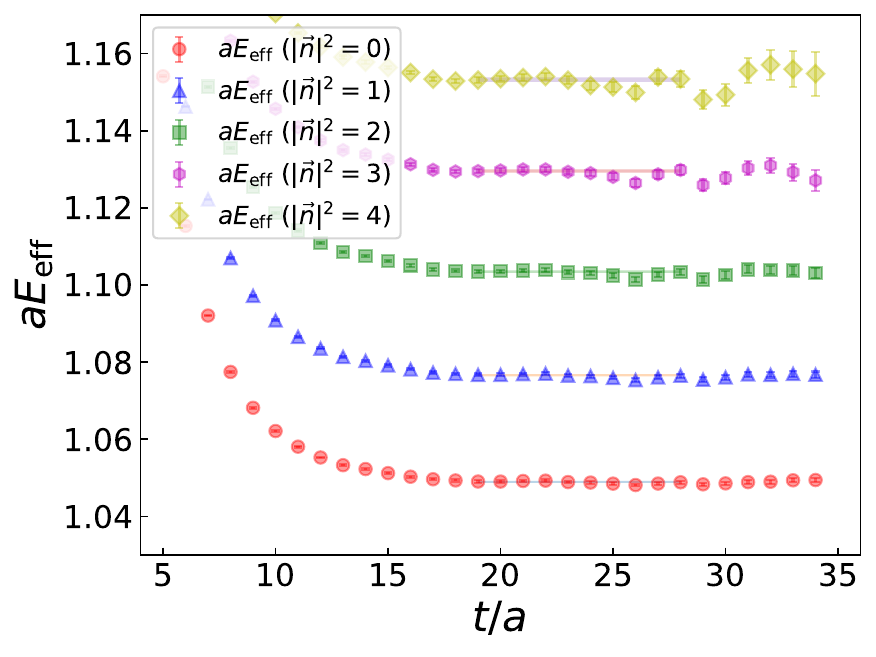}}
\subfigure[For C24P24 $D_s$ meson.]{
\includegraphics[width=7cm]{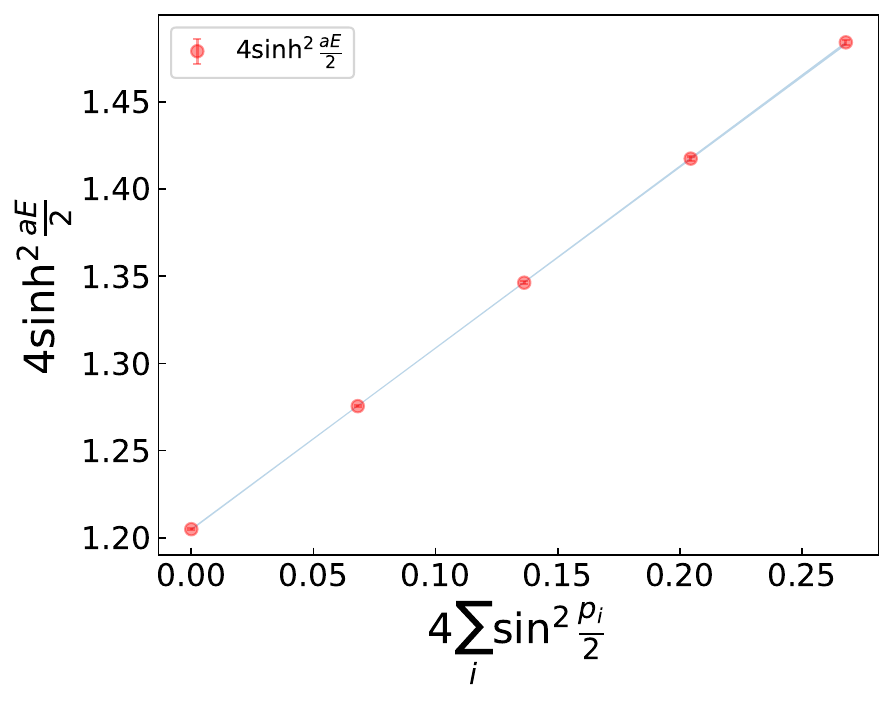}}
\subfigure[For C24P29 $\phi$ meson.]{
\includegraphics[width=7cm]{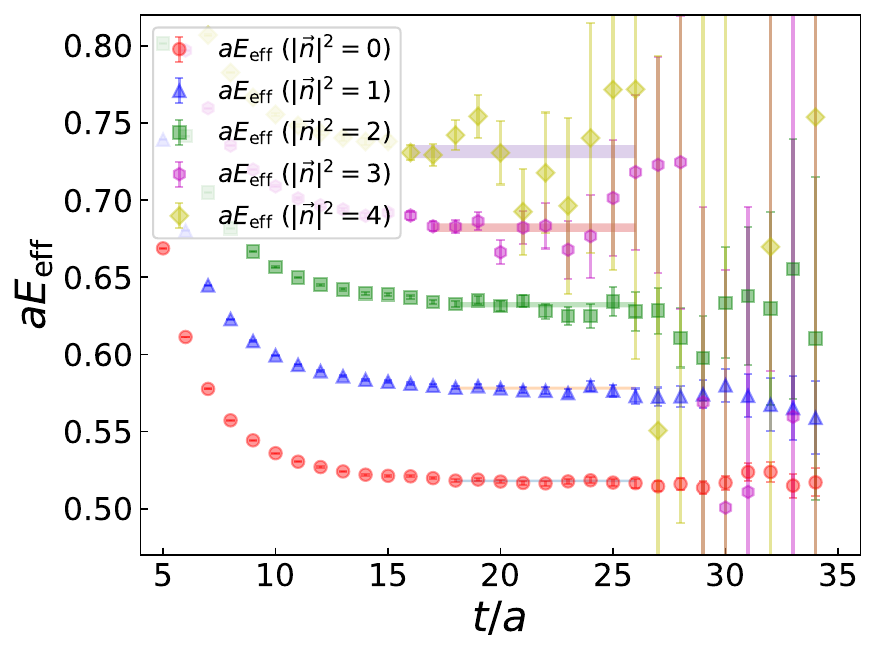}}
\subfigure[For C24P29 $\phi$ meson.]{
\includegraphics[width=7cm]{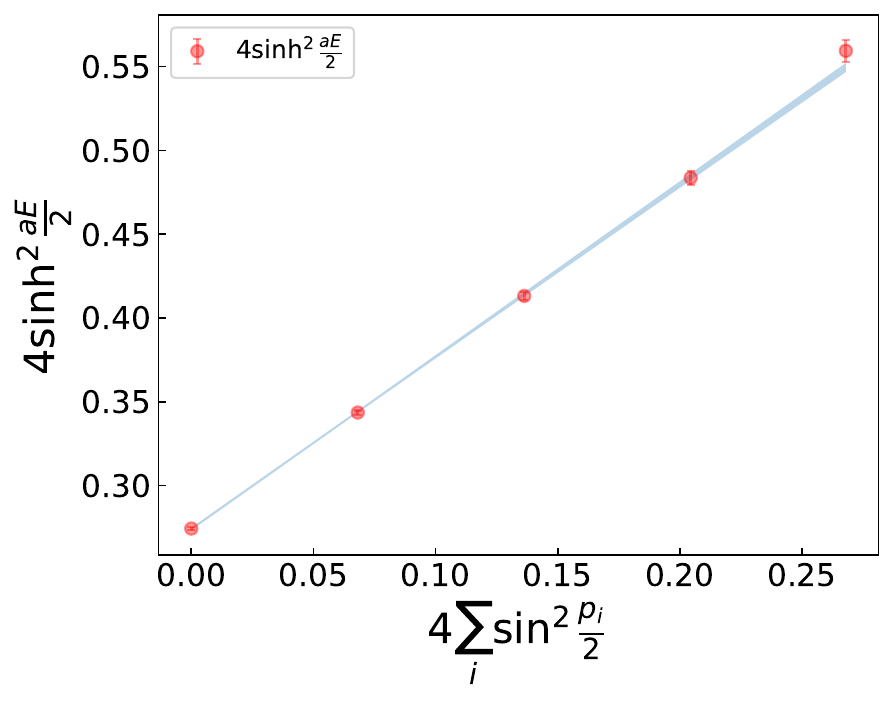}}
\caption{The energy levels of $D_s$ and $\phi$ particles with different momentum $\vec{p}=2\pi\vec{n}/L,~|\vec{n}|^2=0,1,2,3,4$ extracted from two-point functions and the dispersion
relations at ensemble C24P29.}
\label{2ptC24P29}
\end{figure}

\renewcommand{\thesubfigure}{(\roman{subfigure})}
\renewcommand{\figurename}{Figure}
\begin{figure}[htp]
\centering  
\subfigure[For C32P23 $D_s$ meson.]{
\includegraphics[width=7cm]{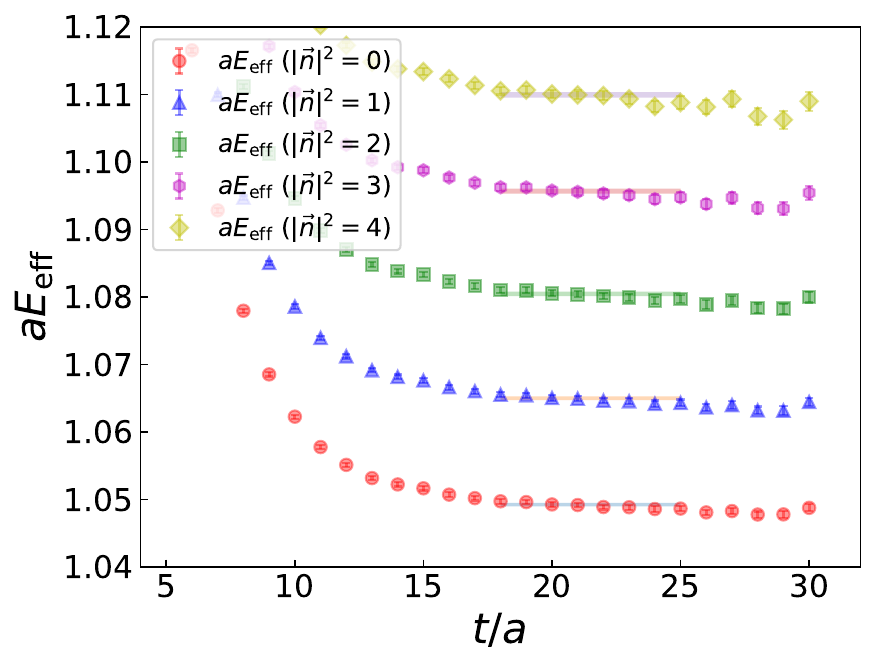}}
\subfigure[For C32P23 $D_s$ meson.]{
\includegraphics[width=7cm]{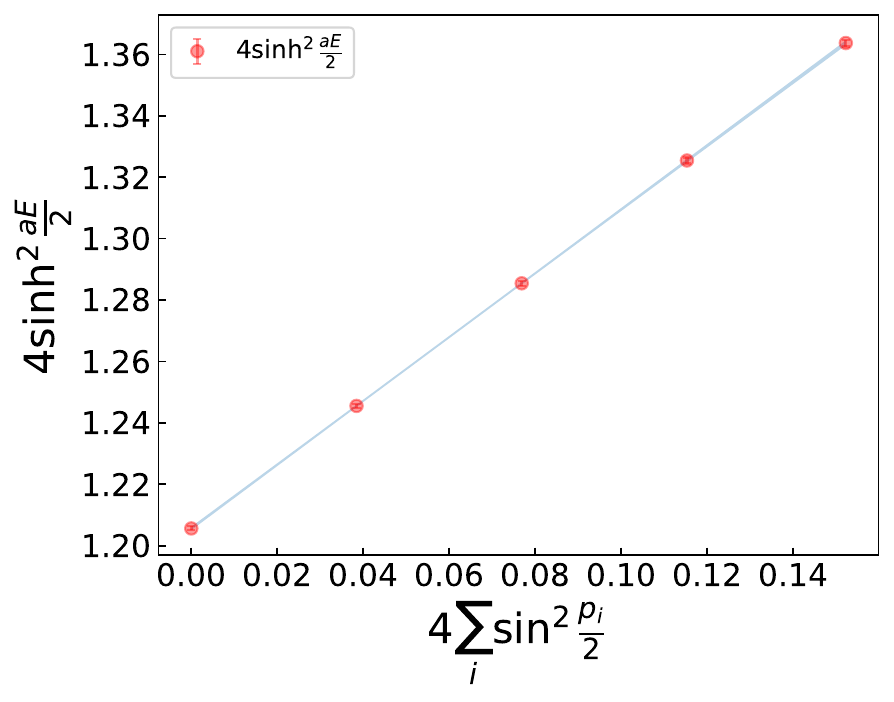}}
\subfigure[For C32P23 $\phi$ meson.]{
\includegraphics[width=7cm]{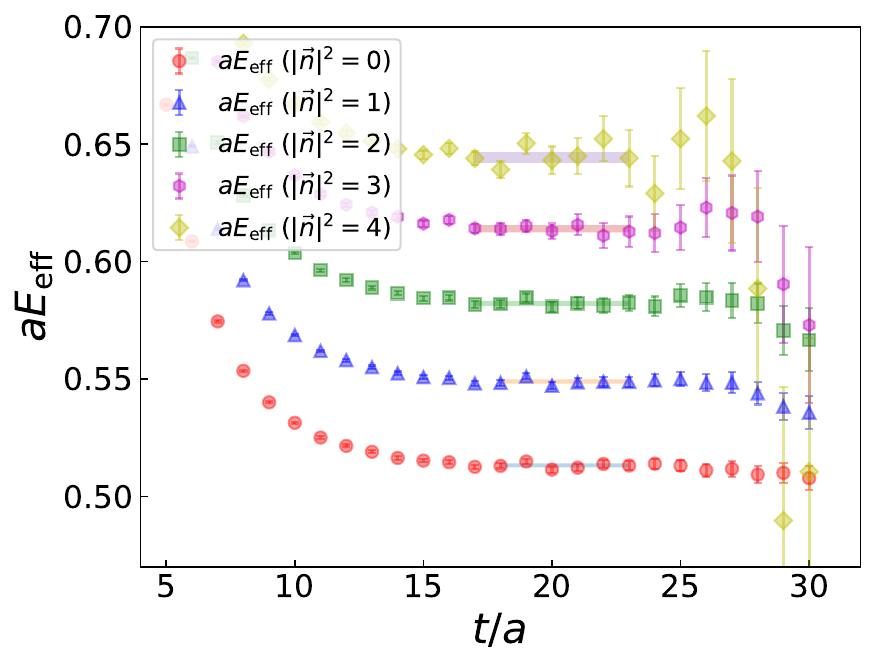}}
\subfigure[For C32P23 $\phi$ meson.]{
\includegraphics[width=7cm]{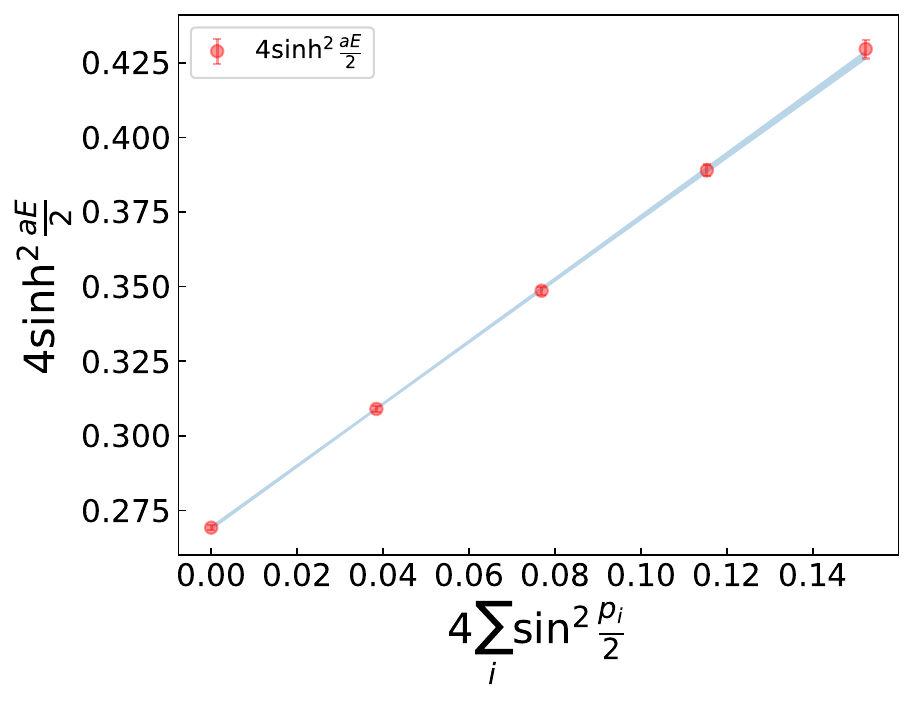}}
\caption{The energy levels of $D_s$ and $\phi$ particles with different momentum $\vec{p}=2\pi\vec{n}/L,~|\vec{n}|^2=0,1,2,3,4$ extracted from two-point functions and the dispersion
relations at ensemble C32P23.}
\label{2ptC32P23}
\end{figure}

\renewcommand{\thesubfigure}{(\roman{subfigure})}
\renewcommand{\figurename}{Figure}
\begin{figure}[htp]
\centering  
\subfigure[For C32P29 $D_s$ meson.]{
\includegraphics[width=7cm]{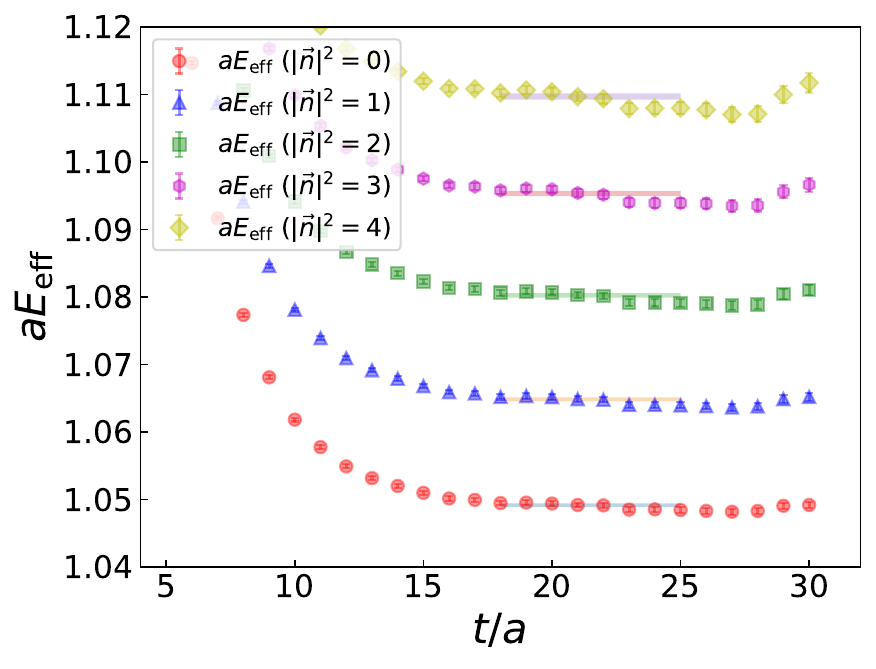}}
\subfigure[For C32P29 $D_s$ meson.]{
\includegraphics[width=7cm]{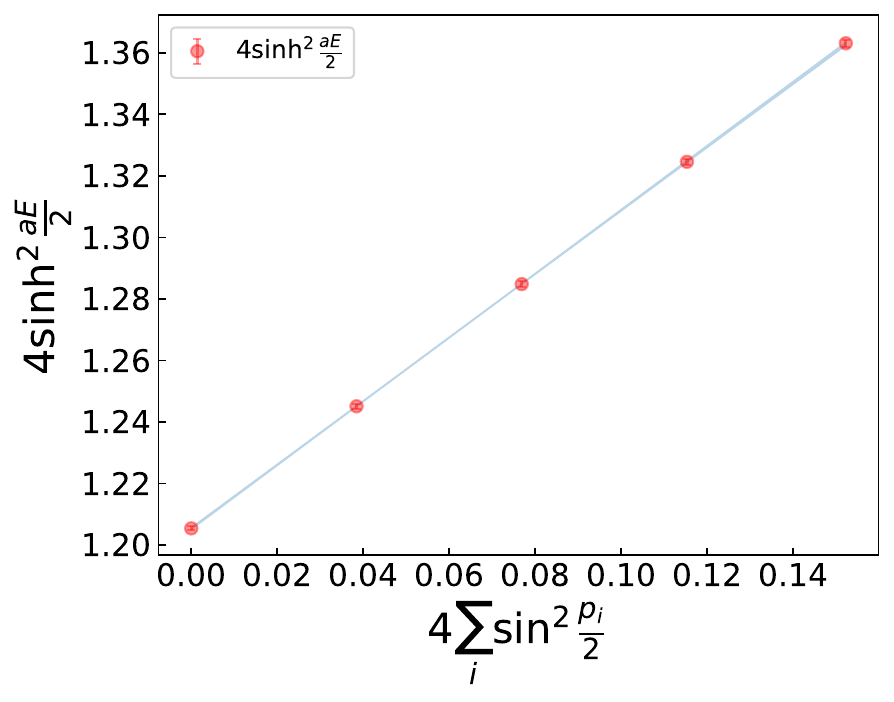}}
\subfigure[For C32P29 $\phi$ meson.]{
\includegraphics[width=7cm]{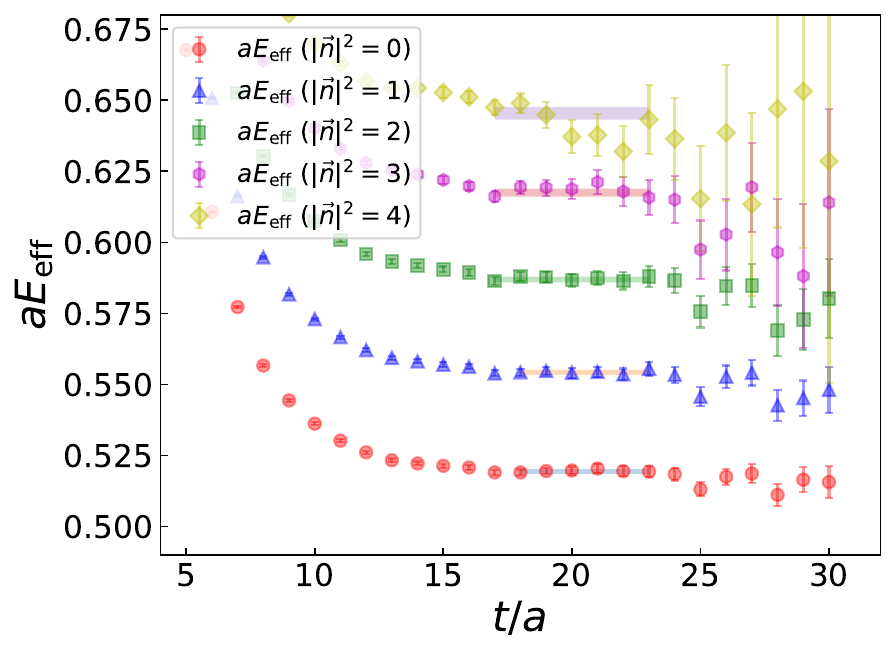}}
\subfigure[For C32P29 $\phi$ meson.]{
\includegraphics[width=7cm]{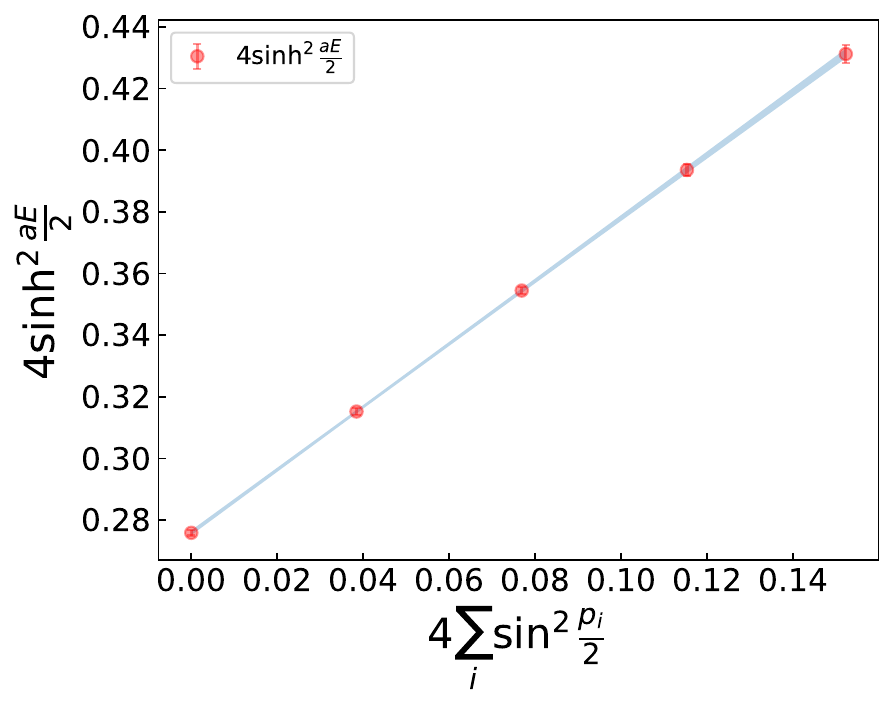}}
\caption{The energy levels of $D_s$ and $\phi$ particles with different momentum $\vec{p}=2\pi\vec{n}/L,~|\vec{n}|^2=0,1,2,3,4$ extracted from two-point functions and the dispersion
relations at ensemble C32P29.}
\label{2ptC32P29}
\end{figure}

\renewcommand{\thesubfigure}{(\roman{subfigure})}
\renewcommand{\figurename}{Figure}
\begin{figure}[htp]
\centering  
\subfigure[For F32P30 $D_s$ meson.]{
\includegraphics[width=7cm]{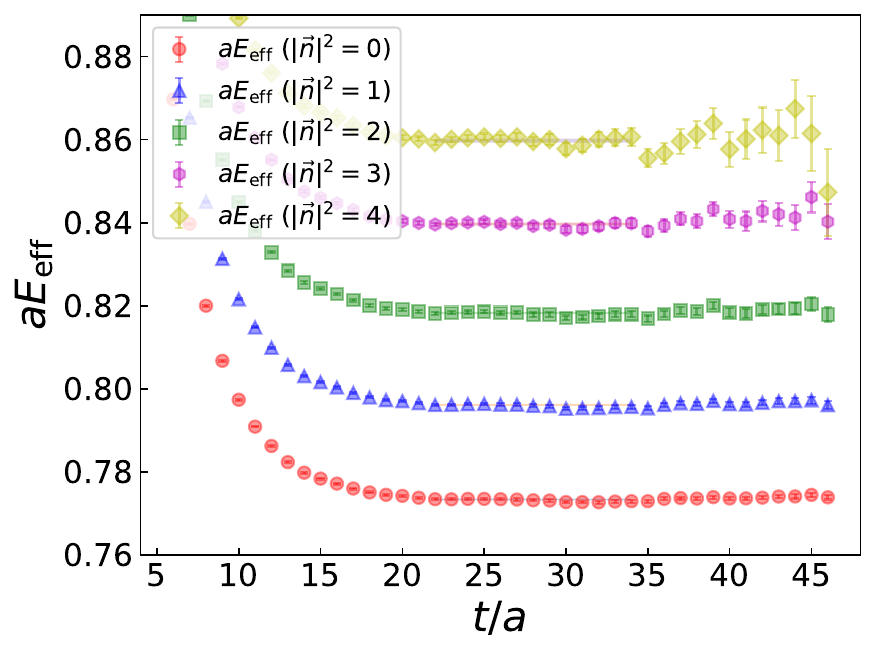}}
\subfigure[For F32P30 $D_s$ meson.]{
\includegraphics[width=7cm]{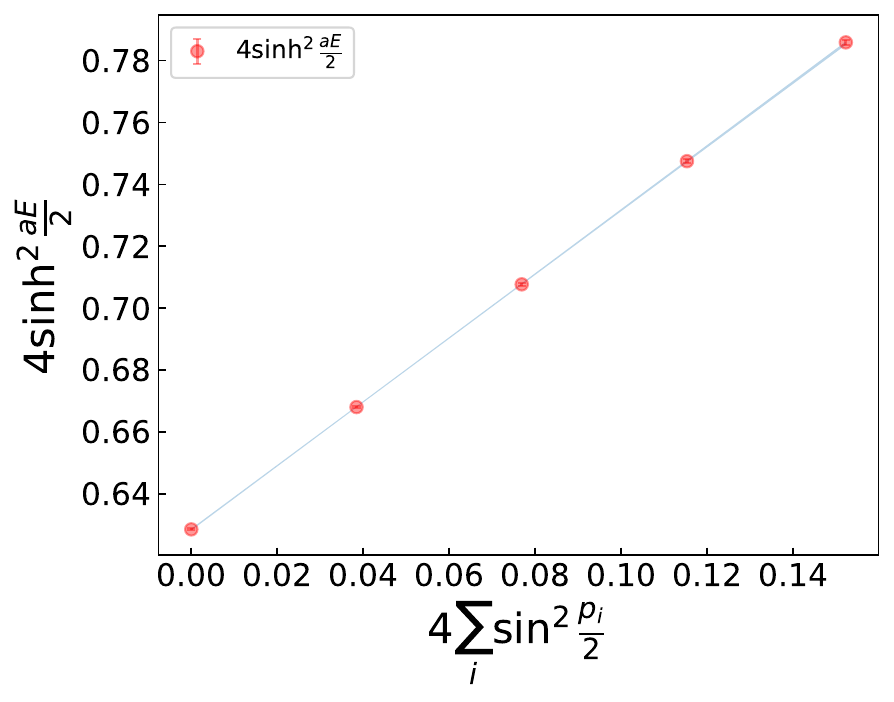}}
\subfigure[For F32P30 $\phi$ meson.]{
\includegraphics[width=7cm]{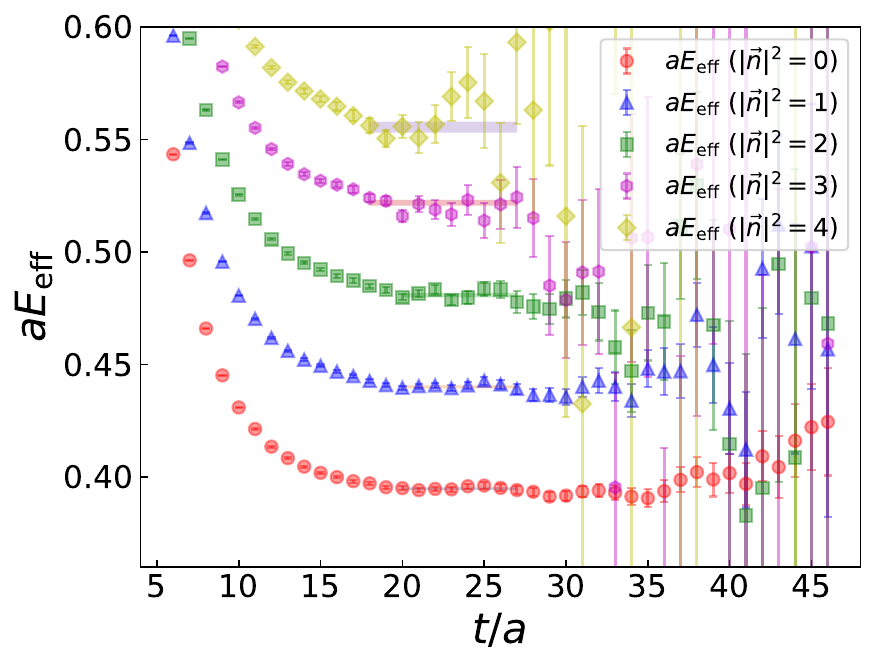}}
\subfigure[For F32P30 $\phi$ meson.]{
\includegraphics[width=7cm]{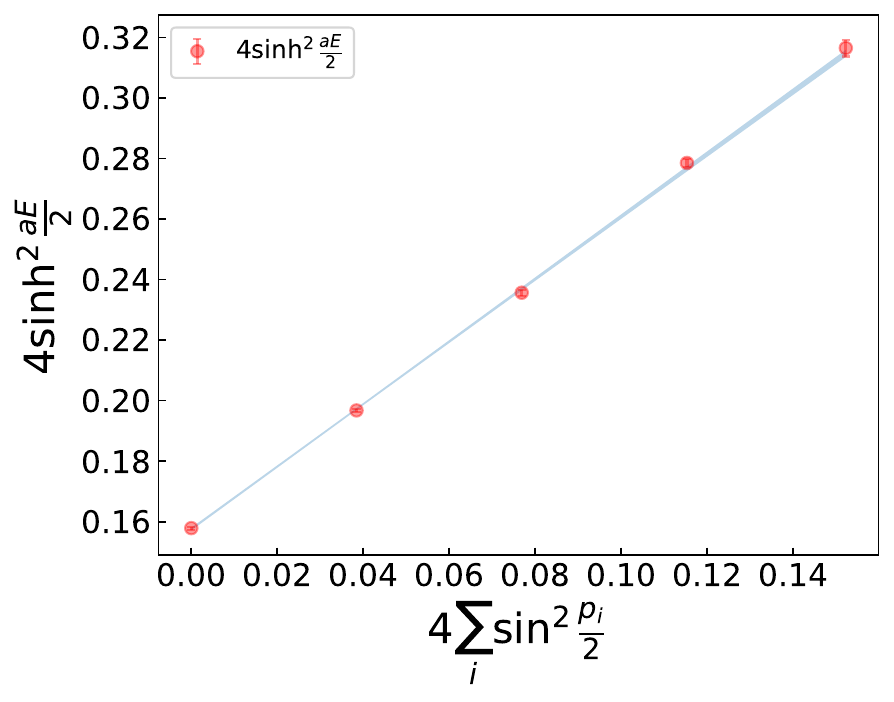}}
\caption{The energy levels of $D_s$ and $\phi$ particles with different momentum $\vec{p}=2\pi\vec{n}/L,~|\vec{n}|^2=0,1,2,3,4$ extracted from two-point functions and the dispersion
relations at ensemble F32P30.}
\label{2ptF32P30}
\end{figure}

\renewcommand{\thesubfigure}{(\roman{subfigure})}
\renewcommand{\figurename}{Figure}
\begin{figure}[htp]
\centering  
\subfigure[For F48P21 $D_s$ meson.]{
\includegraphics[width=7cm]{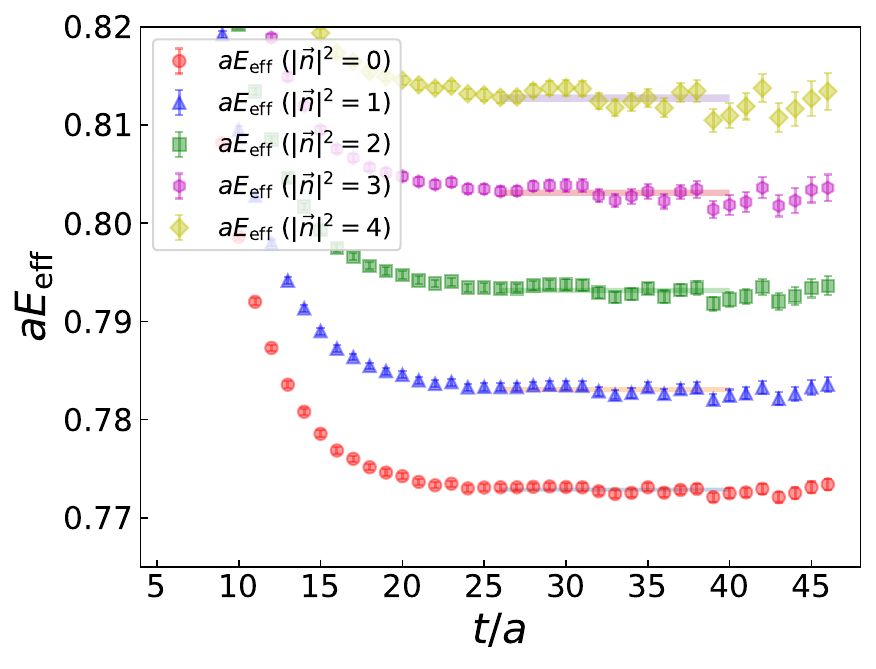}}
\subfigure[For F48P21 $D_s$ meson.]{
\includegraphics[width=7cm]{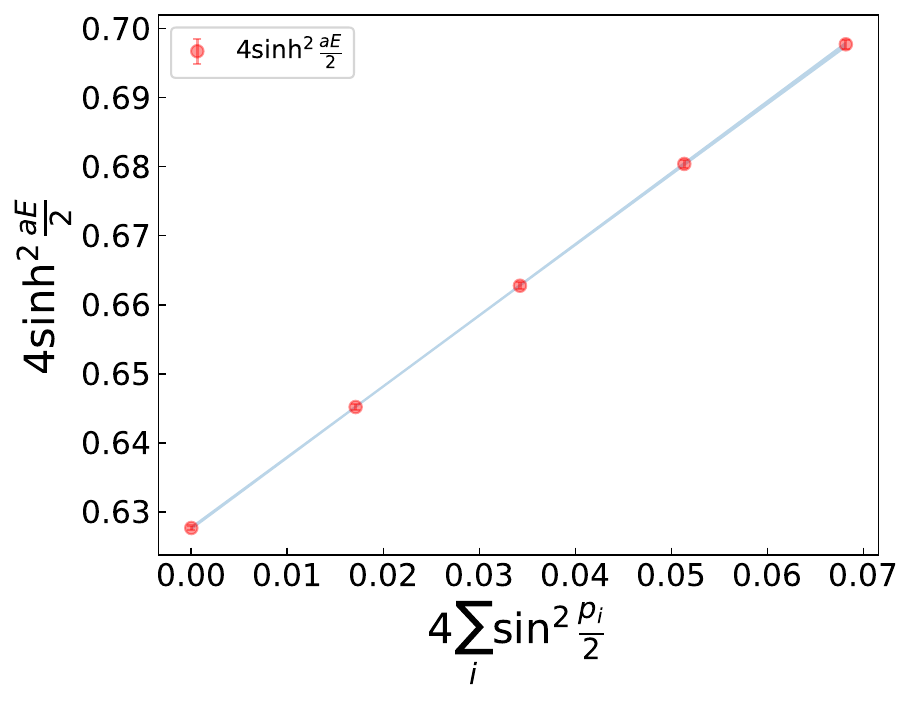}}
\subfigure[For F48P21 $\phi$ meson.]{
\includegraphics[width=7cm]{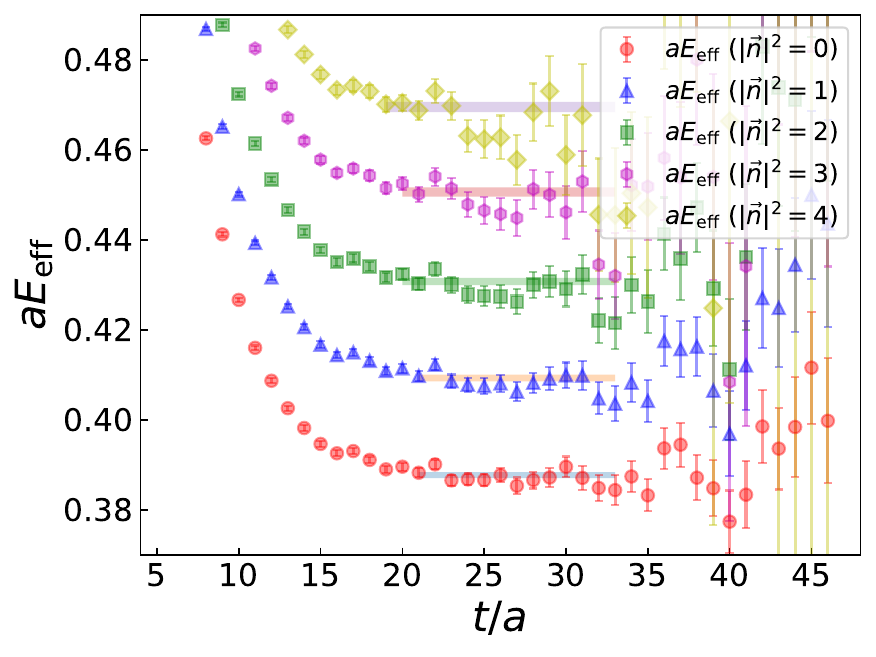}}
\subfigure[For F48P21 $\phi$ meson.]{
\includegraphics[width=7cm]{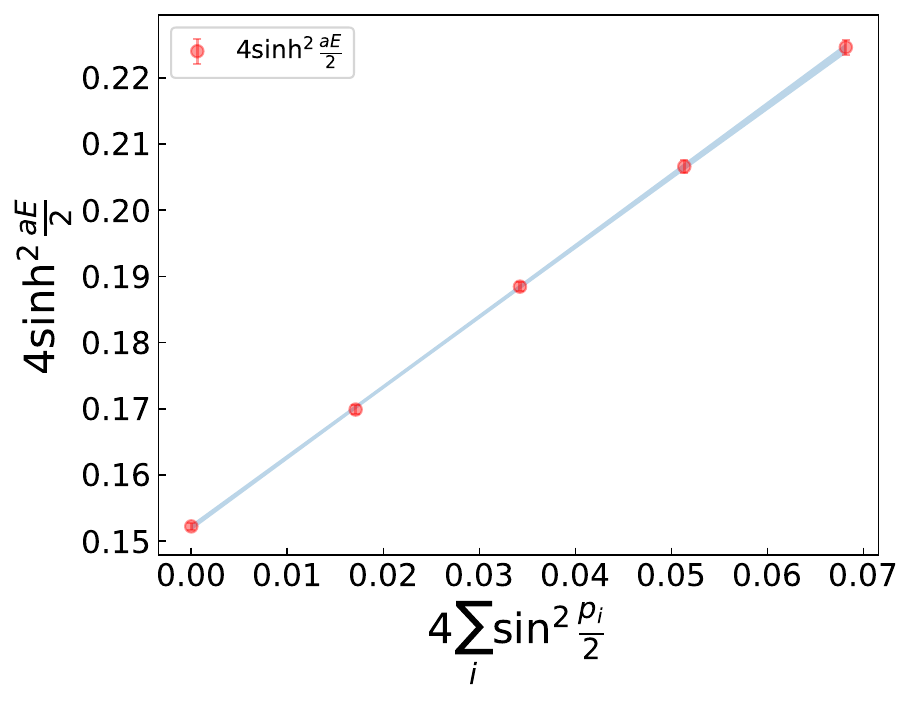}}
\caption{The energy levels of $D_s$ and $\phi$ particles with different momentum $\vec{p}=2\pi\vec{n}/L,~|\vec{n}|^2=0,1,2,3,4$ extracted from two-point functions and the dispersion
relations at ensemble F48P21.}
\label{2ptF48P21}
\end{figure}

\renewcommand{\thesubfigure}{(\roman{subfigure})}
\renewcommand{\figurename}{Figure}
\begin{figure}[htp]
\centering  
\subfigure[For G36P29 $D_s$ meson.]{
\includegraphics[width=7cm]{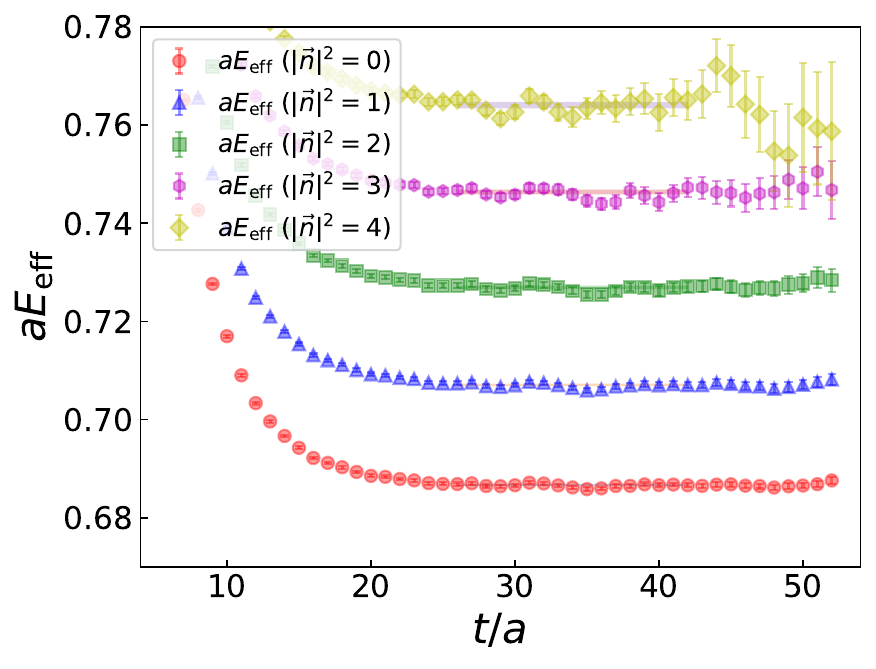}}
\subfigure[For G36P29 $D_s$ meson.]{
\includegraphics[width=7cm]{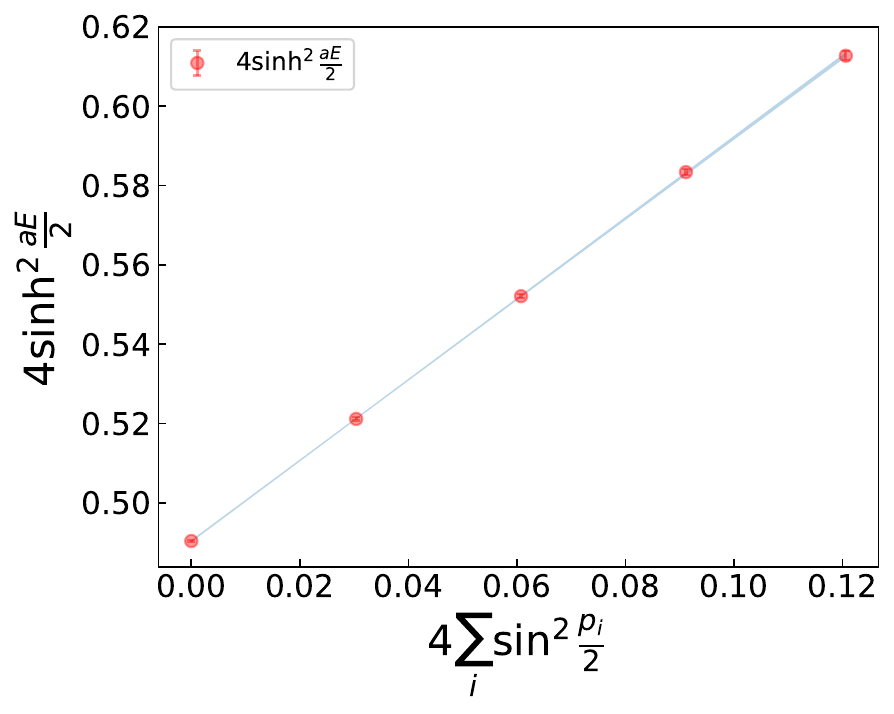}}
\subfigure[For G36P29 $\phi$ meson.]{
\includegraphics[width=7cm]{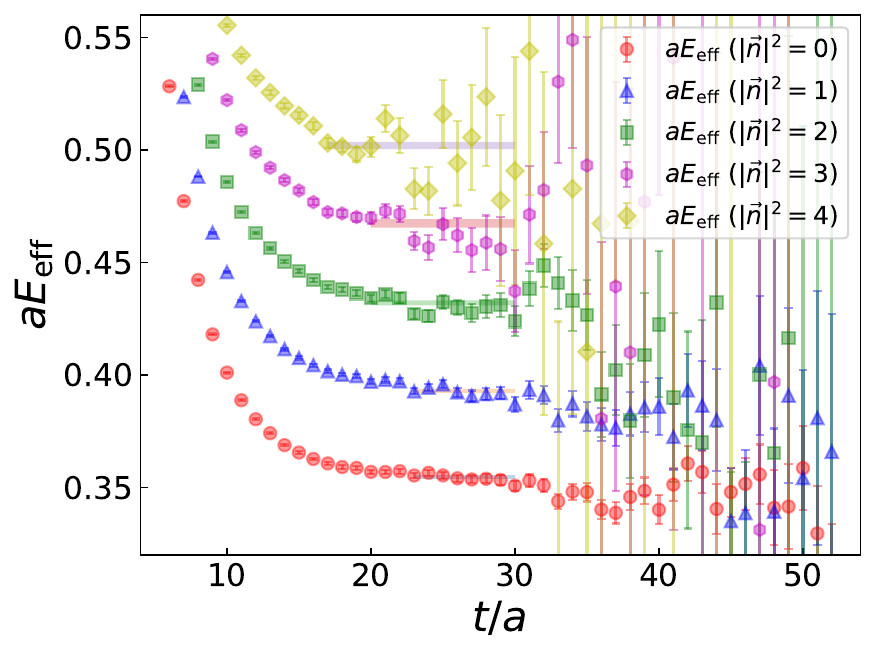}}
\subfigure[For G36P29 $\phi$ meson.]{
\includegraphics[width=7cm]{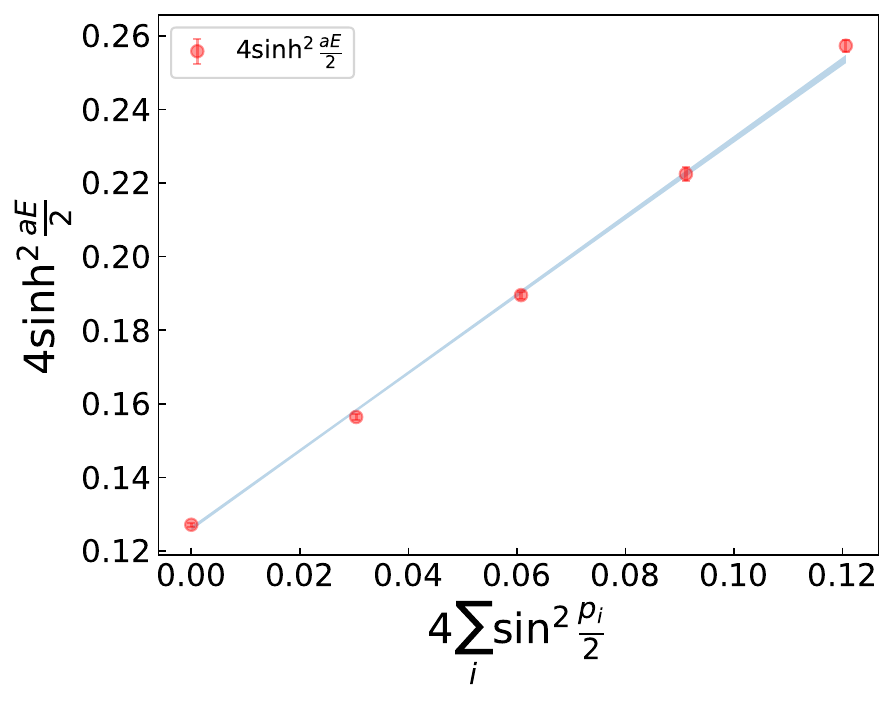}}
\caption{The energy levels of $D_s$ and $\phi$ particles with different momentum $\vec{p}=2\pi\vec{n}/L,~|\vec{n}|^2=0,1,2,3,4$ extracted from two-point functions and the dispersion
relations at ensemble G36P29.}
\label{2ptG36P29}
\end{figure}

\renewcommand{\thesubfigure}{(\roman{subfigure})}
\renewcommand{\figurename}{Figure}
\begin{figure}[htp]
\centering  
\subfigure[For H48P32 $D_s$ meson.]{
\includegraphics[width=7cm]{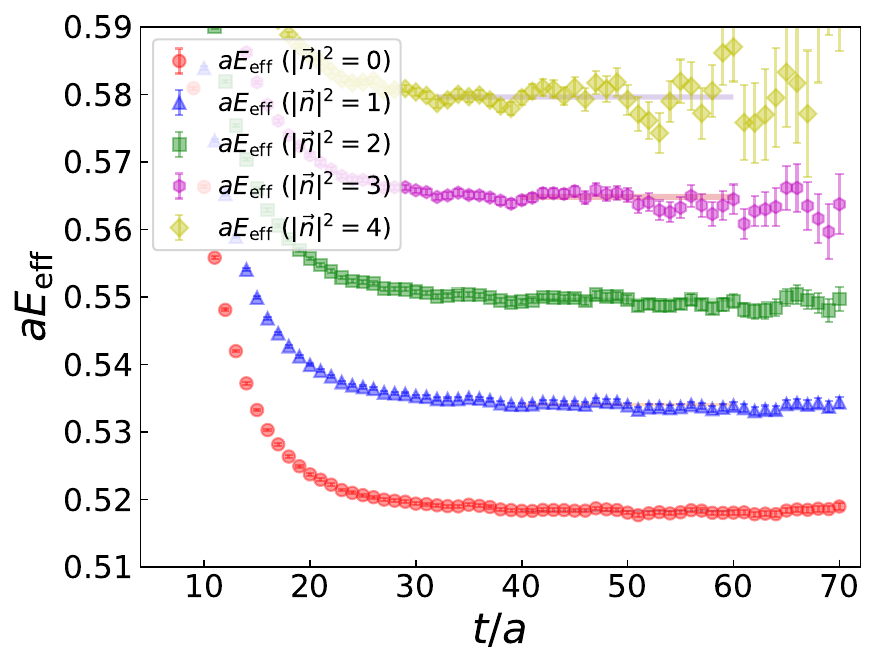}}
\subfigure[For H48P32 $D_s$ meson.]{
\includegraphics[width=7cm]{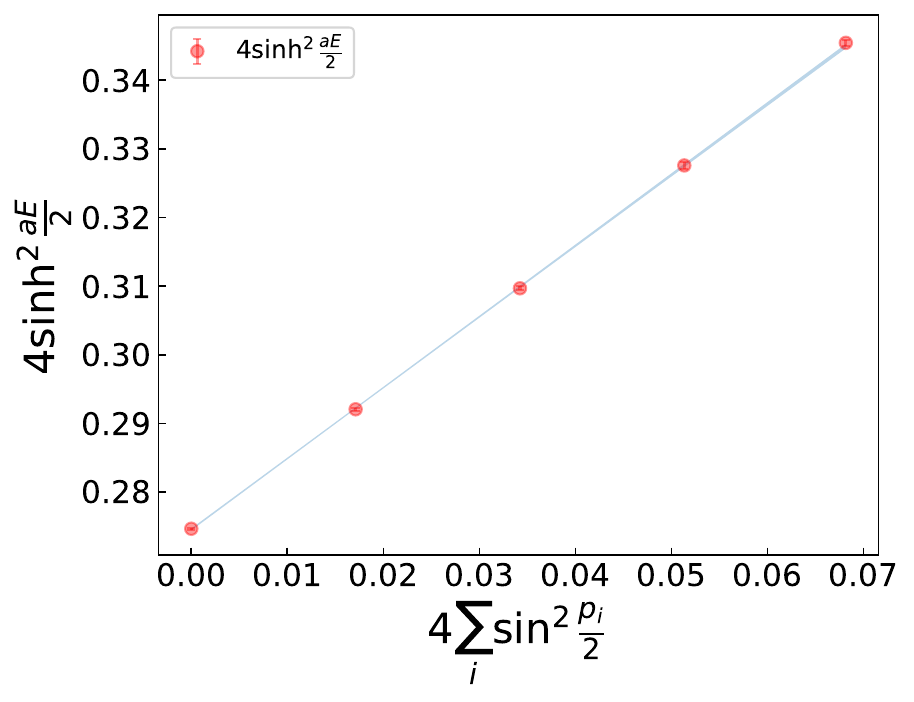}}
\subfigure[For H48P32 $\phi$ meson.]{
\includegraphics[width=7cm]{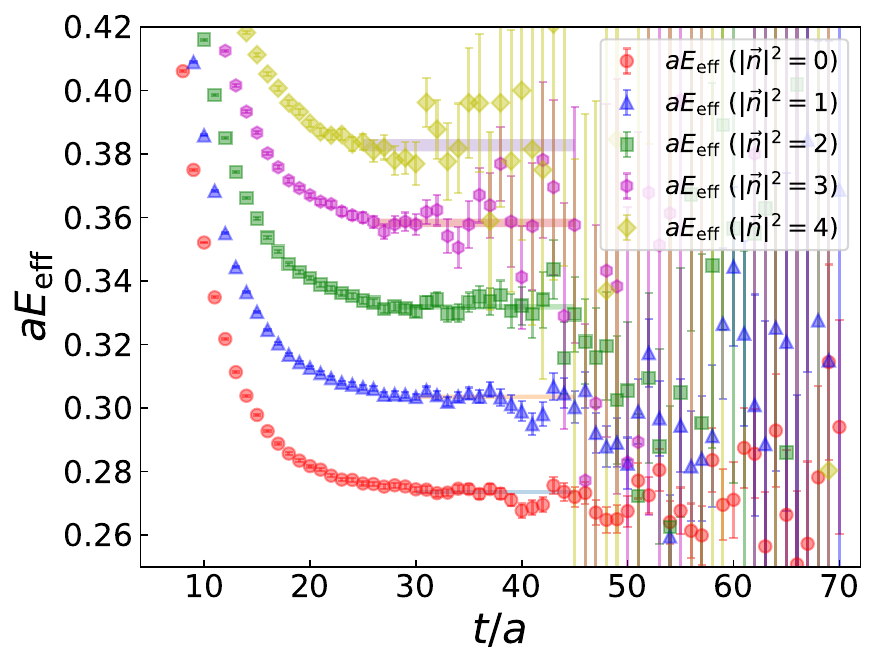}}
\subfigure[For H48P32 $\phi$ meson.]{
\includegraphics[width=7cm]{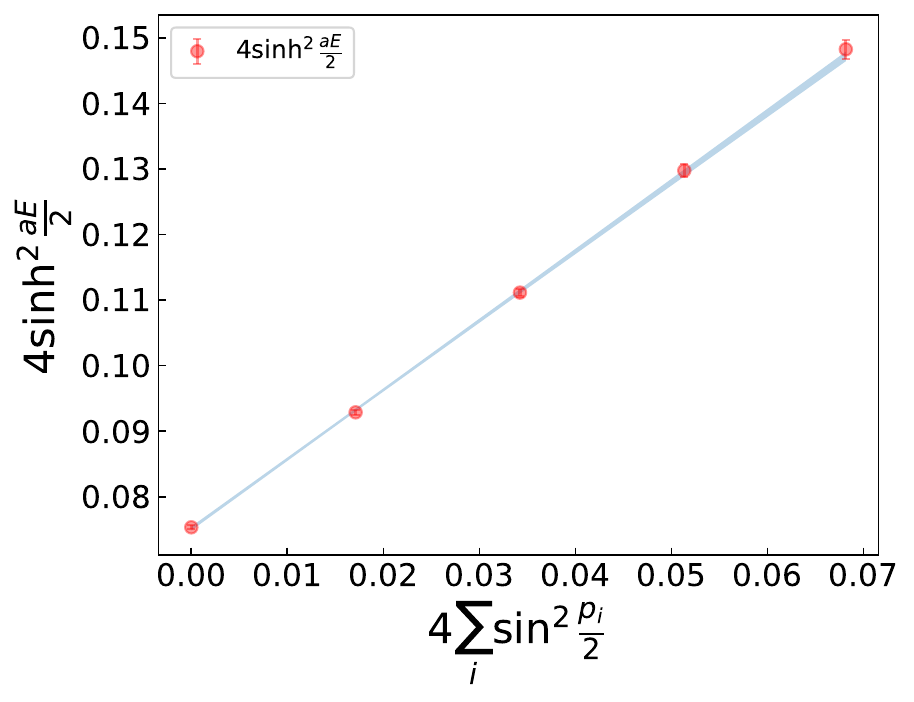}}
\caption{The energy levels of $D_s$ and $\phi$ particles with different momentum $\vec{p}=2\pi\vec{n}/L,~|\vec{n}|^2=0,1,2,3,4$ extracted from two-point functions and the dispersion
relations at ensemble H48P32.}
\label{2ptH48P32}
\end{figure}

\clearpage
\bibliographystyle{JHEP}      
\bibliography{paper}

\end{document}